\PassOptionsToPackage{style=	
numeric            
}{biblatex}	

\documentclass[12pt,oneside,letterpaper,abstract=on,headings=optiontoheadandtoc]{scrartcl}   
\usepackage[utf8]{inputenc} 	
\usepackage[american]{babel}	
\usepackage[autostyle=true]{csquotes}
\usepackage{microtype}  		
\usepackage{geometry}  			
\usepackage{graphicx}			
\usepackage{amsmath}			
\usepackage{marvosym}
\usepackage{bm}					
\usepackage{mathtools}			
\usepackage[ntheorem]{empheq}	
\usepackage{bbm}				
\usepackage{amsfonts} 			
\usepackage{yfonts}				
\usepackage{hyperref}			
\usepackage[thref,amsmath,thmmarks,hyperref]{ntheorem}			
\usepackage[backend=biber]{biblatex}  
\usepackage[shortlabels]{enumitem}
\usepackage{accents}			
\usepackage[whole]{bxcjkjatype}		
						
\newcommand{\abs}[1]{\left\lvert #1 \right\rvert}	
\newcommand{\norm}[1]{\left\lVert #1 \right\rVert}	
\newcommand{\set}[2]{\left\lbrace #1 \middle\vert #2 \right\rbrace}
	
\newcommand{\inp}[2]{\left\langle #1 \text{,} #2 \right\rangle} 

\newcommand{\partd}[2]{\frac{\partial #1}{\partial #2}}	
\renewcommand{\d}{\mathrm{d}}				
\renewcommand{\abs}[1]{\left\lvert #1 \right\rvert}

\newcommand{\tp}{\otimes}				
\newcommand{\cross}{\times}				
\newcommand{\R}{\mathbb{R}}				
\newcommand{\C}{\mathbb{C}}				
\newcommand{\Z}{\mathbb{Z}}				
\newcommand{\N}{\mathbb{N}}				
\newcommand{\K}{\mathbb{K}}				

\newcommand{\Id}{\mathbbm 1}			
\newcommand{\iu}{\mathfrak{i}}		
\DeclareMathSymbol{\varnothing}{\mathord}{AMSb}{"3F}
\renewcommand{\emptyset}{\varnothing}	
\DeclareMathSymbol{\upharpoonright} {\mathrel}{AMSa}{"16}

\DeclareMathOperator{\dom}{dom}
\DeclareMathOperator{\curl}{curl}

\DeclareMathOperator*{\esssup}{ess \, sup}     

\let\Re\relax
\DeclareMathOperator{\Re}{Re}
\let\Im\relax
\DeclareMathOperator{\Im}{Im}

\DeclareMathOperator{\supp}{supp}    		

\DeclareMathSymbol{\square}
{\mathord}{AMSa}{"03}
\DeclareMathSymbol{\blacksquare} {\mathord}{AMSa}{"04}
\newcommand{\evat}[1]{\negthickspace\upharpoonright_{#1}}

\renewcommand{\qedsymbol}{
								$\blacksquare$
							}
\newcommand{\oendmark}{
						$\diamondsuit$
						}

\theoremstyle{break}
\theoremheaderfont{\normalfont\bfseries}
\theorembodyfont{\normalfont 
				}
\theoremseparator{}
\theoremnumbering{arabic}
\theoremsymbol{\oendmark}
\newtheorem{Proposition}{Proposition}[section]	
\newtheorem{Theorem}[Proposition]{Theorem}
\newtheorem{Lemma}[Proposition]{Lemma}
\newtheorem{Corollary}[Proposition]{Corollary}

\theoremsymbol{\oendmark}
\newtheorem{Definition}[Proposition]{Definition}

\theoremsymbol{\oendmark}
\newtheorem{Remark}[Proposition]{Remark}

\newtheorem{Example}[Proposition]{Example}

\theoremheaderfont{\normalfont\scshape}
\theorembodyfont{
		\normalfont
				}
\theoremstyle{nonumberplain}
\theoremseparator{}
\theoremsymbol{\qedsymbol} 
\newtheorem{Proof}{Proof}
\numberwithin{equation}{section}
\allowdisplaybreaks					
									
\renewbibmacro*{journal}{%
  \iffieldundef{shortjournal}
    {%
      \iffieldundef{journaltitle}
        {}
        {%
          \printtext[journaltitle]
            {%
              \printfield[titlecase]{journaltitle}%
              \setunit{\subtitlepunct}%
              \printfield[titlecase]{journalsubtitle}%
             }%
         }%
    }
    {\printtext[journaltitle]{\printfield{shortjournal}}}%
}

\begin{document}


\title{
		Towards a mathematical Theory of the Madelung Equations
		} 	
\subtitle{Takabayasi's Quantization 
			Condition, Quantum Quasi-Irrotationality,  
			variational Formulations, and 
			the Wallstrom Phenomenon}
			
\author{Maik Reddiger%
	\thanks{%
	Department of Physics and Astronomy, and 
	Department of Chemistry and Biochemistry, Texas Tech University, 
	Box 41061, 
	Lubbock, Texas 79409-1061, USA. \, 
	\Letter \, \href{mailto:maik.reddiger@ttu.edu}
	{\texttt{maik.reddiger@ttu.edu}} 
	\, \mbox{\Telefon \, +1-806-742-3067} } 
	\href{https://orcid.org/0000-0002-0485-5044}{
	\includegraphics[width=0.8 em, height=0.8 em]{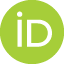}} \\
	\and 
	Bill Poirier%
	\thanks{%
	Department of Chemistry and Biochemistry, and 
	Department of Physics and Astronomy, Texas Tech University, 
	Box 41061, 
	Lubbock, Texas 79409-1061, USA. \,
	\Letter \, \href{mailto:bill.poirier@ttu.edu}
	{\texttt{bill.poirier@ttu.edu}} 
	\, \mbox{\Telefon \, +1-806-834-3099} } 
	\href{https://orcid.org/0000-0001-8277-746X}{
	\includegraphics[width=0.8 em, height=0.8 em]{orcid.png}}\\ 
}  

\date{July 22, 2022} 

\maketitle 

\begin{abstract} 
	\noindent
	Even though the Madelung equations are central 
	to many `classical' approaches to the foundations 
	of quantum mechanics such as Bohmian and stochastic 
	mechanics, no coherent mathematical theory has been developed 
	so far for this system of partial differential equations. 
	Wallstrom prominently raised objections against the 
	Madelung equations, aiming to show 
	that no such theory exists in
	which the system is well-posed and in 
	which the Schrödinger equation is recovered without 
	the imposition of an additional `ad hoc 
	quantization condition'---like the one proposed by 
	Takabayasi. The primary objective of our work is to clarify in which sense  
	Wallstrom's objections are justified and in which sense they are not, with 
	a view on the existing literature. 
	We find that it may be possible to construct a  
	mathematical theory of the Madelung equations which is satisfactory 
	in the aforementioned sense, though more mathematical research is required. 
	
	More specifically, this work makes five main contributions to the 
	subject: First, we rigorously prove that Takabayasi's quantization 
	condition holds for arbitrary $C^1$-wave functions. Nonetheless, we explain why 
	there are serious doubts with regards to its applicability in the general theory 
	of quantum mechanics. Second, we argue that the Madelung equations 
	need to be understood in the sense of distributions. 
	Accordingly, we review a variational formulation due to Gasser and 
	Markowich and suggest a second one based on Nelson's equations. 
	Third, we show that the common examples that motivate 
	Takabayasi's condition do not satisfy one of the Madelung equations 
	in the distributional sense, leading us to introduce the concept 
	of `quantum quasi-irrotationality'. This terminology was inspired 
	by a statement due to Schönberg. 
	Fourth, we construct  
	explicit `non-quantized' strong solutions to the 
	Madelung equations in $2$ dimensions, which were claimed to exist by Wallstrom, 
	and provide an analysis thereof. 
	Fifth, we demonstrate that Wallstrom's argument for non-uniqueness of 
	solutions of the Madelung equations, termed the `Wallstrom phenomenon', 
	is ultimately due to a failure of quantum mechanics to discern 
	physically equivalent, yet mathematically inequivalent 
	states---an issue that finds its historic origins in the 
	Pauli problem. 
\end{abstract}
\noindent

\begin{center}
\footnotesize{\emph{Keywords:} 
Madelung equations \, - \, 
Schrödinger equation  \, - \,
Quantum potential \\	  
Quantum vortices \, - \, 
Stochastic mechanics
\\[0.5\baselineskip]
} 
\end{center} 

\tableofcontents

\pagestyle{headings}    

\section{Introduction and main results}
\label{sec:intro}

The Madelung equations \cite{madelungQuantentheorieHydrodynamischerForm1927} 
are known for having sparked a variety of `classical’ 
approaches to the Schrödinger theory of quantum mechanics. Arguably, the 
most common ones are the de Broglie-Bohm theory  
\cite{bohmSuggestedInterpretationQuantum1952,bohmOntologicalBasisQuantum1987,
hollandQuantumTheoryMotion1993,
durrBohmianMechanicsPhysics2009,sanzTrajectoryDescriptionQuantum2012,
sanzTrajectoryDescriptionQuantum2012a,durrQuantumPhysicsQuantum2013,
nassarBohmianMechanicsOpen2017} and stochastic
mechanics \cite{fenyesWahrscheinlichkeitstheoretischeBegruendungUnd1952,
nelsonDerivationSchrodingerEquation1966,nelsonQuantumFluctuations1985,
nelsonReviewStochasticMechanics2012,delapenaEmergingQuantumPhysics2015,
santosStochasticInterpretationsQuantum2022}. 
For one body of mass $m$ the three (formal) equations in $3$-dimensional space may be stated  
as follows: 
\begin{subequations}
	\label{eq:Madelung}
\begin{gather}
	m \left(\partd{\vec{v}}{t} +\left(\vec{v} \cdot \nabla\right) \vec{v}  \right)= - 
	\nabla V + \frac{\hbar^2}{2m} 
	\nabla \frac{\Delta \sqrt{\varrho}}{\sqrt{\varrho}}  
	\label{eq:Madelung1}
		\\
	\frac{\partial \varrho}{\partial t}	+ 
	\nabla \cdot \left( \varrho \, \vec v \right) = 0 
	\label{eq:Madelung2}
		\\		
	\nabla \cross \vec v = 0   \, . 
	\label{eq:Madelung3}
\end{gather} 
\end{subequations}
The above equations can be generalized to the many-body case (cf. e.g. 
Sec. 7.1.2 in Ref. \cite{hollandQuantumTheoryMotion1993}) and there also 
exist analogue equations within the $1$-body Pauli theory 
(cf. Ref. \cite{janossyHydrodynamicalModelWave1966} and Chap. 11 in Ref. 
\cite{bialynicki-birulaTheoryQuanta1992}). 

Though the Madelung equations and the Schrödinger equation are closely 
related, the mathematically precise relationship between the two 
systems of partial 
differential equations (PDEs) is still a matter of controversy  
(see e.g. \cite{hallIncompletenessTrajectorybasedInterpretations2004,
grublNondifferentiableBohmianTrajectories2011,antonelliFiniteEnergyWeak2008})---a 
controversy that can be traced back to a variety of objections raised by 
Wallstrom \cite{wallstromDerivationSchrodingerEquation1989,
wallstromInitialvalueProblemMadelung1994,wallstromInequivalenceSchrodingerEquation1994} and that 
has recently been linked to a major open problem in the field of `geometric 
hydrodynamics' (cf. Problem 36 in Ref. \cite{khesinGeometricHydrodynamicsOpen2022}). 
The discussion mostly concerns itself with the $1$-body 
equations, but it is also of importance to the more general $N$-body case  
(cf. Ref. \cite{riessNodalStructureNodal1970}). We shall elaborate on Wallstrom's 
contribution later in this section.

The central, underlying question is the following: 
\begin{quote}
	Under which conditions are the Madelung equations and the Schrödinger equation 
	mathematically equivalent and at which point does this equivalence break down?  
\end{quote} 

So far, only three works \cite{gasserQuantumHydrodynamicsWigner1997,vonrenesseOptimalTransportView2012,
reddigerMadelungPictureFoundation2017} 
have made mathematically 
rigorous progress in addressing the above question, 
all with serious limitations. 
In Lem. 2.1 of Ref. 
\cite{gasserQuantumHydrodynamicsWigner1997} Gasser and Markowich 
considered reformulated Madelung equations and showed that, 
under very mild regularity assumptions on the wave function and various 
other conditions, the equations were indeed 
implied by the Schrödinger equation. The caveat is that the authors did not show how to 
reobtain the Schrödinger equation and made overly restrictive 
assumptions on the potential $V$ 
(cf. Eq. A1 therein). In contrast, von Renesse took an optimal transport approach in order to 
find a new reformulation of the Madelung equations via the tools of Wasserstein geometry 
\cite{vonrenesseOptimalTransportView2012}. In the smooth setting  
he showed local equivalence between this formulation, the so called Hamilton-Jacobi-Madelung 
equations (Eqs. 1.4 therein), and the Schrödinger equation 
(Thm. 2.1 and Cor. 2.2 in Ref. \cite{vonrenesseOptimalTransportView2012}). 
By imposing strong assumptions of regularity and restricting himself to time-independent 
simply-connected domains, there was no need to account for Wallstom's objections.%
	\footnote{In a subsequent work \cite{khesinGeometryMadelungTransform2019} Khesin et al. 
				allowed for less regular probability densities and wave functions. However, their 
				work asks for the domain to be compact and connected---which excludes the physically most common 
				cases of $\R^3$ and its unbounded, open subsets.
				}  
In a similar vein, Reddiger showed local equivalence between the Madelung equations 
and the Schrödinger equation in the smooth setting (Thm. 3.2 of Ref. 
\cite{reddigerMadelungPictureFoundation2017}). To assert global equivalence, however, he required an 
additional topological assumption: Namely, 
that at each fixed time every connected component of the domain of $\vec{v}$ 
is simply connected. 

Some approaches, like stochastic mechanics and the theory 
based on Kolmogorovian probability theory laid out in 
Ref. \cite{reddigerMadelungPictureFoundation2017}, assert that the Madelung equations 
are physically more fundamental than the Schrödinger equation. Providing an 
answer to the above question is thus not merely of mathematical 
interest, but is possibly central to the historical and still current 
debate on the foundations of quantum mechanics:%
	   \footnote{We refer, for instance, to p. 143 in 
	   				Ref. \cite{bacciagaluppiQuantumTheoryCrossroads2009}, 
					p. 153 sq. in Ref. \cite{harriganEinsteinIncompletenessEpistemic2010}, 
					p. 327 in Ref. \cite{whitakerEinsteinBohrQuantum2006},  
					Sec. 3.5.1 in Ref. \cite{bacciagaluppiRoleDecoherenceQuantum2020}, 
					and Ref. \cite{bacciagaluppiConceptualIntroductionNelson2005}. 
					} 
Depending on how the mathematical question is resolved, it could shift 
the physical discussion away from the confines of 
quantum mechanics towards the more general question of the foundations 
of non-relativistic quantum theory. 

The importance of Wallstrom's 
objections \cite{wallstromDerivationSchrodingerEquation1989,
wallstromInitialvalueProblemMadelung1994,wallstromInequivalenceSchrodingerEquation1994} 
to this foundational 
discussion is highlighted by the following statement in Sec. 2.6 of a recent book by 
de la Peña et al. \cite{delapenaEmergingQuantumPhysics2015}: 
\begin{quote}
	[...] Wallstrom's work has been considered by many as the definitive blow 
	against Nelson's [stochastic mechanics] and similar theories. 
\end{quote}
In the same sentence the authors state that they consider the issue as 
resolved, but we disagree with this assertion: As Derakshani 
(cf. Refs. 
\cite{derakhshaniStochasticMechanicsAd2017,derakhshaniSuggestedAnswerWallstrom2019} and 
Note 37 in Ref. \cite{bacciagaluppiRoleDecoherenceQuantum2020}) 
and Grössing et al. \cite{grossingClassicalExplanationQuantization2011}, 
de la Peña et al. 
aim to address Wallstrom's objections by imposing assumptions on the stochastic 
processes they consider 
(cf. Eq. 4.37 and Sec. 4.7 in Ref. \cite{delapenaEmergingQuantumPhysics2015}). 
While, for the sake of transparency, we support the idea of 
founding non-relativistic quantum theory on the mathematical theory of stochastic 
processes, addressing Wallstrom's objections in this manner seems unlikely to 
convince contemporary critical voices in the wider physics community. 
Contrarily, those voices might fall silent, if it was shown to be 
possible to address Wallstrom's 
objections on the 
phenomenological level of the Schrödinger theory---that is, without the 
a priori introduction of stochastic processes. As we will argue, the futility 
of such an approach has not been established. 

Wallstrom's substantial objections to non-relativistic 
quantum theories based on the Madelung equations  
\cite{wallstromDerivationSchrodingerEquation1989,
wallstromInitialvalueProblemMadelung1994,wallstromInequivalenceSchrodingerEquation1994} 
are twofold:%
		\footnote{ 	In Sec. 6 in Ref. 
					\cite{wallstromDerivationSchrodingerEquation1989} 
					the further objection was made that particle-based 
					approaches fail to explain wave-like 
					quantum behavior. This argument ought not to be given 
					much weight, for it is the 
					statistics of particles that exhibit wave-like 
					behavior, not the particles themselves. 
					This was already noted by  
					Landé \cite{landeQuantumFactFiction1969} 
					in 1969 (see also p. 341 in Ref. 
					\cite{bellerQuantumDialogueMaking1999} as well as Ref. 
					\cite{landeQuantumMechanicsNew1973}):  
					\begin{quote}
						[W]avelike material phenomena result 
						from the quantum mechanics of matter 
						particles, as we know already since 
						1927 from Born's famous statistical 
						particle interpretation of the 
						de Broglie-Schrödinger waves.
					\end{quote}
					As long 
					as said statistics is described correctly by the theory, 
					the supposed counterargument is devoid of 
					physical content. We also refer to Sec. 2.5.1 in Ref. 
					\cite{delapenaEmergingQuantumPhysics2015}
					and references therein. 
					} 
First, in Refs. \cite{wallstromDerivationSchrodingerEquation1989,
wallstromInequivalenceSchrodingerEquation1994} 
he argued that an additional, supposedly ad-hoc 
`quantization condition' needs to be imposed for the equations to be equivalent. 
Second, in Ref. \cite{wallstromInitialvalueProblemMadelung1994} he asserted that  
the Madelung equations fail to yield a unique 
time-evolution from initial data, at least for the case that 
the connected components on which the density does not vanish 
merge over time. 

The main goal of this article is to elaborate on those objections 
and to encourage further research---with a focus on the first 
objection and in particular with regards to the 
central, underlying question above. The physical relevance of 
the question means that a satisfactory resolution thereof  
has to form the cornerstone of any mathematical theory of the 
Madelung equations. 

The precise contribution of this work to 
the controversy and to the development of such a mathematical theory 
will be given after we have explained the first objection 
in more detail. 

In order to clarify Wallstrom's first objection, recall that at fixed time 
a $C^1$-wave function $\Psi$, which does not vanish on its domain, 
gives rise to the drift field $\vec{v}$ above via
	\begin{equation}
		\label{eq:defv}
			\vec{v} = 
			\frac{\hbar}{m} \Im \left(\frac{\nabla \Psi}{\Psi} \right)
			= \frac{\hbar}{2 \iu m} \left(\frac{\nabla \Psi}{\Psi} - 
			\frac{\nabla \Psi^*}{\Psi^*} \right) \, .  
	\end{equation}		
The aforementioned `quantization' condition then states that  
$\vec{v}$ has to satisfy 
\begin{equation}
	\frac{m}{2 \pi \hbar} \oint \vec{v} \cdot \d \vec{r} \in \Z  
	\label{eq:wallstrom}
\end{equation}
at any given time for it to be physically acceptable. 

While Takabayasi 
\cite{takabayasiFormulationQuantumMechanics1952} is often credited with 
the discovery of condition \eqref{eq:wallstrom}, in the literature 
on quantum vortices%
\footnote{Though there may be links on a mathematical  
			\cite{harveyNavierStokesAnalogQuantum1966,a.jungelDissipativeQuantumFluid2012}
				and even on a physical level \cite{loffredoCreationQuantizedVortex1992}, 
				physically one needs to distinguish between quantum vortices 
				in fundamental physics, as studied here, and those occurring 
				in superfluids. 
				The latter is a macroscopic quantum phenomenon. See e.g. 
				Ref. \cite{tsubotaQuantizedVorticesSuperfluid2006}. 
				}  
\cite{schonbergVortexMotionsMadelung1955,riessNodalStructureNodal1970,bialynicki-birulaMagneticMonopolesHydrodynamic1971,hirschfelderQuantumMechanicalStreamlines1974,
hirschfelderQuantizedVorticesWavefunction1974,hirschfelderAngularMomentumCreation1977,
takabayasiVortexSpinTriad1983,
wuInversesquarePotentialQuantum1994,
bialynicki-birulaMotionVortexLines2000,sanzRoleQuantumVortices2004,
sanzQuantumTrajectoriesAtom2004,imaiEffectsEntanglementVortex2020} 
this credit is mostly given to Dirac 
(cf. p. 67 in Ref. \cite{diracQuantisedSingularitiesElectromagnetic1931}).%
	\footnote{Contrary to what is suggested on p. 278 in 
				Ref. \cite{jammerConceptualDevelopmentQuantum1966} 
				and by Ref. \cite{hushwaterPathQuantizationAction1998}, 
				the article \cite{wentzelVerallgemeinerungQuantenbedingungenFuer1926} 
				by Wentzel should not be credited with the discovery of the 
				condition. Therein, Wentzel considers the $1$-dimensional 
				time-independent Schrödinger equation and a meromorphic 
				continuation 
				of the wave function in order to be able to apply the 
				residue theorem---a wholly different mathematical 
				context. 
				}
We shall nonetheless call condition 
\eqref{eq:wallstrom} \emph{Takabayasi's quantization condition}, 
for he was the first to suggest that it ought to be added to the 
Madelung equations, Eqs. \eqref{eq:Madelung1}-\eqref{eq:Madelung3} 
above 
(cf. p. 155 in Ref. \cite{takabayasiFormulationQuantumMechanics1952}). 

With regards to the greater discussion on the foundations of 
non-relativistic quantum theory, 
Takabayasi \cite{takabayasiFormulationQuantumMechanics1952} 
already voiced the following criticism: 
\begin{quote}
	[Eq. \eqref{eq:wallstrom} above] is so to speak the 
	`quantum condition' for fluidal motion and of \emph{ad hoc} 
	and compromising character for our formulation, just as 
	the [Bohr-Sommerfeld] quantum condition for old quantum theory. 
	
	[reference omitted]
\end{quote}
In essence, the above also constitutes Wallstrom's central objection voiced in 
Refs. \cite{wallstromDerivationSchrodingerEquation1989,
wallstromInequivalenceSchrodingerEquation1994}. 

At this point in time, Takabayasi's condition has become ubiquitous in the literature 
surrounding the Madelung equations (see e.g. Sec. 3.2.2 in Ref. 
\cite{hollandQuantumTheoryMotion1993}, Sec. 3.1 in Ref. 
\cite{bohmUndividedUniverseOntological1993}, Sec. 1.2.1 in Ref. 
\cite{nassarBohmianMechanicsOpen2017}, and Refs. 
\cite{schonbergVortexMotionsMadelung1955,janossyHydrodynamicalModelWave1963,
wilhelmHydrodynamicModelQuantum1970a,senguptaQuantumTheoryMotion1996,
sanzRoleQuantumVortices2004,
poirierReconcilingSemiclassicalBohmian2004,molinaMappingsThermodynamicsQuantum2017}) 
and it appears to have become largely assimilated without 
putting much weight on 
the aforementioned conceptual objection. 
					
The five main contributions of this article to the  subject matter are as follows: 
\begin{enumerate}[1)]
	\item		We rigorously prove that Eq. \eqref{eq:wallstrom} 
				holds for $C^1$-wave functions. 
				Nonetheless, we argue on the basis of the 
				mathematical theory of quantum mechanics that  
				the condition is too restrictive---as already observed by 
				Smolin in this context 
				(cf. Sec. IV in Ref. \cite{smolinCouldQuantumMechanics2006}), 
				not all wave functions are this regular. As we explain in 
				Sec. \ref{ssec:distributions} and 
				Rem. \ref{Rem:Takabayasi} in particular, there is serious 
				doubt whether the condition can be appropriately extended.  
	\item		In Sec. \ref{ssec:distributions} we argue that 
				the relation between the Madelung equations and 
				the Schrödinger equation needs to be understood 
				in an appropriate `distributional' sense.  
				While this is rather obvious for those 
				acquainted with the mathematical theory of 
				quantum mechanics or the modern theory of PDEs, this 
				central point has been largely ignored in the related 
				physics literature. In Sec. 
				\ref{sec:distmadelung} we discuss what constitutes such a
				distributional approach to the Madelung 
				equations, which in the mathematics literature 
				is known as a variational formulation. 
				We discuss two such formulations, one given in seminal work by 
				Gasser and Markowich 
				\cite{gasserQuantumHydrodynamicsWigner1997} 
				and one based on Nelson's equations 
				\cite{nelsonDerivationSchrodingerEquation1966}. There 
				may, however, be other variational formulations of 
				physical relevance.  				
	\item 		Consider the standard solutions $\Psi_{n l \mu}$ 
				of the $1$-body (time-\-independent) Schrö\-ding\-er equation 
				with attractive Coulomb potential in $3$ 
				dimensions---a generally accepted physical model to 
				obtain the gross structure of the electromagnetic 
				absorption spectrum of atomic hydrogen and of `hydrogen-like atoms'.
				
				We show rigorously that -- using `physics notation' for 
				the Dirac delta -- the 
				distributional curl of the drift field $\vec{v}$ 
				corresponding to $\Psi_{n l \mu}$ is given by 
		 		\begin{equation}
					\left(\nabla \cross \vec{v} \right) (t,x,y,z)  
						= \frac{2 \pi \mu \hbar}{m} \, 
								\begin{pmatrix}
									0 \\ 
									0 \\
									\delta(x) \, \delta(y) 
								\end{pmatrix} 
						\label{eq:physicscurlvhydrogen}
				\end{equation}
				(cf. Cor. \ref{Cor:hydrogenvdistder}). 
				
				That is, those solutions exhibit 
					\emph{quantum vorticity} in the sense that the third 
				Madelung equation, Eq. \eqref{eq:Madelung3}, 
				is in general \emph{not} satisfied, if the curl is 
				understood in the distributional sense. 
				Motivated by a statement made by Schönberg 
				\cite{schonbergVortexMotionsMadelung1955}
				in the 1950s, we call such drift fields 
				\emph{quasi-irrotational} (cf. Def. \ref{Def:quasiirrot}). 
				It 
				is also worth noting that here 
				the quantum vorticity depends on the 
				magnetic quantum number $\mu$---whose 
				`quantization' forces the other quantum numbers $n$ and $l$ to 
				take integer values as well. 

	\item		In Sec. \ref{sec:nonquantumsol} 
				we explicitly construct and analyze `non-quantized', stationary, 
				strong solutions of the Madelung equations for the case of the 
				$2$-dimensional isotropic harmonic oscillator. Wallstrom 
				claimed that such solutions exist 
				\cite{wallstromInequivalenceSchrodingerEquation1994}. Yet 
				he neither gave any explicit examples nor is the 
				existence thereof mathematically trivial, since 
				the need to account for other boundary conditions 
				could in principle force the `quantization' of the quantum 
				numbers. We further provide a mathematical 
				analysis of those solutions, taking various 
				perspectives. 
				
	\item 		With regards to Wallstrom's second objection 
				\cite{wallstromInitialvalueProblemMadelung1994}, 
				we argue in Rem. \ref{Rem:Wallstromphenomenon} 
				of Sec. \ref{sec:distmadelung} 
				that it is the failure of the 
				quantum-mechanical Schrödinger theory 
				to discern physically equivalent, yet 
				mathematically inequivalent states 
				that leads to this \emph{Wallstrom phenomenon}. 
				We also show how it is related to the historical 
				Pauli problem (cf. Refs. 
				\cite{stulpeRemarksDeterminationQuantum1990,
				weigertPauliProblemSpin1992,
				weigertHowDetermineQuantum1996,antonelliFiniteEnergyWeak2008})---whose 
				negative resolution may be viewed as a 
				theoretical anomaly within quantum mechanics.  
\end{enumerate} 

With regards to the implication of those results for the general 
analysis of the precise relation between the Schrödinger equation and the 
Madelung equations, we refer to Sec. \ref{sec:conclusion} at 
the end of this article. There we also provide a more in-depth discussion of 
prior attempts in the literature to resolve the controversy. 
	
In this work, we have taken care to provide rigorous 
proofs for the mathematical statements made. In order 
to put the focus on the arguments in the main text, 
those proofs have been pushed to the appendix. 
	
For the convenience of the reader, we shall clarify some notation and 
conventions: 
$\N$, $\Z$, $\R$, and $\C$ denote the set of natural numbers, integers, 
real numbers, and complex numbers, respectively. 
$\N_0$ is $\N \cup \lbrace 0 \rbrace$ and $\R_+$ is the open interval 
$(0,\infty) \subset \R$. For $\alpha \in \C$, the quantities 
$\abs{\alpha}$, $\arg \alpha$, $\Re \alpha$, and $\Im \alpha$ denote 
the absolute value, the principal value of the argument, the 
real part, and the imaginary part of $\alpha$, respectively. By convention, 
we have $\arg{\alpha} \in [0,2 \pi)$. $\iu$ is the imaginary unit and 
the symbol $\sim$ means `asymptotic to'.
As already indicated, we distinguish between the (strictly 
positive) particle mass 
and the magnetic quantum number by using $m$ for the former and 
$\mu$ for the latter. Similarly, we use the symbol $\varrho$ for the probability 
density and $\rho$ for the radial distance in polar coordinates. 
With regards to special functions, we use 
$(n,\mu,x) \mapsto L_n^\mu(x)$ for the associated Laguerre polynomials,%
	\footnote{ 	Note that we use Slater's convention  
				for the associated Laguerre polynomials $L_n^\mu(x)$
				(cf. \S 5.5 in Ref. 
				\cite{slaterConfluentHypergeometricFunctions1960}). The latter  
				coincide with the ones given by Messiah divided by $(n+\mu)!$ 
				(cf. Appx. B, \S I.2 in Ref. \cite{messiahQuantumMechanicsTwo1995}). 
				} 
$(n,\mu,x) \mapsto P_n^\mu(x)$ for the associated 
Legendre polynomials, $(l,\mu,x) \mapsto Y_l^\mu(x)$ for the spherical harmonics, 
$(a,b,x) \mapsto {}_1 F _1(a,b;x)$ and $(a,b,x) \mapsto U(a,b;x)$ 
for the confluent hypergeometric function 
of first and second kind, respectively, and $x \mapsto \Gamma(x)$ 
for the gamma function. 
As for function spaces, for $n,m \in \N$, $k \in \N_0$, 
and $\K \in \lbrace \R, \C \rbrace$, 
we use $C^k(\R^n,\K^m)$ for the space of $k$ times continuously differentiable 
$\K^m$-valued functions in $\R^n$, 
$C_0^\infty(\R^n,\K^m)$ if they are smooth and compactly 
supported, and 
$\mathcal{S}(\R^n,\K^m)$ for the respective space of Schwartz functions 
(all spaces are equipped with their usual topology). $\mathcal{H}$ 
is a general or specified Hilbert space with inner product 
$\inp{\, . \, }{\, . \, }$, antilinear in the first argument and linear in the 
second. We use common multi-variable calculus notation, like $\Delta$ 
for the Laplacian, $\int_{U} \d^n r$ for an integral over 
$U \subseteq \R^n$, $\nabla$ for the del operator, etc. $\partial^\alpha$ 
is a multivariate (strong or weak) derivative with respect to the 
multi-index $\alpha$ and $\abs{\alpha}$ denotes its order. 
We also use the SI-unit system throughout the article, denoting by 
$\varepsilon_0$ the electric constant, 
and by $\operatorname{e}$ the magnitude of the 
electron charge. 

\section{On Takabayasi's quantization condition}
\label{sec:Wallstromcondition}
\subsection{The condition for strong solutions}
\label{ssec:WallstromProp}

Consider the Schrödinger equation for one body in $3$-dimensional space, 
\begin{equation}
	\iu \hbar \partd{\Psi}{t} = - \frac{\hbar^2}{2m} \Delta \Psi
	+ V \, \Psi \, ,
\end{equation}
with initial data $\Psi(t,\vec{r}) \evat{t=0}=\Psi_0(\vec{r})$ 
for $\vec{r}$ in some open subset of 
$\R^3$ and some given, sufficiently `regular' potential $V$ thereon. Naively, 
we find that both $\Psi_0$ and $\Psi$ must be two-times 
differentiable in the variables $\vec r$ and $\Psi$ must be once differentiable in the variable 
$t$ for the equation to make sense on an elementary mathematical level. 
Though, as we shall discuss hereafter, this view is not only naive but 
it runs counter to the axioms of quantum mechanics, we shall nevertheless 
consider such so called `strong' or `classical' solutions here.%
	\footnote{	The word `classical solution' is a term used in the mathematical 
				theory of partial differential equations and does not 
				have any relation to the term `classical physics', apart from 
				signifying historical developments of the respective subject 
				areas. 
				} 

Consequently, if we fix the time $t$ and drop the respective 
dependence in our notation in the remainder of this subsection, 
the wave function $\Psi$ will be twice differentiable 
in each remaining variable---if not on $\R^3$ itself, then on some open subset 
$D \subseteq \R^3$ thereof. As differentiability at a point implies continuity there, we find that 
$\Psi$ is necessarily an element of $C^1 (D, \C)$, i.e. $\Psi$ is continuous on $D$, 
all first order partial derivatives exist thereon and are also continuous. 

We shall show that it is this property alone that forces Takabayasi's condition, 
Eq. \eqref{eq:wallstrom}, to hold. While for the mathematical reader acquainted with 
the concept of the winding number as well as degree theory this may 
not be too surprising, no proof of this fact or reference thereto 
has so far been provided in the literature on Takabayasi's condition. 
The most detailed `derivation' was given by 
Hirschfelder et al. \cite{hirschfelderQuantizedVorticesWavefunction1974}, which, due to the 
nature of the work, lacked mathematical rigor. As we shall see in the coming sections, the 
mathematics surrounding Takabayasi's condition can become quite subtle, so that it is important 
to clarify when it holds and when it may fail. 
\begin{Proposition}
	\label{Prop:C1quantization}
\begin{subequations}
	For $n \in \N$ with $n \geq 2$, let $D$ be open in $\R^n$ 
	and let $\Psi$ be in $C^1(D,\C)$ such that $\Psi$ does not 
	vanish on $D$. Define $\vec{v}$ as in 
	Eq. \eqref{eq:defv} above. 
	
	If the integral in Eq. \eqref{eq:wallstrom} is 
	taken over any $C^1$-curve $\gamma$
	in $D$ defined on some closed interval $[a,b]$ 
	with $\gamma(a)=\gamma(b)$, then the condition, Eq. \eqref{eq:wallstrom}, holds true. 
\end{subequations}
\end{Proposition}

Weakening the assumptions of Prop. \ref{Prop:C1quantization} above 
is non-trivial: Clearly, $\Psi$ is not allowed to vanish for $\vec{v}$ 
from Eq. \eqref{eq:defv} to be generally well-defined. Furthermore, continuous 
differentiability of $\Psi$ assures that $\vec{v}$ is continuous, which in turn 
allows us to apply the fundamental theorem of calculus without any 
further complications. Furthermore, not every $C^1$-wave function 
$\Psi$ is a so-called `WKB-state' or `JWKB-state', named after the authors 
of the historical Refs. \cite{jeffreysCertainApproximateSolutions1925,
wentzelVerallgemeinerungQuantenbedingungenFuer1926,
kramersWellenmechanikUndHalbzahlige1926,
brillouinMecaniqueOndulatoireSchrodinger1926} 
(see e.g. 
Chap. VI, Sec. II in Ref. \cite{messiahQuantumMechanicsTwo1995},  
Refs. 
\cite{sparberWignerFunctionsWKB2003,figalliWKBAnalysisBohmian2014} and 
references therein): 
As shown by countless examples, it is in general not possible 
to find even a continuous function $S$ on $D$ such that
 $\Psi= \abs{\Psi} \, e^{\iu S / \hbar}$.%
 	\footnote{ 	In the mathematical literature 
 				a function $S$ satisfying this equality 
 				(up to the constant $1/\hbar$) 
 				is called a \emph{lifting} of the 
 				$\mathbb{S}^1$-valued map 
 				$\Psi/\abs{\Psi}$. See e.g. Refs.  
 				\cite{bourgainLiftingSobolevSpaces2000,brezisW11MapsValues2005}.} 
That there is no such function $S$ appearing in the proof of Prop. 
\ref{Prop:C1quantization} is therefore not accidental. 

Next consider the domain $D$, as given in Prop. \ref{Prop:C1quantization}, and assume 
that $\vec{v}$ is a $C^1$-vector field. 
If $D$ is simply-connected or consists of countably many simply-connected (connected) 
components, then for a given smooth curve $\gamma$ in $D$ as above 
we can always find an (oriented) bounded, connected smooth surface such that $\gamma$
is its boundary. Then 
Stokes' theorem implies that the integral 
Eq. \eqref{eq:wallstrom} vanishes (cf. e.g. 
Sec. 13.3.3.2 in Ref. \cite{bronshteinHandbookMathematics2015}). 
This is the topological condition that Reddiger imposed in his proof of 
the (local) equivalence of the Schrödinger equation and the Madelung equations
(Thm. 3.2 in Ref. \cite{reddigerMadelungPictureFoundation2017}). Contrarily, if 
this topological assumption does not hold -- for instance due to 
a `line singularity' of $\vec{v}$ -- then Stokes' theorem cannot be employed, 
as the respective surface is not compact. Roughly speaking, this is how the 
integral in Eq. \eqref{eq:wallstrom} can fail to be zero.%
	\footnote{ 	Note that there exist several examples of 
				wrongful application of Stokes' theorem in the literature 
				related to Takabayasi's condition. Though there exist modern 
				generalizations of Stokes' theorem 
				(cf. Refs. \cite{harrisonStokesTheoremNonsmooth1993,
				harrisonGeometricRepresentationsCurrents2004} and Thm. 8.9 in 
				Ref. \cite{harrisonOperatorCalculusDifferential2015}), they 
				are also not applicable to the special cases 
				considered in Sec. \ref{ssec:vorticity} below.}
				
The paramount examples for Takabayasi's condition are the  
standard solutions $\Psi_{n l \mu}$ of the aforementioned model for 
atomic hydrogen and of `hydrogen-like atoms'. In that instance, the 
integer in Eq. \eqref{eq:wallstrom} is always given by the quantum number $\mu$. 

In order to illustrate how the condition can hold despite the possibility 
to superpose the respective states, we shall consider the following example.
\begin{Example}
	\label{Ex:wallstoempseudocounterex}
\begin{subequations}
	\begin{figure}
		\centering
		\includegraphics[width=0.5 \textwidth]{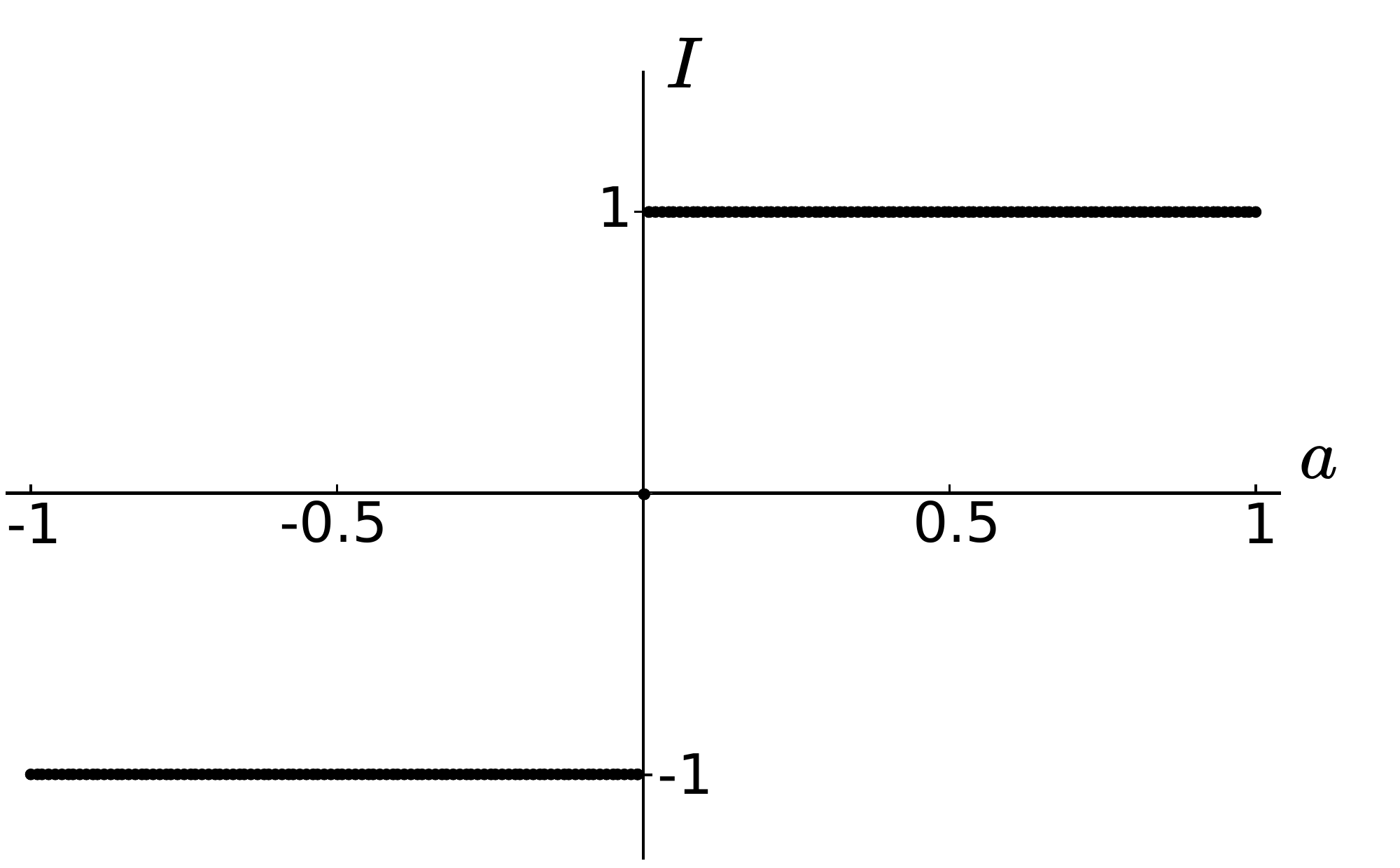}
		\caption{A plot of $100$ equidistant, 
					numerically evaluated values of the integral  
					$I(a)$ from Eq. \eqref{eq:Ipseudocounterex} in 
					the interval $[-1,1]$ is shown. 
					For $a \in [-1,0)$ we have $I(a)=-1$. For $a =0$ the wave function 
					$\Phi_a$ from Eq. \eqref{eq:wallstoempseudocounterexphi} 
					becomes real and a nodal surface develops in the $x$-$z$-plane. 
					Though $\vec{v}_0=0$ and thus $I(0)$  
					formally evaluates to $0$, $\gamma$ does not lie in $D_0$ 
					due to the nodal 
					surface, so that,  
					strictly speaking, $I(0)$ is not defined. For 
					$a \in (0,1]$ the value $I(a)$ then jumps to $+1$.}
		\label{Fig:1}
	\end{figure}
	The textbook result for the above model is that 
	the energy eigenfunctions $\Psi_{n l \mu}$ in spherical coordinates $(r, \theta, \phi)$ 
	centered at the position of the
	proton read as
	\begin{equation}
		\Psi_{n l \mu} ( r, \theta, \phi) = R_{n l} (r) \, Y_l^\mu(\theta, \phi) \, ,
	\end{equation}
	with $n \in \N$, 
	$l \in \lbrace 0, \dots, n-1 \rbrace$ and 
	$\mu \in \lbrace -l, \dots, l \rbrace$. 
	The energy eigenvalues $E_n$ are degenerate in the sense that they do not depend 
	on $l$ or $\mu$. Therefore, any linear combination of the possible 
	$\Psi_{n l \mu}$ with same $n$ 
	is an energy eigenfunction and consequently gives rise to 
	a stationary state.

	Fixing $n>1$, we may now consider the family of energy eigenfunctions 
	$\Phi_a$ with $a \in [-1,1]$ and 
		\begin{equation}
				\Phi_a(r,\theta, \phi) 
				= R_{n 1}(r) 
				\left(  \frac{1-a}{\sqrt{2 (1 + a^2)}} \, Y_1^{-1} (\theta, \phi) 
				+ \frac{1+a}{\sqrt{2 (1 + a^2)}} \, Y_1^{+1} (\theta, \phi)\right)
				\, .   
				\label{eq:wallstoempseudocounterexphi}
		\end{equation} 
	As one easily checks, all $\Phi_a$ are normalized and they reduce to 
	$\Psi_{n 1 \pm 1}$ for $a = \pm 1$. As in Prop. \ref{Prop:C1quantization}, 
	denote by $D_a$ the 
	respective maximal domain on which $\Phi_a$ is $C^1$ and nowhere vanishing. 

	It is worthwhile to check for which values of $a$ the respective drift field 
	$\vec{v}_a$ is well-defined and continuous. 
	For convenience, set $\hbar$ and $m$ equal to $1$. As $R_{n 1}$ is real-valued, 
	$\vec{v}_a$ does not depend thereon. As the	
	angular part of $\Phi_a$ is well-defined and smooth on $\R^3 \setminus 
	\lbrace 0\rbrace$, we only need to check where it vanishes. 
	The latter happens if and only if both its real part and its imaginary part 
	vanishes. Using
		\begin{equation}
			Y_1^{+1} (\theta, \phi) = - \frac{3}{8 \pi} \, \sin \theta 
					\, e^{\iu \phi} = - \left( Y_1^{-1} (\theta, \phi) \right)^* \, , 
		\end{equation}
 	and excluding $\theta \in \lbrace 0 , \pi \rbrace$, we find that this is only the 
 	case for 
 	$a= 0$ in the limits of $\phi$ tending to an element of the set 
 	$\lbrace 0, \pi, 2 \pi\rbrace$. 
 	
 	In this instance, $\Phi_0$ is real and the 
 	$x$-$z$-plane is a nodal surface. If this nodal surface had not `cut 
 	off' the curve 
 	\begin{equation}
		\gamma \colon \quad [0,2\pi] \to D_a \quad \colon 
		\quad \phi \mapsto \left(r_0 \cos \phi, r_0 \sin \phi, 0\right)
	\end{equation}
	for some admissible $r_0 \in \R_+$, then the function 
	\begin{equation}
		I\colon \quad  [-1,1] \quad \to  \quad \R \quad \colon \quad 
		a \mapsto I(a) = 
		\frac{m}{2 \pi \hbar} \oint_\gamma \vec{v}_a \cdot \d \vec{r}
			\label{eq:Ipseudocounterex}
	\end{equation}
 	would have been well-defined and continuous. In turn, 
 	we could have applied the intermediate value theorem to show that $I$ 
 	takes every value between $-1$ and $+1$---thus contradicting 
 	Prop. \ref{Prop:C1quantization}. 
 	
 	Due to the formation of the nodal surface for $a=0$, this is, of course, not what 
 	occurs. Numerically calculated values of the function $I$ are 
 	depicted in Fig. \ref{Fig:1}. 
\end{subequations}
\end{Example}

\subsection{On distributions in quantum theory}
\label{ssec:distributions}

Mathematically, Takabayasi's condition, Eq. \eqref{eq:wallstrom}, 
suffers from the problem that not all wave functions are as regular 
as the ones considered in the prior section. This was already observed 
by Smolin \cite{smolinCouldQuantumMechanics2006}. While the fact that not all 
wave functions are $C^1$ is insufficient to adequately address Wallstrom's criticism 
\cite{wallstromDerivationSchrodingerEquation1989,
wallstromInequivalenceSchrodingerEquation1994}, it is nonetheless a central point we 
wish to make in this work. 

In this spirit, this section serves to motivate the use of distributions and distributional 
derivatives in the context of this article. We shall find that this distributional point 
of view is necessary due to the mathematical axioms of quantum mechanics. 

To begin with, consider the first order, linear differential equation 
\begin{equation}
	u' = 0 
	\label{eq:samplePDE}
\end{equation}
for a function $u \colon \R \to \R$. 
This is arguably the simplest equation one can consider as an introduction to 
the modern mathematical theory of PDEs.%
	\footnote{We refer, for instance, to 
				Refs. \cite{follandIntroductionPartialDifferential1995,
				leoniFirstCourseSobolev2017,brezisFunctionalAnalysisSobolev2011} 
				for an introduction 
				to this theory. More advanced material can be found in 
				Ref. \cite{bahouriFourierAnalysisNonlinear2011}, which 
				also defines common notation employed in this field. 
				} 
Nonetheless, we shall see that there are a number of a priori 
inequivalent ways in which this formal equation can be mathematically interpreted. 

Approaching Eq. \eqref{eq:samplePDE} from a naive perspective, the perhaps 
most obvious solution is $u(x) = c$ for some constant $c \in \R$. 
However, any real-valued step function solves \eqref{eq:samplePDE}, 
if we are free to take out isolated points from the domain $\R$ so that 
$u$ is defined on $\Omega \subset \R$ instead. The derivative will 
then satisfy the PDE on $\Omega$---and we may even smoothly extend it to $\R$. 
Of course, one could force the former solution by asking $u$ to be well-defined and 
smooth (or $C^1$) on the entirety of $\R$. Yet 
for more complicated PDEs an analogous requirement may be too rigid, as we often 
do not know an appropriate maximal domain for each solution beforehand. 

The common resolution to this problem, that is employed 
in the modern theory of PDEs, is to first consider Eq. \eqref{eq:samplePDE} 
as a distributional differential equation. 

To do this, we first need to posit 
the existence of a distribution $T$. The precise definition of 
distributions depends on 
the choice of space of `test functions' $\mathcal{D}$ 
as well as the topology on it, we shall only define it for the standard 
choice of $\mathcal D = C_0^\infty(\R,\R)$ that is appropriate to 
Eq. \eqref{eq:samplePDE} above and we refer the 
reader to the literature for other cases. For this 
choice of $\mathcal{D}$, we say that a sequence 
$\left(\varphi_k \right)_{k \in \N}$ of 
test functions, each of which is supported on a common compact subset of 
$\R$, converges to 
$\varphi$ in $\mathcal D$, if for every multi-index $\alpha$ the 
sequence $\left(\partial^\alpha \varphi_k\right)_{k \in \N}$ 
converges to $\partial^\alpha \varphi$ uniformly. A 
distribution $T$ is then a linear functional 
\begin{equation}
	T \colon \quad \mathcal{D} \to \R \quad \colon \quad \varphi \mapsto T (\varphi) 
\end{equation}
that is continuous in the sense that for every such sequence 
$\left(\varphi_k \right)_{k \in \N}$ converging to $\varphi$ in $\mathcal D$ 
we have 
\begin{equation}
	\lim_{k \to \infty} T \left(\varphi_k\right) = T ( \varphi) 
\end{equation}
(cf.  Sec. 0.E in Ref. \cite{follandIntroductionPartialDifferential1995}).

In order to express Eq. \eqref{eq:samplePDE} in the language of distributions, 
we also recall that the derivative $T'$ of a distribution $T$ is 
defined via  
\begin{equation}
	T'(\varphi) = - T \left( \varphi' \right)
	\label{eq:Tprime}
\end{equation}
for all $\varphi \in \mathcal{D}$. As we shall see below, this 
definition is motivated by the integration by parts formula. 

Using this definition, Eq. \eqref{eq:samplePDE} simply reads $T' = 0$ or, equivalently, 
\begin{equation}
	\forall \varphi \in \mathcal{D} \colon 
	\quad T'(\varphi) = 0 \, . 
	\label{eq:TsamplePDE}
\end{equation}
It is noteworthy that, at least for our choice of $\mathcal{D}$ above, 
derivatives of distributions are always well-defined---which need not 
be the case for `ordinary functions' $u$. 

As a solution ansatz to Eq. \eqref{eq:TsamplePDE}, 
we may make the further 
assumption that $T$ is a \emph{regular distribution}, 
i.e. that there exists a locally integrable function%
	\footnote{	For $p \in [1,\infty)$ 
				the function $u\colon \R \to \R$ is \emph{locally $L^p$-integrable}, 
				if for any compact subset $K$ of $\R$ the integral 
				\begin{equation*}
					\int_K \abs{u(x)}^p \, \d x
				\end{equation*}
				is well-defined and converges. 
				Then the integral in Eq. \eqref{eq:defregdist} 
				converges as well due to Hölder's inequality. 
				The case $p=1$ is the most important one here, so 
				`local integrability' refers to local $L^1$-integrability. 
				The condition as well as the argument 
				is easily generalized to higher dimensions.
				\label{fn:L1loc}
				}
$u$ such that 
\begin{equation}
	T (\varphi) = \int_{\R} u(x) \, \varphi(x)  
	\d x 
	\label{eq:defregdist}
\end{equation}
for all $\varphi \in \mathcal{D}$. Due to the 
implicit dependence on $u$, we shall use the notation $T_u$ instead of $T$.%
	\footnote{Not all distributions are regular. The most common counterexample 
				is the \emph{Dirac (delta) distribution}: In $1$ dimension it is 
				defined via the equation $\delta_x( \varphi) = \varphi(x)$ 
				for any $x \in \R$ and $ \varphi \in \mathcal{D}$. 
				\label{fn:Diracdelta}	
				} 

The differential equation, Eq. 
\eqref{eq:samplePDE}, then reads $T'_u = 0$ or, equivalently, 
\begin{equation}
	\forall \varphi \in \mathcal{D} \colon \quad 
	{T'}_u(\varphi) = - 
	\int_{\R} u(x) \, \varphi'(x)  \, 
	\d x  = 0 \, . 
	\label{eq:preweakder}
\end{equation}
If it were possible to integrate by parts, we would obtain 
\begin{equation}
	\forall \varphi \in \mathcal{D} \colon 
	\quad \int_{\R} u'(x) \,  \varphi (x)  
	\d x  = 0 \, .  
	\label{eq:weakder}
\end{equation}
The fundamental theorem of calculus of variations implies that the only 
functions $u'$ that satisfy \eqref{eq:weakder} are those that 
vanish almost everywhere---that is, $u'$ can only be non-zero on a 
subset of Lebesgue measure zero.%
	\footnote{Intuitively, those are the sets that the 
				(Lebesgue) integral 
				`cannot see'.}
Indeed, it can be shown that, up to a set of 
measure zero, 
$u(x) = c$ is the only solution to Eq. \eqref{eq:preweakder} 
(cf. Lem. 8.1 in Ref. \cite{brezisFunctionalAnalysisSobolev2011}), so that 
the step to obtain Eq. \eqref{eq:weakder} is justified. 

Reformulating Eq. \eqref{eq:samplePDE} into Eq. \eqref{eq:preweakder} or 
Eq. \eqref{eq:weakder} has two advantages: 
\begin{enumerate}[i)]
	\item 	It is enough for $u$ to be \emph{weakly differentiable} 
			in the sense that there exists some locally integrable function 
			$v$ such that $T'_u = T_{v}$. In that case, we may use the 
			\emph{notation} $u'$ for $v$. In particular, the function $u'$ 
			may differ from $c$ on an arbitrary set of  
			measure zero. 
	\item 	Eq. \eqref{eq:preweakder} excludes step 
			functions as a solution. In fact, the 
			distributional derivative of a step function is not a regular distribution, 
			i.e. step functions are not weakly differentiable. Roughly 
			speaking, even though at isolated points 
			the values of a function are not of relevance   
			in this distributional framework, the distributional derivative 
			can still `see' some discontinuities and singularities. 
\end{enumerate}

The example above 
exemplifies that, in the context of solving differential equations, 
the distinction between interpreting the differential equation in the 
strong sense versus interpreting it in the distributional sense 
can be crucial in terms of what counts as a solution and 
what does not. This applies, in particular, to the dynamical equations of 
quantum theory.  

Of course, while the theory of distributions may provide a 
mathematically more satisfying -- as well as more elaborate -- way of 
formulating and solving PDEs, its relevance in the context of 
this article and quantum mechanics in general still requires justification. 

Before providing such a justification, we shall recall some 
central aspects of the mathematical theory 
of quantum mechanics: 

The reader may recall that in quantum mechanics observables, such as
the Hamiltonian $\hat{H}$, are given by linear operators on a 
separable Hilbert space $\mathcal{H}$. This sentence is, 
however, only true in general if we allow linear operators $\hat{A}$ to be 
\emph{unbounded} in the sense that $\hat{A}$ is only defined on a 
linear subspace of $\mathcal{H}$. This definition of an unbounded 
operator has historical origins and can be somewhat confusing, 
for it paradoxically allows $\hat{A}$ to be bounded. By definition 
an `unbounded' operator $\hat{A}$ is \emph{bounded}, if 
its domain $\dom \hat{A}$ is $\mathcal{H}$ and its operator norm 
\begin{equation}
	\lVert \hat{A} \rVert = \sup_{\Psi \in \mathcal{H}} \frac{\lVert \hat{A}\Psi 
	\Vert}{\norm{\Psi}}
\end{equation}
is finite. Contrarily, the stereotypical example of an 
unbounded operator is defined on a dense linear subspace 
$\dom \hat{A}$ of $\mathcal{H}$ and  
\begin{equation}
	\lVert \hat{A} \rVert = \sup_{\Psi \in \dom \hat{A}}  
	\frac{\lVert \hat{A}\Psi 
	\Vert}{\norm{\Psi}}
	\label{eq:normAsubspace}
\end{equation}
does not exist. 
\begin{Example}
	\label{Ex:freeHam1}
\begin{subequations}
	We consider a single free body in $n$-dimensional space, $n \in \N$.  
	The Hilbert space $\mathcal{H}$ is $L^2(\R^n,\C)$, 
	the space of complex-valued, square integrable functions 
	on $\R^n$. While the mathematical definition of $L^2(\R^n,\C)$
	is slightly more subtle, as we shall address below, this is the 
	appropriate space to consider due to the fact that 
	square-integrability assures that 
	any wave function $\Psi$ in $\mathcal{H}$ gives rise to 
	a corresponding probability density. 
	
	The free body Hamiltonian $\hat{H}_0$ is proportional to the (negative) 
	Laplacian and, accordingly, the expression 
	$\hat{H}_0 \Psi$ is not well-defined for arbitrary 
	$\Psi \in \mathcal{H}$---even if the derivatives are understood in the 
	weak sense. It is, of course, possible to choose a variety of 
	different domains $\dom \hat{H}_0$ on which $\hat{H}_0$ is well-defined. 
	A natural choice is the space of 
	Schwartz functions $\mathcal{S} \left(\R^n,\C\right)$. 
	
	As  $\mathcal{S} \left(\R^n,\C\right)$ contains 
	$C_0^\infty \left(\R^n,\C\right)$ and the latter is dense in 
	$L^2(\R^n,\C)$, $\mathcal{S} \left(\R^n,\C\right)$ is also dense 
	(see e.g. Cor. 4.23 in Ref. 
	\cite{brezisFunctionalAnalysisSobolev2011}).  
	Accordingly, $\hat{H}_0$ is a densely defined linear operator 
	on $\mathcal{H}$. Moreover, by considering a sequence 
	$\left(\Psi_k\right)_{k \in \N }$ of 
	normalized Gaussians with standard deviation $1/k$  
	one finds that the expression $\lVert \hat{H}_0 \Psi_k \rVert / 
	\norm{\Psi_k}$ diverges as $k \to \infty$. 
\end{subequations}
\end{Example}

As a theory of mechanics, a central question in 
quantum mechanics 
is how to find the time evolution of an appropriate, given initial state 
$\Psi_0 \in \mathcal{H}$. If we naively consider the generalized 
Schrödinger equation
\begin{equation}
	\iu \hbar \partd{}{t} \Psi_t = \hat{H} \Psi_t \, ,
	\label{eq:generalSE}
\end{equation}
we find that the state $\Psi_t = U_t \Psi_0$ with 
$U_t = \exp\bigl(- \iu \hat{H} t / \hbar\bigr)$ 
provides a solution for all $t \in \R$. While 
this illustrates the general idea, the actual mathematical 
situation is, of course, more 
subtle since $\hat{H}$ is in general unbounded. 
In particular, even if $\Psi_0$ is in the domain of $\hat{H}$, 
$\hat{H} \Psi_0$ need not be, so that it is not 
a priori clear how to interpret $U_t$. 

The mathematical machinery needed to resolve this problem 
was mainly developed by Stone and von Neumann in the years 
around 1930: It is possible to define $U_t$ on the entirety of $\hat{H}$, 
provided that $\hat{H}$ is not only \emph{symmetric}, i.e. 
\begin{equation}
	\forall \Psi, \Phi \in \dom \hat{H}\colon \quad 	
		\inp{\Psi}{\hat{H}\Phi}=\inp{\hat{H} \Psi}{\Phi}
	\, , 
\end{equation}
but \emph{self-adjoint}, i.e. that in addition $\dom \hat{H}$ 
is equal to the domain of its adjoint%
	\footnote{Following the definition in 
				Eq. \eqref{eq:domhatHstat}, the 
				adjoint $\hat{H}^\dagger$ of 
				$\hat{H}$ is then defined by setting 
				$\tilde{\Psi}=\hat{H}^\dagger \Psi$. }
\begin{equation}
	\dom \hat{H}^\dagger
		=  \set{\Psi \in \mathcal{H}}{\exists \tilde{\Psi} \in \mathcal{H} \, 
					\forall \Phi \in \dom \hat{H}\colon 
					\quad \langle \Psi, \hat{H} \Phi \rangle 
						= \langle \tilde{\Psi}, \Phi \rangle}
			\label{eq:domhatHstat}
\end{equation}
(cf. Secs. VIII.1 and VIII.4 in Ref. \cite{reedMethodsModernMathematical1972}). 
The generalized Schrödinger equation, Eq. \eqref{eq:generalSE}, 
is then only a formal PDE, that, if taken literally, only holds for 
initial states $\Psi_0 \in \dom \hat{H}$ 
(cf. Thm. VIII.7 (c) in Ref. \cite{reedMethodsModernMathematical1972} 
and Prop. \ref{Prop:ReedSimondomaininvariance} in Appx. 
\ref{appx:A})---despite the fact that $U_t \Psi_0$ is defined for 
any $\Psi_0 \in \mathcal{H}$. 

In principle this reduces the problem of solving 
Eq. \eqref{eq:generalSE} to determining the operator $U_t$. Yet in
practice one faces a number of hurdles. 

First, one needs to show 
that $\hat{H}$ is not merely 
symmetric but also self-adjoint on an appropriately chosen domain 
$\dom \hat{H}$. Heuristically, the smaller one chooses  
$\dom \hat{H}$, the larger $\dom \hat{H}^\dagger$ will become, so 
that the domain on which $\hat{H}$ is self-adjoint -- if it exists -- is the one 
on which $\hat{H}$ can still be sensibly defined and $\dom \hat{H}^\dagger$ 
has become small enough to be contained by $\dom \hat{H}$. 
Finding this domain might be too difficult, but, fortunately, there exist 
a number of theorems that allow one to prove the weaker condition of 
\emph{essential self-adjointness} on a much smaller domain. If a symmetric 
operator is essentially self-adjoint, then it 
can be uniquely extended to a larger domain on which it is self-adjoint. 
As this is only a sketch of the mathematical theory of quantum mechanics, 
we shall neither define essential self-adjointness 
here nor provide any such theorems. 
The reader is referred, for instance, to Sec. VIII.2 in Ref. 
\cite{reedMethodsModernMathematical1972} for the former and 
Ref. \cite{reedMethodsModernMathematical1975} for the latter. 
\begin{Example}
	\label{Ex:freeHam2}
\begin{subequations}
	In Ex. \ref{Ex:freeHam1}, $\hat{H}_0$ 
	is essentially self-adjoint on 
	$\dom \hat{H}_0 = \mathcal{S} \left(\R^3,\C \right)$. 
	We refer to Thm. IX.27 in Ref. 
	\cite{reedMethodsModernMathematical1975}. The domain 
	on which $\hat{H}_0$ is self-adjoint will be given in Rem. 
	\ref{Rem:Takabayasi} below.  
\end{subequations}
\end{Example}

The second hurdle one faces is that of finding `eigenfunctions' 
of $\hat{H}$, so that one can indeed compute $U_t$. 
The word is put in quotation marks, 
since -- even for self-adjoint $\hat{H}$ -- 
a maximal collection of linearly independent 
(orthonormal) eigenfunctions $\Phi_k \in \dom \hat{H}$, 
\begin{equation}
	\hat{H} \Phi_k = E_k \Phi_k
\end{equation}
with $E_k \in \R$ and $k$ in some countable index set $I$, 
need not be a (Schauder) basis for $\mathcal{H}$---i.e. they need 
not `span' the entire Hilbert space. Only if this is the case,  
the time evolution of a given $\Psi_0$ in $\mathcal{H}$ can be determined 
via 
\begin{equation}
	\Psi_t \equiv U_t \Psi_0 = 
	\sum_{k \in I}  
			\inp{\Phi_k}{\Psi_0} \, e^{- \iu E_k t/ \hbar} 
			\, \Phi_k
	\, . 
	\label{eq:timeevolHp}
\end{equation}
Still, in the general case of a self-adjoint, unbounded Hamiltonian $\hat{H}$
the space $\mathcal{H}$ can be decomposed 
into three, mutually orthogonal Hilbert subspaces of $\mathcal{H}$: 
\begin{equation}
	\mathcal{H} = \mathcal{H}_\text{p} \oplus \mathcal{H}_\text{s.c.} 
					\oplus \mathcal{H}_\text{a.c.} \, .
\end{equation}
The respective subspaces correspond to the decomposition of the 
spectrum of $\hat{H}$ into its point spectrum, singularly continuous 
spectrum, and absolutely continuous spectrum, respectively (see e.g. Sec. 28.2 
in Ref. \cite{blanchardMathematicalMethodsPhysics2015}). 
If the Hamiltonian $\hat{H}$ is `well-behaved' in some mathematical sense 
-- as it is the case for the model of `hydrogen-like atoms' -- 
then $\mathcal{H}_\text{s.c.} = \lbrace 0 \rbrace$,  the space 
$\mathcal{H}_\text{p}$ may be identified with the 
bound states, and $\mathcal{H}_\text{a.c.}$ with the scattering states 
in $\mathcal{H}$ (see e.g. Sec. 
28.3 in Ref. \cite{blanchardMathematicalMethodsPhysics2015}).%
	\footnote{We refer to Ref. \cite{pearsonSingularContinuousMeasures1978} 
				for a physical interpretation of 
				$\mathcal{H}_\text{s.c.} $. }
\begin{Example}
	\label{Ex:freeHam3}
			Continuing Ex. \ref{Ex:freeHam2}, we first recall that, due to 
			Plancherel's theorem, the Fourier transform%
				\footnote{As it is common in quantum mechanics, on 
							$L^2 \left(\R^3,\C\right)$ we 
							define the Fourier transform via 
							\begin{equation*}
							\left(\mathcal{F} \Psi\right) 
								(\vec{p} ) 
								= \frac{1}{{\sqrt{2 \pi \hbar}}^{
										\phantom{.}3}} 
									\int_{\R^3} \d^3 r 
									\, e^{-\iu \vec{p} \cdot \vec{r}/\hbar} 
									\, \Psi (\vec{r})
								\, . 
							\end{equation*}
							The (sign) convention is that the 
							Fourier 
							transform takes `position space wave 
							functions' to `momentum space wave 
							functions' 
							(cf. Chap. IV, \S 2 in Vol. 1 of Ref. 
							\cite{messiahQuantumMechanicsTwo1995}). 
							}
			$\mathcal{F}$ is a continuous, linear automorphism 
			of $\mathcal H = L^2 \left(\R^3,\C \right)$ 
			that is also unitary. Since for 
			all $\Psi \in \dom \hat{H}_0$ and $\vec{p} \in \R^3$ we have 
			\begin{equation}
				\left( \mathcal{F}
					\bigl(\hat{H}_0 \Psi \bigr) \right) \left(\vec{p}\right) = 
				\frac{\vec{p}^{\phantom{.} 2}}{2 m} \, \left( \mathcal{F} \Psi \right) 
				\left(\vec{p}\right) \, ,
				\label{eq:FourierfreeHam}
			\end{equation}
			the free Hamiltonian $\hat{H}_0$ is unitarily equivalent to a 
			multiplication operator. Since the spectrum of the latter is 
			easily shown to be 
			$[0, \infty)$, this is also the spectrum of $\hat{H}_0$. 
			By Thm. 28.7 in Ref. 
			\cite{blanchardMathematicalMethodsPhysics2015}, we have 
			$\mathcal{H}= \mathcal{H}_\text{a.c.}$. In fact, 
			all elements of $L^2 \left(\R^3,\C \right)$ 
			are scattering states of $\hat{H}_0$. 
\end{Example}	
		 
Even if $\hat{H}$ is not `well-behaved', the restriction of the 
self-adjoint 
Hamiltonian to the intersection of its domain with the respective Hilbert 
subspaces again defines a self-adjoint operator on said subspace 
(cf. Thm. 28.3 in Ref. \cite{blanchardMathematicalMethodsPhysics2015}). Furthermore, 
the time evolution operator $U_t$ preserves those subspaces and 
its restriction thereto is the time-evolution operator of the 
respective restricted Hamiltonian. Only in the subspace $\mathcal{H}_p$ 
this time evolution is given by Eq. \eqref{eq:timeevolHp} above---a fact 
that is sometimes glossed over in the physics literature 
(cf. Sec. XI.II in Ref. \cite{messiahQuantumMechanicsTwo1995} 
and Sec. 5.2 in Part 1 of 
Ref. \cite{takhtajanQuantumMechanicsMathematicians2008}). 

So assuming that $\hat{H}$ is well-behaved in the above sense, 
how do we determine $U_t$ for the scattering states? 
One possible approach is via the use of so-called  
\emph{generalized eigenfunctions}  
(see e.g. \S 1.6 and Supplement 1.2 in Ref. 
\cite{berezinSchrodingerEquation1991} as well as Chap. 29 in Ref. 
\cite{blanchardMathematicalMethodsPhysics2015}). 

This forces one to go beyond standard 
Hilbert space theory and to consider 
\emph{rigged Hilbert spaces} or \emph{Gelfand triples} 
instead---thus giving the aforementioned 
link to the mathematical theory of distributions. Without 
going too much into detail, the general idea is that one 
chooses an `appropriate space' of test functions 
$\mathcal{D}$ contained in the domain of $\hat{H}$, which 
in turn gives rise to a space of distributions $\mathcal{D}'$. 
Ignoring topological subtleties, the Gelfand triple 
is then given by $\left(\mathcal{D},\mathcal{H}, \mathcal{D}'\right)$.  
Each former space is viewed as being contained in the latter: 
$\mathcal{D}$ is trivially contained in $\mathcal{H}$, and 
$\mathcal{H}$ is contained in $\mathcal{D}'$, provided we view 
elements of the former as regular distributions in the latter. 
A generalized eigenfunction for a real number $\lambda$ in the 
spectrum of $\hat{H}$ is then a distribution $T_\lambda \in \mathcal{D}'$ 
such that for all $\varphi \in \mathcal{D}$ we have 
\begin{equation}
	T_\lambda \left( \hat{H} \varphi \right) 
	= \lambda \, T_\lambda \left(\varphi \right) \, .
	\label{eq:generalizedeigeneq}
\end{equation}
This construction allows one to allow for `eigenfunctions' 
that are distributions as opposed to merely elements of $\mathcal{H}$. 

We shall continue Ex. \ref{Ex:freeHam3} above to illustrate the 
central ideas. 
\begin{Example}
	\label{Ex:freeHam4}
\begin{subequations}
	We may formally compute 
	\begin{equation}
		\hat{H}_0 \, e^{\iu \vec{p} \cdot \vec{r}/\hbar} 
		= \frac{\vec{p}^{\phantom{.} 2}}{2 m} \, 
		 e^{\iu \vec{p} \cdot \vec{r}/\hbar} \, . 
	\end{equation}
	This suggests that for all $\vec{p} \in \R^3$ the function 
	\begin{equation}
		\Phi_{\vec{p}} \colon \quad \R^3 \to \C \quad
		\colon \quad
		\vec{r} \mapsto  \Phi_{\vec{p}} (\vec{r}) =
		e^{\iu \vec{p} \cdot \vec{r}/\hbar} 
	\end{equation}
	is a kind of `eigenfunction' of $\hat{H}_0$. Yet 
	$\Phi_{\vec{p}}$ is not square-integrable, so it cannot be 
	an eigenfunction in the strict sense. This means, in particular,  
	that $\Phi_{\vec{p}}$ is not a physically realized state. 
	
	Yet we may use the complex conjugate of 
	$\Phi_{\vec{p}}$ to define a regular distribution 
	$\tilde{\Phi}_{\vec{p}}$ 
	on $\mathcal{D}=\dom \hat{H}_0 = \mathcal{S}\left(\R^3,\C\right)$. 
	The complex conjugate is taken, as we heuristically think of the function  
	as being put in the first slot of $\inp{\, . \, }{\, . \,}$. 
	Up to a constant factor, $\tilde{\Phi}_{\vec{p}}$ 
	is the restriction of 
	$\mathcal{F} \left(\, . \, \right) (\vec{p})$ to 
	$\mathcal{S}\left(\R^3,\C\right)$.%
		\footnote{Note that the Fourier transform is also 
					a continuous 
					linear automorphism on the space of Schwartz 
					functions.}
	Eq. \eqref{eq:FourierfreeHam} 
	then shows that $\tilde{\Phi}_{\vec{p}}$ is a generalized 
	eigenfunction in the sense of Eq. \eqref{eq:generalizedeigeneq} 
	with eigenvalue $E(\vec{p})=\vec{p}^{\phantom{.}2}/2m$. 
	
	Using the above generalized eigenfunctions, for any 
	`initial' $\varphi_0 \in \mathcal{D}$ we may compute 
	\begin{equation}
		\tilde{\Phi}_{\vec{p}} \left( 
				\varphi_t \right)
			=\tilde{\Phi}_{\vec{p}} \left( \exp\bigl( 
							- \iu t \hat{H}_0/ \hbar \bigr)
				\varphi_0 \right)
			= \exp \left( - \iu t E ( \vec{p}) / \hbar\right)
					\tilde{\Phi}_{\vec{p}} \left( 
					\varphi_0 \right) \, .
				\label{eq:timeevolgeneigen1}
	\end{equation}
	Recalling the relation to the Fourier transform, we thus 
	find for all $\vec{r} \in \R^3$ that 
	\begin{equation}
		\varphi_t ( \vec{r}) 
			= 	
				\frac{1}{{\sqrt{2 \pi \hbar}}^{
										\phantom{.}3}} 
									\int_{\R^3} \d^3 p 
									\, \left(\mathcal{F}\varphi_0 \right)
										(\vec{p})  
									\,
									e^{- \iu t E ( \vec{p}) / \hbar} 
									\, 
									e^{\iu \vec{p} \cdot \vec{r}/\hbar}
									\, .
				\label{eq:timeevolgeneigen2}
	\end{equation}
	As Eqs. \eqref{eq:timeevolgeneigen1} and 
	\eqref{eq:timeevolgeneigen2} also make sense for 
	$\varphi_0 \in \mathcal{H}$ and the respective extension 
	is continuous, we find that Eq. \eqref{eq:timeevolgeneigen2}
	defines the time evolution on 
	$\mathcal{H}=\mathcal{H}_\text{a.c.}$
	in analogy to Eq. \eqref{eq:timeevolHp} 
	(cf. Sec. 1.6 in Ref. \cite{berezinSchrodingerEquation1991}). 
\end{subequations}
\end{Example}

We refer to Supplement 
1 in Ref. \cite{berezinSchrodingerEquation1991}, 
Chap. 29 in Ref. 
\cite{blanchardMathematicalMethodsPhysics2015}, as well as 
the works \cite{madridmodinoQuantumMechanicsRigged2001,
madridRoleRiggedHilbert2005,madridRiggedHilbertSpace2004} by de la Madrid 
for further reading. Secs. 8 and 9 in Chap. V of Ref. 
\cite{messiahQuantumMechanicsTwo1995}
provide a heuristic introduction to the concept of generalized 
eigenfunctions. 

Apart from the need for generalized eigenfunctions 
in scattering problems, we would like to mention two more points 
that underscore the statement that quantum mechanics relies on the theory 
of distributions in its mathematical formulation: 

First, one may recall that the use of 
so called `delta-potentials' is common in quantum-mechanical 
models 
(see e.g. Secs. 2.5 and 5.3.2 as well as 
Problems 2.50, 6.1, 6.8, and 11.4 
in Ref. \cite{griffithsIntroductionQuantumMechanics2005}). 
Though such potentials are certainly idealizations 
of more realistic (still idealized)  
potentials, such as the one in the Kronig-Penney model 
\cite{kronigQuantumMechanicsElectrons1931}, they nevertheless require 
a mathematically rigorous treatment. 
 
The second argument is that in quantum theory 
weak and distributional derivatives -- not strong derivatives -- 
are the appropriate notion of derivative to use: 

Recall that the common Hilbert spaces in quantum mechanics are  
$L^2(\R^n,\C)$ and its relatives, as exemplified by Ex. 
\ref{Ex:freeHam1} above. Strictly speaking, 
in order to assure that $L^2(\R^n,\C)$ is indeed complete with 
respect to the natural norm (and thus a Hilbert space), elements 
of $L^2(\R^n,\C)$ are not functions $\Psi$ but equivalence classes 
thereof, denoted by $[\Psi]$. 
By definition, we consider two square-integrable functions 
$\Psi$ and $\Psi'$ to be equivalent 
-- i.e. they correspond to the same element 
$[\Psi]=[\Psi']$ in $L^2(\R^n,\C)$ -- 
if they differ at most on a set of measure zero. The importance 
of this mathematical subtlety can for instance be observed 
in the theory of Fourier series, where the 
Fourier series of a function need not equal the function at 
every point---the so called Gibbs phenomenon. Similarly, an energy eigenfunction 
expansion of a square integrable function in quantum mechanics, if it exists,  
may differ from the original function on a set of measure zero. 

Indeed, the consideration of equivalence classes of functions as opposed to 
functions themselves is not merely  
mathematical pedantry: Events of probability zero, 
that is changes of the probability density 
on a set of measure zero, should neither change our statistical 
description nor how it evolves in time. Considering equivalence 
classes of functions is an appropriate mathematical expression thereof.

Therefore, if a wave function is supposed to be an element of 
$\mathcal{H}$, it is -- strictly speaking -- not admissible to speak 
of the value of a wave function at an individual point, let alone 
evaluate its strong derivative.
Yet, provided we have 
specified an appropriate space $\mathcal{D}$ of `test functions', 
we may consider its distributional derivative, and, if it exists, 
its weak derivative. 

Surely, in many cases of practical interest the aforementioned 
mathematical subtlety is not of much relevance. Yet it cannot be glossed 
over in discussions concerning 
the mathematical structure of quantum mechanics. 
In this spirit, we shall make the following remark on 
Takabayasi's condition from Sec. 
\ref{ssec:WallstromProp}. 
\begin{Remark}[Limitations of Takabayasi's quantization condition]
	\label{Rem:Takabayasi}
\begin{subequations}
	In Sec. \ref{ssec:WallstromProp} we explained why 
	the generalization of Prop. \ref{Prop:C1quantization} 
	to wave functions that are less regular than $C^1$ 
	is not trivial. We shall show here why there 
	are serious doubts on whether the condition in Eq.  
	\eqref{eq:wallstrom} can be appropriately 
	extended to be sensible within 
	the general mathematical theory of quantum mechanics. 
	
	We shall start off by noting that the above statement that 
	wave functions may not be evaluated at a point 
	does have an important limitation: 
	It may happen that an equivalence class 
	$[\Psi] \in \mathcal{H}$ has a representative $\Psi$
	that satisfies certain continuity or even 
	differentiability assumptions. Generally speaking, 
	such representatives are unique. In that case, we 
	can, of course, evaluate the function $\Psi$ and 
	ask if its strong derivatives exist.
	If those derivatives indeed exist everywhere, then 
	$[\Psi]$ is weakly differentiable up 
	to the same order and 
	the (the equivalence classes of the) 
	strong derivatives of $\Psi$ are equal to the 
	respective 
	weak derivatives (as equivalence classes). 

	An important mathematical result that should be mentioned 
	in this context is the \emph{Sobolev embedding theorem}. 
	We refer, for instance, to Chap. V in Ref.
	\cite{adamsSobolevSpaces1975}. Thm. 1.1 in Supplement 2 of 
	Ref. \cite{berezinSchrodingerEquation1991} provides a 
	result on special cases of interest to quantum mechanics, 
	namely for the Sobolev spaces 
	$H^k \left(\R^n,\C\right)$:  
	
	For $k \in \N_0$ and $n \in \N$ the space 
	$H^k \left(\R^n,\C\right)$ is defined as the set of 
	all equivalence classes $[\Psi]$ 
	of functions $\Psi \colon \R^n \to \C$ such that 
	both $[\Psi]$ and all of its $k$th order weak derivatives are 
	in $L^2(\R^n,\C)$. Each $H^k \left(\R^n,\C\right)$ is 
	equipped with the inner product 
	\begin{equation}
		\inp{\, . \,}{\, . \, }_{H^k}
		\colon \colon \quad 
			(\Phi,\Psi) \mapsto 
			 \inp{\Phi}{\Psi}_{H^k}
			= \sum_{\abs{\alpha}\leq k} 
			\inp{\partial^\alpha \Phi}{\partial^\alpha \Psi}\, ,
			\label{eq:Hkinp}
	\end{equation}
	$\partial^\alpha$ denoting the partial derivative with respect to the 
	multi-index $\alpha$. 
	With respect to this inner product and 
	the induced norm, every $H^k \left(\R^n,\C\right)$ is a separable 
	Hilbert space (cf. Sec. 1.4.1 in Ref. 
	\cite{bahouriFourierAnalysisNonlinear2011}). 
	
	It is, however, important to understand that the inner product 
	in Eq. \eqref{eq:Hkinp} is only of direct interest in quantum 
	mechanics for $k = 0$, i.e. for the case of 
	$H^0 \left(\R^n,\C\right)=L^2 \left(\R^n,\C\right)$. 
	While for $k \geq 1$ the $H^k$-spaces are not appropriate Hilbert spaces 
	of quantum-mechanical states, they are nonetheless of use 
	for mathematical analysis within the theory. 
	This is due to the fact that, for the 
	case of $1$-body in $\R^3$, the respective free body 
	Hamiltonian $\hat{H}_0$ with domain 
	$H^2 \left(\R^3, \C\right)$ 
	is self-adjoint on $L^2 \left(\R^3, \C\right)$ 
	(cf. Thm. IX.27 in Ref. \cite{reedMethodsModernMathematical1975})
	and that, 
	in many cases, Hamiltonians with potentials 
	may be viewed as perturbations of $\hat{H}_0$ 
	that preserve self-adjointness (cf. 
	Sec. X.2 in Ref. \cite{reedMethodsModernMathematical1975}).  			
	
	The Sobolev 
	embedding theorem now states that every $[\Psi] \in 
	H^2 \left(\R^3, \C\right)$ there exists a representative 
	$\Psi$ that is bounded and Lipshitz-continous 
	(cf. Thm. 
	5.4 in Ref. \cite{adamsSobolevSpaces1975}, Part I Case C 
	and Part II Case). It does not, however, imply that 
	a $C^1$ representative exists---which would be needed to 
	establish Takabayasi's quantization 
	condition for elements in the domain of such perturbed 
	Hamiltonians. 
	
	If we are considering general elements of the Hilbert 
	space $L^2 \left(\R^3,\C \right)$, the situation is even worse: 
	Functions whose equivalence class lies in that space 
	are generally very ill-behaved and not even 
	continuous. Moreover, they can be changed on 
	an arbitrary set of measure zero, so even if 
	the integral in Takabayasi's condition, 
	Eq. \eqref{eq:wallstrom}, is well-defined for 
	a given representative 
	it can be chosen to yield \emph{any} number in 
	$\R$ without changing the equivalence class. 
\end{subequations}
\end{Remark}

In conclusion and returning to the main subject of this work, 
if we intend to partially reproduce or even generalize 
quantum mechanics on the basis of the Madelung equations, 
then we need to consider the latter as a system of PDEs that is to  
be understood in the sense of distributions. 
The existence of solutions to the latter 
as well as the (in)equivalence to the Schrödinger equation 
therefore fundamentally depends on how we formulate the 
equations, our choice of solution spaces for the quantities in 
question as well as for their initial values, and also the boundary 
conditions we may wish to encode in those solution spaces. In  
Sec. \ref{sec:distmadelung} below we shall give explicit 
examples of such `distributional Madelung equations'. 

\subsection{Quantum vorticity and quantum quasi-irrotationality}
\label{ssec:vorticity}
  
As argued in the previous section, the Madelung equations need to 
be understood as distributional differential equations. Due to a lack of 
regularity of the respective drift fields or wave functions -- 
depending, of course, on the choice of function spaces 
-- Takabayasi's condition, Eq. \eqref{eq:wallstrom}, 
may therefore not be universally applicable 
(cf. Rem. \ref{Rem:Takabayasi}). 

One may, however, ask the question, if one can generalize 
Takabayasi's condition in a manner that does not depend on the  
regularity of those (stationary) 
solutions of the Madelung equations it was motivated by. 
Given the discussion in Sec. \ref{ssec:distributions} above, 
a distributional approach to this question seems natural. 
Furthermore, the considerations in Sec. \ref{ssec:WallstromProp} 
imply a close relationship between Takabayasi's condition and the third 
Madelung equation, Eq. \eqref{eq:Madelung3}. We are thus lead to investigate 
whether those model solutions 
indeed satisfy the third Madelung equation in a distributional sense 
or not. 

The particular model solutions we consider here 
are stationary solutions of the isotropic harmonic oscillator 
in $2$ dimensions, as well as 
the functions $\Psi_{n l \mu}$ in the aforementioned model for 
hydrogen-like atoms. Consideration of the former model in this context 
was suggested by Wallstrom in Ref. 
\cite{wallstromInequivalenceSchrodingerEquation1994}, the 
latter model we chose as we view it to be of more direct physical 
relevance.

We will find that for both of those classes of solutions the third 
Madelung equation does  
in general not hold. Moreover, the results suggest the possibility 
to generalize Takabayasi's condition to the wider class of 
drift fields $\vec{v}$ whose (Euclidean) components are locally 
($L^1$-)integrable. 

In the first model we 
consider the time-independent Schrödinger equation for a single isotropic  
harmonic oscillator 
in $2$ dimensions with frequency $\omega \in \R_+$ and in polar coordinates 
$(\rho, \phi)$: 
\begin{equation}
	E \, \Psi = - \frac{\hbar^2}{2 m}
	\left( \frac{1}{\rho} \partd{}{\rho}\left( \rho \, \partd{\Psi}{\rho} \right) 
	+ \frac{1}{\rho^2} \partd{^2 \Psi}{\phi^2}\right)
	+ \frac{m \omega^2}{2} \, \rho^2 \, \Psi \, .
	\label{eq:oscillatortindepSE}
\end{equation}
Though the problem of finding energy eigenstates 
is arguably easier to solve in Cartesian coordinates, 
one commonly approaches the above problem via a
separation ansatz
\begin{equation}
	\Psi(\rho, \phi) = R(\rho) \, \Phi(\phi) \, . 
	\label{eq:sepansatz}
\end{equation}
The latter leads to an angular equation
\begin{equation}
	\frac{\d^2 \Phi}{\d \phi^2} + \mu^2 \Phi = 0
	\label{eq:angularSE}
\end{equation}
as well as a radial equation%
	\footnote{ Note that the radial equation, Eq. 3.1 in Ref. 
				\cite{wallstromInequivalenceSchrodingerEquation1994}, 
				is incorrect.}
\begin{equation}
	E \, R = - \frac{\hbar^2}{2 m}
	\left( \frac{\d^2 R}{\d \rho^2} + \frac{1}{\rho} 
	\frac{\d R}{\d \rho}
	- \frac{\mu^2}{\rho^2} R \right)
	+ \frac{m \omega^2}{2} \, \rho^2 \, R
	\label{eq:radialSE}
\end{equation}
for some constant $\mu \in \C$. 

It is not the focus of this section to consider the question of if and why 
$\mu$ has to be `quantized' here, i.e. if and why $\mu \in \Z$. 
We refer to 
Sec. \ref{sec:nonquantumsol} for a discussion of this point. 
Instead we use stationary solutions of the respective Schrödinger equation with 
$\mu \in \Z$, take the respective densities $\varrho$ and drift fields 
$\vec{v}$ and then use those to look at the third Madelung 
equation from a distributional point of view. 

For $n \in \N_0$, $\mu \in \Z$, and some normalization 
factor $1/A_{n \mu}$ the functions $\varrho$ and $\vec{v}$,  
taking values 
	\begin{equation}					
			\varrho(x,y)	= 
							\frac{1}{A_{n \mu}}
							\,	(x^2+y^2)^{\abs{\mu}}  \, 
								\left( 
									L^{\abs{\mu}}_n \left(  
									\frac{m \omega}{ \hbar} \, (x^2+y^2)
									\right) \right)^2
									\, e^{- \frac{m \omega}{ \hbar} 
									\, (x^2+y^2)} 
									\phantom{ab}
					\label{eq:varrhostandardoscillator}
	\end{equation}
and 
	\begin{equation}
						\vec{v}(x,y) = \frac{\mu \hbar}{m} \, 
								\frac{1}{ x^2 + y^2}
								\begin{pmatrix}
									- y \\ 
									x
								\end{pmatrix} 
					\label{eq:vstandardoscillator}
	\end{equation}	
for $(x,y) \in \R^2 \setminus \lbrace 0 \rbrace$, 
provide strong, stationary solutions to the Madelung equations 
for this problem.%
	\footnote{To avoid cluttered notation and also due to the 
				lack of relevance of the 
				trivial time-dependence here, we drop the latter.
				\label{fn:tdependence}}  
We refer to Sec. \ref{ssec:strongnonquantumsol}  
below for the corresponding solutions of the Schrödinger 
equation. 

To analyze the above strong solutions from a distributional 
perspective and also to prepare for a similar analysis of the 
physically more interesting, aforementioned hydrogen model, 
we shall make the following definition: 
\begin{Definition}
	\label{Def:dcurl}
\begin{subequations}	
	\begin{enumerate}[1)]
		\item 
		Let $\vec{v}$ be a (real-valued) 
		vector field on $\R^2$ (or an equivalence class thereof, 
		as discussed in Sec. \ref{ssec:distributions})
		such that each component, $v^1$ and $v^2$, is locally 
		$L^1$-integrable. Let $\mathcal{D}$ 
		be either $C_0^\infty(\R^2,\R)$ or 
		$\mathcal{S}(\R^2,\R)$. 
		The \emph{distributional curl of $\vec{v}$} is 
		the distribution $\operatorname{curl} \vec{v}$ 
		defined via 
		\begin{equation}
			\left( \operatorname{curl} \vec{v} \right) (\varphi)
			= - \int_{\R^2} \d x \, \d y \, 
				\left(v^2 \partd{}{x} \varphi - 
				 	v^1 \partd{}{y} \varphi \right)
		\end{equation}
		for all $\varphi \in \mathcal{D}$. 
		\item 
		Let $\vec{v}$ be a (real-valued) 
		vector field on $\R^3$ 
		(or an equivalence class thereof) 
		such that each component is locally 
		$L^1$-integrable. Let $\mathcal{D}$ 
		be either $C_0^\infty(\R^3,\R^3)$ or 
		$\mathcal{S}(\R^3,\R^3)$. 
		Then the \emph{distributional curl of $\vec{v}$} is 
		the distribution 
		\begin{equation}
			\nabla \cross \vec{v} 
			\colon \quad \mathcal{D} \quad \to \quad \R^3
		\end{equation} 		
		defined via 
		\begin{equation}
			\left( \nabla \cross \vec{v}\right) (\vec{\varphi})
			= \int_{\R^3} \d^3 r \, 
				\left(\vec{v} \cross \nabla \right)
				\vec{\varphi}
				\label{eq:defcurlv3d}
		\end{equation}
		for all $\vec{\varphi} \in \mathcal{D}$. 
	\end{enumerate}
\end{subequations}
\end{Definition}

Of course, the definitions in Def. \ref{Def:dcurl} were 
chosen such that they provide an adequate distributional 
generalization of the respective curl operator. That this is indeed the case 
can be checked by assuming that $\vec{v}$ is weakly 
differentiable in the sense that all component functions 
$v^i$ are weakly differentiable (`integrating by parts'). 

Let us now compute the distributional curl of $\vec{v}$ from 
Eq. \eqref{eq:vstandardoscillator} above. We shall find that it is 
proportional to the $2$-dimensional Dirac 
distribution at the origin $\delta_0$
(cf. Footnote \ref{fn:Diracdelta}). 
\begin{Proposition}
	\label{Prop:wallstromv}
	The distributional curl of $\vec{v}$ from 
	Eq. \eqref{eq:vstandardoscillator} above equals 
		\begin{equation}
			\operatorname{curl} \vec{v} = 
			\frac{2 \pi \mu \hbar}{m} \, \delta_0 \, . 
			\label{eq:wallstromv}
		\end{equation}
	In particular, $\vec{v}$ is not weakly differentiable. 
\end{Proposition}

Generally, we say that 
a drift field $\vec{v}$ exhibits \emph{quantum vorticity} (at a given time), 
if its distributional curl does not vanish (at that time). 

According to 
Prop. \ref{Prop:wallstromv}, the drift field 
$\vec{v}$ from Eq. \eqref{eq:vstandardoscillator} does indeed 
exhibit quantum vorticity. This vorticity is concentrated at the 
origin in the sense that the distributional support of 
$\operatorname{curl} \vec{v}$ is the singleton 
$\lbrace 0 \rbrace \subset \R^2$.%
		\footnote{	
					For $n \in \N$ and a given space $\mathcal{D}$ of 
					test functions on $\R^n$ a distribution $T$ is 
					said to \emph{vanish} on an open set $U \subseteq \R^n$, 
					if for all $\varphi \in \mathcal{D}$ with 
					support $\supp \varphi$ contained in $U$ we have 
					$T(\varphi) = 0$. Then the 
					\emph{support of a distribution} 
					is defined as the complement of the union of all open 
					sets on which $T$ vanishes. 
					\label{fn:supportdist}
					} 
This observation motivates the following definition. 

\begin{Definition}
	\label{Def:quasiirrot}
\begin{subequations}
	A locally integrable vector field $\vec{v}$, as given 
	in Def. \ref{Def:dcurl} above, 
	is \emph{quasi-irrotational}, if its distributional 
	curl is supported on a non-empty set of 
	Lebesgue measure zero. It is \emph{irrotational}, if 
	its distributional curl vanishes entirely. 
\end{subequations}
\end{Definition}

The above terminology was inspired by 
a statement made by Schönberg \cite{schonbergVortexMotionsMadelung1955} 
in the 1950s: 	
		\begin{quote}
				The presence or absence of vorticity is not 
				the fundamental fact, since the Schrödinger 
				equation may be applicable even when there is 
				vorticity, provided the motion be 
				quasi-irrotational. 
		\end{quote}
While Def. \ref{Def:quasiirrot} does not capture Schönberg's 
original intentions behind those words, our results suggest 
that 
the statement nonetheless holds true if reinterpreted in this sense 
(even in the absence of magnetic fields, cf. 
Ref. \cite{schonbergVortexMotionsMadelung1955}).  

Given a quasi-irrotational drift field 
$\vec{v}$ in the sense of Def. \ref{Def:quasiirrot}, 
we call a connected component of the support of 
its distributional curl a \emph{quantum vortex}. Note that the above 
definition does not imply any restriction on 
the number of such connected components, so, at least in principle,  
$\vec{v}$ could have infinitely many quantum vortices. 

The consequences of Prop. \ref{Prop:wallstromv} 
for the $2$-dimensional ($1$-body) 
Madelung equations are as follows: 

If we view the above solutions $(\varrho,\vec{v})$ 
from Eq. \eqref{eq:varrhostandardoscillator} and 
\eqref{eq:vstandardoscillator} as physically acceptable -- 
which is not beyond dispute -- 
then we cannot simply carry the third Madelung equation 
over to the 
distributional setting. As Prop. \ref{Prop:wallstromv} 
shows then, it does not hold for such solutions, for $\vec{v}$ is only 
quasi-irrotational. 

Of course, one could have constructed a basis of 
real-valued wave functions for this problem instead 
and still obtained corresponding stationary solutions 
of the Madelung equations. As the 
respective drift field is zero in this instance, there are 
not any quantum vortices and the third Madelung equation is 
also satisfied in the distributional sense. In fact, 
in the absence of magnetic fields%
	\footnote{In the presence of magnetic fields 
				the third Madelung equation does not even hold 
				in the strong sense. See e.g. Refs.   
				\cite{schonbergVortexMotionsMadelung1955,
				janossyHydrodynamischenModellQuantenmechanik1962,
				janossyHydrodynamicalModelWave1964}.} 
a given basis of energy eigenfunctions can always be 
transformed into a basis of real-valued energy eigenfunctions, 
so that this behavior is generic  
(cf. Sec. 5.6 in Ref. 
\cite{shankarPrinciplesQuantumMechanics1994}). We shall elaborate 
on this point below as well as in Sec. \ref{ssec:distnonquantumsol}. 

With regards to the issue of `quantization', we observe 
that $\operatorname{curl} \vec{v}$ 
is proportional to the magnetic quantum number $\mu$ 
in this instance---just like the respective 
integral in Takabayasi's condition,  Eq. \eqref{eq:wallstrom}. 
Therefore, in this distributional setting 
the expression for 
$\operatorname{curl} \vec{v}$ allows one to 
distinguish between the different stationary solutions 
$(\rho,\vec{v})$ above  on the level of the Madelung equations---albeit 
the expressions for $\vec{v}$ and 
$\operatorname{curl} \vec{v}$, Eqs. \eqref{eq:vstandardoscillator} and 
\eqref{eq:curlvcompute2}, in principle also make sense 
for non-integer $\mu$. 

We shall not address the general question of how the distributional curl 
behaves under superposition, as this would go beyond the scope 
of this work. Whether it is possible to generalize Prop. 
\ref{Prop:wallstromv} to general $2$-dimensional,  
not necessarily stationary, states therefore remains to be shown. 

There remains the question of 
the physical meaning of the distribution 
$\operatorname{curl} \vec{v}$. As this is 
not a purely mathematical question, we shall move on 
to an arguably more physical example instead and take 
up this discussion there again.  

As already indicated, this example is the 
$1$-body Schrödinger theory with attractive 
Coulomb interaction for `hydrogen-like atoms'. 
While it is true that this is more adequately 
treated as a $2$-body problem, it is known 
that the high mass of the 
proton allows one to treat it as effectively 
fixed in space, so that we may consider the 
physical situation as a $1$-body problem 
instead.

Denote by $n$, $l$, and 
$\mu$ the principal, angular momentum, and 
magnetic quantum number, respectively, and by 
\begin{equation}	
	a_0 = \frac{4 \pi \varepsilon_0 \hbar^2}{m \operatorname{e}^2}
	\label{eq:Bohrradius}
\end{equation}
the Bohr radius (in SI units) for this model. $a_0$ is the 
Bohr-radius for hydrogen, for hydrogen-like atoms one needs to further
divide this number by the number of protons in the nucleus. For   
$n \in \N_0$, $l \in \lbrace 0, \dots, n-1 \rbrace$, 
$\mu \in \lbrace -l, \dots, l \rbrace$, 
and $\vec{r}=(x,y,z)$ in $\R^3$ excluding the $z$-axis, we set 
\begin{equation}
	\varrho(\vec{r}) 
		= \frac{1}{A_{n l \mu}} 
			\, 
			\left(\lambda(\vec{r}) \right)^{2 l} 
			\, 
			\left( L_{n-l-1}^{2 l +1} (\lambda(\vec{r})) \right)^2
			\,
			e^{-\lambda(\vec{r})}
			\,
				\bigl( P_l^{\abs{\mu}} \left(z / \abs{\vec{r}} 
			\right)\bigr)^2 
			\label{eq:varrhostandardhydrogen}
\end{equation}
with $\lambda(\vec{r}) = 2 \abs{\vec{r}} / (n a_0)$ and some normalization 
factor $1/A_{n l \mu} \in \R_+$, and we set 
\begin{equation}
	\vec{v}(\vec{r}) =  
						\frac{\mu \hbar}{m} \, 
								\frac{1}{ x^2 + y^2}
					\begin{pmatrix}
							- y \\ 
							x \\
							0
					\end{pmatrix} \, .
			\label{eq:vstandardhydrogen}
\end{equation}
If one makes the quantities $\varrho$ and $\vec{v}$ 
trivially dependent on time (see footnote \ref{fn:tdependence}), then 
they provide 
stationary strong solutions of the respective Madelung 
equations (cf. Vol. I, Chap. XI, \S 6 and Appx. B, \S I.2 in Ref. 
\cite{messiahQuantumMechanicsTwo1995},  
Eqs. 7.34 and 10.31 in Ref. \cite{ballentineQuantumMechanicsModern1998}, 
Eqs. A.5.7 and A.6.3 in Ref. \cite{sakuraiModernQuantumMechanics1985}), 

Given the close relationship between the drift fields from Eq. 
\eqref{eq:vstandardoscillator} 
and Eq. \eqref{eq:vstandardhydrogen}, we can easily compute 
the distributional curl of the latter. 

\begin{Corollary}
	\label{Cor:hydrogenvdistder}
\begin{subequations}	
		\label{eq:Corhydrogenvdistderequs}
		Let $\vec{v}$ be as in Eq. \eqref{eq:vstandardhydrogen}. 
		Denote by $\varphi^3$ the third component of 
		a given vector-valued test function $\vec{\varphi} \in \mathcal{D}$. 
		
		Then the distributional curl of $\vec{v}$ is given by 
		\begin{equation}
			\nabla \cross \vec{v}  
				= \frac{2 \pi \mu \hbar}{m} \, 
								\begin{pmatrix}
									0 \\ 
									0 \\
									\xi 
								\end{pmatrix} \, , 
							\label{eq:dcurlvhydrogen}
		\end{equation}
		with 
		\begin{equation}
			\xi \left(\vec{\varphi}\right)= \int_{- \infty}^\infty \d z \, 
					\varphi^3(0,0,z) 
					\label{eq:deltazaxis}
		\end{equation}
		for all $\vec{\varphi} \in \mathcal{D}$. 
		In particular, $\vec{v}$ is not weakly differentiable. 
\end{subequations}
\end{Corollary}

Cor. \ref{Cor:hydrogenvdistder} implies that the drift field from Eq. 
\eqref{eq:vstandardhydrogen} also exhibits quantum vorticity and that 
this vorticity is concentrated on the $z$-axis. Therefore, 
the drift field is again quasi-irrotational in the sense of 
Def. \ref{Def:quasiirrot}. 

Thus, in this physical model and for the solutions above, 
the third Madelung equation does not hold---as it was the case for 
the previous $2$-dimensional model. Much of the above 
discussion on the $2$-dimensional model carries over to this case, 
so we shall not repeat those points here. 

A notable difference is that 
the additional degree of freedom in the $3$-dimensional 
case suggests a larger diversity of quantum vortices compared to the 
$2$-dimensional case: According to Bialynicki-Birula et al. 
\cite{bialynicki-birulaMotionVortexLines2000}, the drift field of 
the normalized state 
\begin{equation}
	\frac{1}{\sqrt{2}} \left( \Psi_{2 0 0 } + e^{\iu \pi / 2} \, 
				\Psi_{2 1 0 } \right)
\end{equation}
has a `vortex ring' of radius $2 a_0$ in the $z=0$ plane. 
Whether this is indeed a quantum vortex in the sense of 
Def. \ref{Def:quasiirrot} would need to be shown. In an earlier article 
\cite{bialynicki-birulaMagneticMonopolesHydrodynamic1971} the same first two 
authors suggest that all quantum vortices in $3$ dimensions are 
`singularities of the $\delta^{(2)}$ type', yet this, too, remains to 
be rigorously defined and proven. In Ref. 
\cite{hirschfelderAngularMomentumCreation1977} Hirschfelder 
implicitly suggested that quantum vortices in $3$ dimensions 
are topological $1$-manifolds (cf. p. 5478 therein), on 
which he based the assertion that every such quantum vortex is 
homeomorphic to either the real line 
(`axial vortex') or to the unit circle 
$\mathbb{S}^1$ (`toroidal vortex'). Due to the 
Level Set Theorem (cf. Thm. 1.2.1 in Ref. 
\cite{rudolphDifferentialGeometryMathematical2013}), this is 
indeed the case if $0$ is a regular value of a $C^1$ 
wave function (possibly restricted in domain). It is an open 
question whether it holds, 
for instance, if we significantly weaken the assumption on 
the wave function -- say, if $[\Psi]$ is an element of the 
Sobolev spaces $H^1(\R^3,\C)$ or $H^2(\R^3,\C)$. 

Let us now consider the physical interpretation 
of the distributional curl in this $3$-dimensional model. 
We suggest two closely related interpretations. 

First we may view Eq. \eqref{eq:dcurlvhydrogen} as an 
idealized expression representing a limit case. A good analogy is the 
view of a Dirac delta as an idealization of a Gaussian function 
with a very narrow peak in the limit where the width tends to zero 
and the volume integral over the function is kept constant. Indeed, 
in an affirmation to the above comment made by 
Bialynicki-Birula and Bialynicka-Birula 
\cite{bialynicki-birulaMagneticMonopolesHydrodynamic1971},   
one may view the distribution $\xi$ from Eq. 
\eqref{eq:deltazaxis} as a Dirac delta on the $z$-axis, 
so that in `physics notation' one formally obtains 
Eq. \eqref{eq:physicscurlvhydrogen}. 
If one is willing to accept the 
interpretation of Eq. \eqref{eq:physicscurlvhydrogen} as a limiting 
expression, then it forces 
one to view the third Madelung equation, interpreted in the 
distributional sense, not as a general 
equation, but one that only holds in particular cases. 

This view is indeed suggested by the the first Madelung equation, Eq. 
\eqref{eq:Madelung1}: If, in the smooth setting, one takes the curl thereof 
and denotes the vorticity $\nabla \cross \vec{v}$ by $\vec{\omega}$, then 
the following evolution equation is obtained: 
\begin{equation}
	\partd{\vec{\omega}}{t} = \nabla \cross \left( \vec{v} \cross \vec{\omega} 
	\right) 
	\label{eq:omegaevolution}
\end{equation}
(see Eqs. 26 and 27 in Ref. 
\cite{tessarottoInverseKineticTheory2007} and 
Eq. 92 in Ref. \cite{nottaleGeneralizedQuantumPotentials2009}). 
An equivalent formulation is given by 
\begin{equation}
	\partd{\vec{\omega}}{t} + \left(\vec{v} \cdot \nabla \right) \vec{\omega} 
	= \left(\vec{\omega} \cdot \nabla \right) \vec{v} - 
		\left(\nabla \cdot \vec{v} \right) \, \vec{\omega} \, . 
\end{equation}
As the constraint $\omega = 0$ trivially solves Eq. \eqref{eq:omegaevolution}, it is 
at least consistent with the first Madelung equation. Yet, in principle 
-- even in the strong sense -- the initial data $\vec{v}_0$ does not need to be vorticity-free, and neither does $\vec{v}$ as part of   
a (strong) solution $\left(\varrho,\vec{v} \right)$ of the first two Madelung equations, 
Eqs. \eqref{eq:Madelung1} and \eqref{eq:Madelung2}. Furthermore, 
if one takes the heuristic view \cite{reddigerMadelungPictureFoundation2017} 
that the Madelung equations are evolutionary equations for 
ensembles of single point masses with probability density $\varrho(t, \, . \,)$ at 
time $t$ with the evolution of $\varrho$ being 
governed by the flow of the vector field $\vec{v}$, 
then the constraint imposed by the third Madelung equation, 
Eq. \eqref{eq:Madelung3}, seems somewhat ad hoc. Why should only irrotational 
drift fields be allowed? 

The idea of dropping the third Madelung equation 
is far from novel: Takabayasi already suggested such 
`vorticial flow' in 1952 (cf. \S 13 in Ref. 
\cite{takabayasiFormulationQuantumMechanics1952} and Appx. E in 
Ref. \cite{takabayasiRemarksFormulationQuantum1953}). Freistadt has credited  
a 1954 article by Bohm and Vigier 
\cite{bohmModelCausalInterpretation1954} for suggesting 
"that only the \flqq{}smeared out\frqq{} flow might be irrotational 
	[...] while there might be vortices on a small scale" (cf p. 14. in 
	Ref. \cite{freistadtCausalFormulationQuantum1957}). Schönberg 
also suggested the consideration of rotational drift fields in 
1954 
(cf. p. 119 in Ref. \cite{schonbergHydrodynamicalModelQuantum1954} 
and p. 567 in Ref. \cite{schonbergVortexMotionsMadelung1955}). 

Since there are no universally agreed upon `distributional Madelung 
equations', it is an open question, whether an equation analogous to Eq. 
\eqref{eq:omegaevolution} holds in a distributional sense. 
With regards to the results of this section, 
it is possibly the case, that a consideration of 
the vorticity of locally integrable drift fields $\vec{v}$ 
will play a role in establishing a precise mathematical relationship 
between such distributional Madelung equations and the Schrödinger 
equation. 

In this respect, it is worth mentioning that an 
even more general approach was suggested by Loffredo 
and Morato in the context of stochastic mechanics (cf. 
Refs. \cite{loffredoLagrangianVariationalPrinciple1989,
loffredoCreationQuantizedVortex1992} as well as 
Ref. \cite{moratoPathwiseStochasticCalculus1985}): They suggested  
that in the case of non-vanishing vorticity $\vec{\omega}$ an additional 
$\vec{\omega}$-dependent 
term needs to be added to the first Madelung equation. The authors 
state that ``any solution to these equations, with a generic 
rotational velocity field, relaxes towards a standard 
solution with irrotational velocity field associated to a 
quantum state'' 
(cf. p. 209 in Ref. \cite{loffredoCreationQuantizedVortex1992}). 
While they did allow drift fields such as the ones in Eq. 
\eqref{eq:vstandardhydrogen}, they did not elaborate on 
Takabayasi's condition. Nonetheless, their work shows that 
allowing for `vorticial flows' may not be as simple as 
dropping Eq. \eqref{eq:Madelung3}, so that a 
modification of the first Madelung equation, 
Eq. \eqref{eq:Madelung1}, may need to be considered as well.%
	\footnote{The second Madelung equation assures probability 
				conservation. At least in the absence of particle 
				creation and annihilation, there is no 
				justification for modifying it 
				(see also Sec. 6 in Ref. 
				\cite{reddigerMadelungPictureFoundation2017}).}

As further way to generalize the third Madelung equation is to account for 
spin and magnetic fields (cf. e.g. Refs. 
\cite{bohmCausalInterpretationPauli1955,
bohmCausalInterpretationPauli1955a,janossyHydrodynamicalModelWave1966} 
and Sec. 9.3.2 in Ref. \cite{hollandQuantumTheoryMotion1993}).
Indeed, Gurtler and Hestenes \cite{gurtlerConsistencyFormulationDirac1975} 
noted that for electrons or other particles with 
non-zero magnetic moment  
it is generally more appropriate to view the Schrödinger equation as 
a special case of the Pauli equation. In the non-relativistic limit 
the latter describes such `spin-1/2 particles' in the presence of 
general electromagnetic fields. We shall, however, not dive 
into this any further. 

We shall return to the question of the physical interpretation 
of the distributional curl for the $3$-dimensional example 
considered in this section. A second such interpretation 
would be to view the distributional curl and 
other distributions in this context 
as \emph{generalized random variables} in the sense that one can apply the 
respective distribution to the probability density $\varrho$ and 
interpret the result as the corresponding expectation value. If 
$\tilde{A}$ is a distribution on say $\mathcal{S}(\R^3,\C)$ 
and one is willing to employ `physics 
notation' to formally construct a `random variable' $A$, then this 
translates to the formal equation%
	\footnote{ 	The expression is formal because not every 
				distribution $\tilde{A}$ is regular. 
				Still, such formal expressions 
				are abundant in physics, as exemplified by the formula 
				\begin{equation*}
					\delta_0(\varphi) 
					= \int_{-\infty}^\infty \d x \, \delta(x) 
						\varphi(x) = \varphi(0) 
						\, . 
				\end{equation*}
				} 
\begin{equation}
	\tilde{A} (\varrho) = \int_{\R^3} \varrho \, \d^3 r \, A \, . 
	\label{eq:distvsexpectation}
\end{equation}
Eq. \eqref{eq:distvsexpectation} 
justifies the view of $\tilde{A} (\varrho)$ as the expectation 
value of the random variable $A$. 

A mathematical problem one faces with this approach is that 
a given probability density $\varrho$ -- especially in 
quantum mechanics -- need not be a Schwartz function.

In the $2$-dimensional example above, $\varrho$ is indeed 
a Schwartz function, so that $\curl \vec{v}(\varrho)$ is 
well-defined. One easily checks that the 
corresponding expectation value vanishes. 

In our $3$-dimensional example we find, however, that $\varrho$
from Eq. \eqref{eq:varrhostandardhydrogen} fails to be smooth at the 
origin. Thus $\varrho$ is not a test function in any 
of the senses above and so, strictly speaking,
\begin{equation}
	(\nabla \cross \vec{v})(\varrho) := 
		(\nabla \cross \vec{v})\left(\varrho,\varrho,\varrho\right)
			\label{eq:genrandomcurlvonvarrho}
\end{equation}
is not a mathematically 
sensible expression. Still, one may ask if one can canonically extend 
the domain of $\nabla \cross \vec{v}$ to include $\varrho$. 

As the next lemma shows, this is indeed the case. 

\begin{Lemma}
	\label{Lem:curlvextenddomain}
	Let $\vec{v}$ be given by Eq. \eqref{eq:vstandardhydrogen}. 
	
	Then its distributional curl $\nabla \cross \vec{v}$ 
	can be canonically extended to the domain of 
	all $\vec{\varphi} \in C^1(\R^3 \setminus \lbrace 0 \rbrace, \R)$
	for which $\varphi^i$ is $L^1$-integrable, vanishes at infinity, 
	and admits a continuous 
	extension to $\R^3$ for each $i \in \lbrace 1, 2, 3 \rbrace$. 
	Eqs. 
	\eqref{eq:Corhydrogenvdistderequs} still hold on this extended domain. 
	
	In particular, the expression \eqref{eq:genrandomcurlvonvarrho} is well-defined 
	for any $\varrho$ given by Eq. \eqref{eq:varrhostandardhydrogen}. 
\end{Lemma}

Lem. \ref{Lem:curlvextenddomain} also provides us with a simple 
method to compute the respective expectation values. 

\begin{Proposition}
	\label{Prop:Expectationvorthydrogen}
\begin{subequations}
	Let $\nabla \cross \vec{v}$ be the distributional curl 
	of the vector field $\vec{v}$, as given by Eq. 
	\eqref{eq:vstandardhydrogen}, and 
	let $\varrho$ be given by 
	Eq. \eqref{eq:varrhostandardhydrogen}. 
	
	Then the quantity $(\nabla \cross \vec{v})(\varrho)$ 
	vanishes for all admissible 
	$n$, $l$, and $\mu$.  
\end{subequations}
\end{Proposition}

Therefore, even though in this model 
the distribution $\nabla \cross \vec{v}$ 
does not vanish for $\mu \neq 0$, the corresponding expectation value 
always does---as it was the case for the $2$-dimensional example above. 

Physically, this suggests that, at least for those two 
models and for the solutions considered here, 
the appearance of quantum vorticity cannot be measured (directly) 
even when the solutions exhibit it on a mathematical level. 

If one subscribes to the priorly described view that those models 
represent limiting cases of more general models in which the quantum vorticity 
is supported on a set of positive Lebesgue measure, then in 
such more general models the vorticity is measuable `in principium'. 

\section{Remarks on the distributional Madelung equations}
\label{sec:distmadelung} 

The purpose of this section is to clarify the loose term 
`distributional Madelung equations' and make some remarks thereon 
in relation to the published literature. While the topic has 
not received much attention in the mathematical literature, 
we do provide some suggestions and considerations that we expect to 
be of assistance in future considerations of the problem. 
Nonetheless, it should be kept in mind that other variational formulations 
of the Madelung equations may exist and that there might even 
be physical reasons to prefer such a formulation over the ones presented here. 

In Sec. \ref{sec:intro} we already mentioned that Gasser and Markowich 
\cite{gasserQuantumHydrodynamicsWigner1997}, with some caveats, 
showed that the Schrödinger equation with initial data 
in the Sobolev space $H^1(\R^3,\C)$ implies a reformulated version 
of the Madelung equations in the distributional sense: 
\begin{subequations}
	\label{eq:gassermarko}
\begin{gather}
	\partd{\vec{j}}{t} + \nabla \cdot \biggl( 
			\frac{\vec{j} \tp \vec{j}}{\varrho} \biggr) 
		= - \varrho \, \nabla V + \frac{\hbar^2}{4 m} 
		\left( \Delta \left( \nabla \varrho \right) 
				- \nabla \cdot \left( \frac{\nabla \varrho \tp 
				\nabla \varrho}{\varrho}\right)\right)  
			\label{eq:gassermarko1}
		\\ 
	\partd{\varrho}{t} + \nabla \cdot \vec{j} = 0 \, . 
			\label{eq:gassermarko2}
\end{gather}
\end{subequations}
More specifically, let $\Psi$ in%
	\footnote{See e.g. p. 55 in Ref. \cite{bahouriFourierAnalysisNonlinear2011} 
				for an explanation of this notation. 
				We note that, strictly speaking, an element $\Psi$ in 
				$L^\infty \left(\R; H^1(\R^3,\C)\right)$ 
				is an equivalence class of an equivalence class of 
				functions. 
				}  
\begin{equation}
	L^\infty \left([0, \infty); H^1(\R^3,\C)\right)
		= \set{\Psi \colon 
					[0, \infty) \to H^1(\R^3,\C) 
					\colon t \mapsto \Psi_t 
				}{\esssup_{t \in [0, \infty)} \, \norm{\Psi_t}_{H^1} 
		< \infty} 
\end{equation}
be a solution of the Schrödinger equation with potential $V$ and
initial data $\Psi_0 \in H^1(\R^3,\C)$, 
where $\Psi_0$ and $V$ satisfy some further technical assumptions%
	\footnote{It is worthy of note that potentials 
				can be time-dependent, though they are often of 
				electromagnetic origin so that 
				the magnetic field and the spin may need to be taken 
				into account as well in that instance. 
				See e.g. Ref. \cite{penzRegularityEvolutionEquations2018} for 
				a mathematical treatment of time-dependent 
				potentials in the mathematical theory of quantum mechanics. 
				  } 
(cf. Lem. 2.1 in Ref. \cite{gasserQuantumHydrodynamicsWigner1997}).  
Denoting the weak gradient of $\Psi$ by $\nabla \Psi$, 
define the quantities 
\begin{equation}
	\varrho = \abs{\Psi}^2 \quad \text{and} \quad \vec{j} = \frac{\hbar}{m} 
					\Im \left(\Psi^* 
					\nabla \Psi \right) \, . 
\end{equation}
Then $\varrho$ and $\vec{j}$, with initial data 
$\varrho_0$ and $\vec{j}_0$ obtained from $\Psi_0$, 
solve the integral equations%
	\footnote{For sufficiently regular 
				vector fields $\vec{a}$, $\vec{b}$ on 
				$\R^3$, we set  
				$\left( \vec{a} \tp \vec{a} \right) \bullet
					\nabla \vec{b} 
					:= \vec{a} \cdot \left( \left(\vec{a} \cdot \nabla \right)
						\vec{b}  \right)$
			}
\begin{subequations}
\label{eq:gassermarkovar}
\begin{multline}
		0= 
		\int_0^\infty 
			\d t \int_{\R^3} \d^3 r 
			\Biggl( 
					\vec{j} \cdot \partd{\vec{\varphi}}{t}
					+ \biggl( 
					\frac{\vec{j} \tp \vec{j}}{\varrho} \biggr) 
						\bullet \nabla \vec{\varphi} 
					- \varrho \, \nabla V  \, \vec{\varphi} 
		\\ 					
					+ \frac{\hbar^2}{4 m} 
				\left(  \nabla \varrho \cdot \Delta \vec{\varphi} 
				+ \left( \frac{\nabla \varrho \tp 
				\nabla \varrho}{\varrho}\right) \bullet 
				\nabla \vec{\varphi}  \right)  
			\Biggr)
			+ \int_{\R^3} \d^3 r \, \vec{j}_0 \cdot \vec{\varphi} (0, \, . \,) 
			\label{eq:gassermarkovar1}
\end{multline}
and 
\begin{equation}
	0 = \int_0^\infty 
	\d t \int_{\R^3} \d^3 r 
	\left( \varrho \, \partd{\xi}{t} 
			+ \vec{j} \cdot \nabla \xi 
			\right) 
			+ \int_{\R^3} \d^3 r \,  
		\varrho_0 \, \xi(0, \, . \,) 
			\label{eq:gassermarkovar2}
\end{equation}
\end{subequations}
for all $\vec{\varphi} \in C^\infty_0 
	\left([0,\infty) \times \R^3, \R^3 \right)$ and 
	all $\xi \in C^\infty_0 
	\left([0,\infty) \times \R^3, \R \right)$. 

Eqs. \eqref{eq:gassermarkovar} constitute  
	a \emph{variational formulation} of the (formal)  
	equations 
	\eqref{eq:gassermarko}. 
That is, if one states that some given  
$\varrho \in 
L^\infty\left([0,\infty),W^{1,1}(\R^3,\R)\right)$ and 
$\vec{j}$ $\in L^\infty([0,\infty),$ $L^{1}(\R^3,\R^3))$ 
solve Eqs. \eqref{eq:gassermarko} 
in a distributional sense,%
	\footnote{For $p \in [1,\infty]$, $k$, $n$, $m$ $\in \N$, and 
				$\K \in \lbrace \R, \C \rbrace$, 
				the space 
				$W^{k,p}(\R^n,\K^m)$ denotes the Sobolev space of 
					equivalence classes of $\K^m$-valued 
					$L^p$-functions on $\R^n$
					whose weak derivative up to $k$th order are 
					also $L^p$. Every $W^{k,p}(\R^n,\K^m)$ is a Banach 
					space, and for finite $p$ it is separable.
					\label{fn:defSobolevW}
					}
then it is implicit 
that each function in the respective equivalence classes 
satisfies the respective 
variational formulation, Eqs. \eqref{eq:gassermarkovar}, 
for all $\vec{\varphi} \in C^\infty_0 
	\left([0,\infty) \times \R^3, \R^3 \right)$ and 
	all $\xi \in C^\infty_0 
	\left([0,\infty) \times \R^3, \R \right)$. 
	
Heuristically, one obtains a variational formulation for 
a given set of formal PDEs by multiplying the equations with 
an appropriate test function, integrating over the 
space-time region of interest, and applying integration 
by parts and the divergence theorem to `shift' 
the derivatives to the test function. The idea is that, 
if given weak (or even distributional) solutions are 
`sufficiently regular' on the entire domain, one can 
reverse those steps and apply the fundamental theorem of 
the calculus of variations 
(cf. Cor. 4.24 in Ref. \cite{brezisFunctionalAnalysisSobolev2011}) 
to show that one has in fact 
strong solutions of the original set of PDEs. We 
refer to Sec. 3.4 in Ref. 
\cite{evansPartialDifferentialEquations1998} for an 
introduction to variational formulations of initial value 
problems. Sec. 5.1 in Ref. 
\cite{bahouriFourierAnalysisNonlinear2011} discusses 
the variational formulation of the incompressible Navier-Stokes 
equations, which is a system of PDEs that is mathematically 
similar to the ones considered here. 

The significance of the article \cite{gasserQuantumHydrodynamicsWigner1997} 
by Gasser and Markowich is twofold: First, to our knowledge, the 
authors provided the 
first variational formulation of the Madelung equations, Eqs. 
\eqref{eq:Madelung},  
in the literature. 
Second, they used this formulation 
to make major contributions to the 
mathematical theory of the classical limit 
(see also Sec. 3 in Ref. \cite{gasserMacroscopicTheoryCharged1995}). 
\begin{Remark}[On the classical limit]
	\label{Rem:gassermarkolimit}
\begin{subequations}
	In the contemporary physics literature the 
		theory of `(quantum) decoherence' is 
		a popular approach to the question
		of the classical limit of quantum mechanics.  
		We refer, for instance, to 
		Refs. \cite{zurekDecoherenceTransitionQuantum2007,
		schlosshauerDecoherenceQuantumToClassicalTransition2007,
		bacciagaluppiRoleDecoherenceQuantum2020} for an 
		introduction. 
		
		Contrarily, the work by Gasser and Markowich 
		\cite{gasserQuantumHydrodynamicsWigner1997}  
		shows that for JWKB-initial data and 
		in the limit $\hbar \to 0$,
		the time evolution of the 
		probability density $\varrho$ and 
		the drift field $\vec{v}$ is 
		governed by solutions of the respective 
		Newtonian point particle equations 
		(cf. Eqs. 3.13 and 3.17, as well as Lems. 3.2 and 3.3
		therein).%
			\footnote{Note that on p. 106 the authors define 
						$\vec{v}$ as 
						the Radon-Nikodym derivative of the 
						vector-valued measure 
						$A \to \int_A\vec{j} \, \d^3 x$ with 
						respect to the 
						measure $A \to \int_A 
						\varrho \, \d^3 x$.} 
		
		To understand the significance of their result, recall that, 
		in accordance with the ensemble interpretation of 
		quantum mechanics 
		\cite{ballentineStatisticalInterpretationQuantum1970}, 
		$\varrho$ is to be understood as a 
		particle detection probability 
		density. Moreover, in the stochastic approach 
		mentioned in the introduction, $\varrho$ is 
		a phenomenological quantity 
		obtained from a stochastic process that describes the 
		more fundamental evolution of samples (i.e. mass points) 
		in the ensemble. 
		Roughly speaking, if the characteristic lengths and masses of the physical 
		system are large as compared to $\hbar$, then the system 
		is approximately governed by equations in the mathematical 
		limit $\hbar \to 0$. In this sense, the 
		work by Gasser and Markowich 
		\cite{gasserQuantumHydrodynamicsWigner1997} may be 
		understood as saying that, in this instance, 
		the samples behave in accordance with the laws of 
		Newtonian mechanics and that their motion, in turn, 
		determines the evolution of the ensemble 
		quantities $\varrho$ and $\vec{v}$. No 
		additional `theory of 
		measurement' is needed to achieve this remarkable 
		result. 
		
		So far, the significant contribution of Ref. 
		\cite{gasserQuantumHydrodynamicsWigner1997} to the 
		literature on the classical limit 
		has not received much acknowledgement in  
		the physics community.   

		It is also worthy of note that their work 
		lends further credence to the hypothesis 
		expressed in Postulate 1 of Ref. 
		\cite{reddigerMadelungPictureFoundation2017}: 
		Namely, that Kolmogorovian probability
		theory ought to be used as an axiomatic basis 
		of non-relativistic quantum theory, not the 
		Dirac-von Neumann axioms. The reason is that, 
		in passing to the  
		classical limit, there is nothing in the formalism that would 
		justify a change of 
		the theory of probability employed. 
\end{subequations}
\end{Remark}

With regards to Takabayasi's condition, we note that 
in Ref. \cite{gasserQuantumHydrodynamicsWigner1997}  
Gasser and Markowich did not consider the question of 
whether (a variant of) the Schrödinger equation can be 
recovered from Eqs. \eqref{eq:gassermarkovar}.  
Indeed, one does not expect this to be the case without any further 
assumptions: In the smooth setting the equations 
are equivalent to Eqs. \eqref{eq:Madelung1} and \eqref{eq:Madelung2}, 
yet the former one only implies the vorticity equation,  
Eq. \eqref{eq:omegaevolution}, not the third Madelung 
equation, Eq. \eqref{eq:Madelung3}. 

Hence, if one wanted to show that 
Eqs. \eqref{eq:gassermarko1} and \eqref{eq:gassermarko2} are 
equivalent to the Schrödinger equation in some sense, we expect 
an additional constraint on the vorticity to be 
required.  

In the literature on super-fluidity such a condition was 
recently suggested by Antonelli et al. 
\cite{antonelliResultsQuantumHydrodynamical2018}: 

The authors have shown 
that in any dimension $n \in \N$ any (equivalence class 
of) wave function(s) $\Psi$ in the Sobolev space 
$H^1(\R^n,\C)$ can be split into 
$\Psi = \abs{\Psi} \, Q$, that 
$\sqrt{\rho}:= \abs{\Psi}$ is in 
$H^1\left(\R^n,\C\right)$, that the vector field 
\begin{equation}
	\vec{\lambda} = \frac{\hbar}{m} \, \Im \left(Q^* \, \nabla \Psi \right)
\end{equation}
is square-integrable, and -- most notably -- 
that the current density 
\begin{equation}
	\vec{j} = \sqrt{\rho} \, \vec{\lambda} 
	= \frac{\hbar}{m} \, \Im \left(\Psi^* \, \nabla \Psi \right)
\end{equation}
satisfies a \emph{generalized irrotationality condition} 
\begin{equation}
	\nabla \cross \vec{j} = 
			2 \, \left( \nabla \sqrt{\rho}\right) \cross \vec{\lambda} 
		\label{eq:AntonelliMarcaticond}
\end{equation}
in the distributional sense (cf. Lem. 3.1 therein). 

The condition, 
Eq. \eqref{eq:AntonelliMarcaticond}, as well as the `polar 
decomposition' $\Psi = \abs{\Psi} \, Q$ 
was first described in a prior article by Antonelli and Marcati 
\cite{antonelliFiniteEnergyWeak2008}. In particular, 
for an elaboration on the authors' polar decomposition method, 
see Sec. 3 in Ref. \cite{antonelliFiniteEnergyWeak2008}, 
the appendix of Ref. \cite{antonelliQuantumHydrodynamicsSystem2012}, 
as well as the foundational 
article by Brenier \cite{brenierPolarFactorizationMonotone1991}. 

In Rem. 1.3 in Ref. \cite{antonelliResultsQuantumHydrodynamical2018} 
the authors state the following: 
\begin{quote} 
	In the case of a smooth solution 
	$(\varrho, \vec{j})$, for which we can write $\vec{j} = \varrho \, \vec{v}$, 
	for some smooth velocity field $\vec{v}$, the Generalized Irrotationality condition 
	defined above [Eq. \eqref{eq:AntonelliMarcaticond}] 
	is equivalent to $\varrho \, \nabla \cross \vec{v}=0$, 
	i.e. the velocity field $\vec{v}$ is irrotational $\varrho \, \d^3 r$ 
	almost everywhere. It shows that the previous definition 
	is the right weak formulation 
	of the classical irrotationality condition 
	$\nabla \cross \vec{v} = 0$ valid away from vacuum [...] 
	
	[notation adapted]
\end{quote}

As we do not expect the condition \eqref{eq:AntonelliMarcaticond} 
to yield any `quantization condition', we consider it an open question of whether 
it is `the right weak formulation of the classical irrotationality 
condition' in this different, foundational context. In any case, 
such an additional constraint needs to be shown to be compatible with the 
evolution equations, so 
as to not make the full system of PDEs overdetermined. 

There might, however, be another approach. 

A potential problem that Eqs. 
\eqref{eq:gassermarko1} and \eqref{eq:gassermarko2} 
exhibit is that this choice of formulation of the 
Madelung equations was motivated by the study of 
similar equations in the mathematical theory of 
fluid-dynamics. Yet, contrary to Madelung's 
original interpretation 
\cite{madelungQuantentheorieHydrodynamischerForm1927}, 
the interpretation of the Madelung-equations as describing 
the evolution of a `quantum fluid' has been discarded 
by the vast majority of 
physicists. According to the Born rule -- 
which is today considered a standard part of 
quantum mechanics -- at some 
given time $t$ the function $\varrho(t,\,. \,)$ is a probability density 
for detecting the body in an arbitrary region of 
space. It is therefore a misinterpretation 
to view $m \varrho(t,\,. \,)$ as the mass density of a 
fluid. 

Hence, at least on physical grounds, there is a substantial difference 
between the Madelung equations and fluid-dynamical equations, 
be they `classical' or `quantum' fluids. Carrying over ideas and 
methods from the mathematical theory of fluid dynamics 
in a one-to-one manner can therefore lead to incorrect physics. 
A particular example is that in the context of 
microscopic physics, as described for instance by the 
(linear) Schrödinger 
equation, the potential $V$ is not directly coupled to the 
density $\varrho$ (as it would be for charged fluids). Also, 
fluid-dynamical concepts such as temperature, pressure, and 
constitutive equations have no place here, as those are of 
macroscopic nature.%
	\footnote{Such considerations do not apply to 
				drift-diffusion equations. A 
				relationship between the latter and the 
				Madelung equations is suggested by the 
				theories of stochastic mechanics and 
				stochastic electrodynamics. See e.g. Ref. 
				\cite{santosStochasticInterpretationsQuantum2022}
				for a recent review.} 
 
Still, using Eqs. \eqref{eq:gassermarko} for a 
variational formulation of the Madelung equations 
does have mathematical 
justification: Roughly speaking, if one does not  
multiply Eq. \eqref{eq:Madelung1} 
by $\varrho$ to obtain 
Eq. \eqref{eq:gassermarko1}, 
the division by $\sqrt{\varrho}$ in the quantum force  
term makes it difficult to find a workable variational 
formulation. 

In spite of this, it is possible to find such a  
variational formulation---though, as we shall see, 
it is not entirely without limitations. 

First, we reformulate the 
problematic term with the help of Nottale's 
identity 
\begin{equation}
	\frac{1}{\alpha} 
	\, \nabla \left( \frac{\Delta \varphi^\alpha}
			{\varphi^\alpha}\right)
	= \Delta \left( \nabla \ln \varphi \right) 
		+ 2 \alpha \bigl( \left(\nabla \ln \varphi\right)\cdot 
			\nabla \bigr) \left(\nabla \ln \varphi\right) \, ,
			\label{eq:Nottaleid}
\end{equation}
which holds for all $\alpha>0$ and sufficiently regular 
$\varphi$ (see e.g. Eq. 153 
in Ref. \cite{nottaleGeneralizedQuantumPotentials2009}).%
	\footnote{Eq. \eqref{eq:Nottaleid} 
				can be shown by pulling $\alpha$ into the exponent of 
				each logarithm and then using various vector identities like  
				Eq. \eqref{eq:Webercurlv0} below.} 
Thereafter, one defines the osmotic velocity%
	\footnote{Different sign conventions are used in the literature. 
				The terms `diffusive velocity' and 
				`stochastic velocity' are also in use 
				(cf. p. 123 in Ref. \cite{delapenaEmergingQuantumPhysics2015}). 
				}
\begin{equation}
	\vec{u}(\varrho) = \frac{\hbar}{2 m} \frac{\nabla \varrho}{\varrho} 
	\label{eq:defu}
\end{equation}
(cf. Eqs. 2.27 and 2.64 as well as p. 123 in 
Ref. \cite{delapenaEmergingQuantumPhysics2015}) and 
reformulates the continuity equation in terms of 
$\vec{u}(\varrho)$ and $\vec{v}$. 

Ultimately, one arrives at \emph{Nelson's equations} 
(cf. Eqs. 34 in Ref. 
\cite{nelsonDerivationSchrodingerEquation1966} and 
Eq. 1 in Ref. \cite{casatiAharonovBohmEffectHydrodynamical1979}):  
\begin{subequations}
	\label{eq:Nelson}
\begin{gather}
	m \left(\partd{\vec{v}}{t} +\left(\vec{v} \cdot \nabla\right) \vec{v}  \right)= - 
	\nabla V + m \left(\vec{u} \cdot \nabla \right) \vec{u} 
	+ \frac{\hbar}{2} \Delta \vec{u} 
	\label{eq:Nelson1}
	\\
	\partd{\vec{u}}{t} 
	+ \nabla \left( \vec{v} \cdot \vec{u} + \frac{\hbar}{2m} \, 
		\nabla \cdot \vec{v} \right) = 0 \, . 
	\label{eq:Nelson2}
\end{gather}
\end{subequations}
The advantage of Nelson's reformulation of the Madelung equations is that 
it allows one to express the latter entirely in terms of $\vec{v}$ and 
$\vec{u}$, so that, via the Eq. \eqref{eq:defu}, one can 
treat $\varrho$ as a derived quantity.%
	\footnote{The fact that the Madelung equations can be reformulated 
				in this manner suggest that the introduction 
				of $\vec{u}$ via \eqref{eq:defu}
			 	is not merely a mathematical convenience, 
				but that the quantity is of potential 
				physical significance as well. 
				} 
Note that one does not require irrotationality of $\vec{v}$ 
to derive Eqs. \eqref{eq:Nelson}. 

In looking for a variational formulation of Nelson's equations, 
we need to find a way to reformulate the two terms of the form 
$\left(\vec{w} \cdot \nabla\right) \vec{w}$ in Eq. \eqref{eq:Nelson1}. 
The terms 
are inconvenient, since, following the aforementioned 
procedure, we ultimately intend to `shift' as many derivatives as possible 
to the test functions. For the incompressible Navier-Stokes equations 
the term is handled using the divergence-freeness of 
$\vec{w}$ (cf. Sec. 5.1 in Ref. 
\cite{bahouriFourierAnalysisNonlinear2011}). While this does not work 
in this instance, 
we can impose the condition of irrotationality instead, which 
in turn allows us to use the identity%
		\footnote{Using a common physics notation, we set $\vec{w}^2 = \vec{w} \cdot \vec{w}$.} 
\begin{equation}
	\left( \vec{w} \cdot \nabla \right) \vec{w} = \nabla 
	\left(\frac{\vec{w}^2}{2}\right) \, . 
	\label{eq:Webercurlv0}
\end{equation}

We make use of Eq. \eqref{eq:Webercurlv0} 
to suggest a variational formulation of the Madelung 
equations that differs from Eqs. \eqref{eq:gassermarkovar}: 

Given initial data $\vec{u}_0$ and $\vec{v}_0$, 
	we require that $\vec{u}$ and $\vec{v}$ satisfy 
\begin{subequations}
	\label{eq:Nelsonvar}
\begin{gather}
	\label{eq:Nelsonvar1}
	\begin{split}
		0 =  \int_0^\infty \d t \int_{\R^3} \d^3 r \, 
				\biggl( m \vec{v} \cdot 
				\partd{\vec{\varphi}}{t} 
				+ 
				&
				\left( \frac{m}{2} \, \vec{v}^2
				+ V - \frac{m}{2} \, \vec{u}^2 \right) \, \nabla  \cdot 
				\vec{\varphi} 
				\\ 
				&
				+ 
				\frac{\hbar}{2} \, \vec{u} \cdot \Delta \vec{\varphi} 
				\biggr) 
				+ \int_{\R^3} \d^3 r \, m \vec{v}_0 \cdot 
				\vec{\varphi}(0, \, . \, ) 
	\end{split} 
	\\
	\label{eq:Nelsonvar2}
	\begin{split}
	0 =  \int_0^\infty \d t \int_{\R^3} \d^3 r \, 
		\biggl(  \vec{u} \cdot 
				\partd{\vec{\xi}}{t} 
				+ \vec{u} \cdot \vec{v} \, \,  \nabla \cdot \vec{\xi}
				-
				&
				\, 
				\vec{v} \cdot \nabla \left( \nabla \cdot \vec{\xi}
				\right)
				 \biggr) 
				\\
				&
				+ \int_{\R^3} \d^3 r \, \vec{u}_0 \cdot 
				\vec{\xi}(0, \, . \, ) 
	\end{split}
\end{gather}
\end{subequations}
for all $\vec{\varphi}$, $\vec{\xi} \in C^\infty_0 
	\left([0,\infty) \times \R^3, \R^3 \right)$.
Now let $\vec{u}_0$, $\vec{v}_0$ be in 
in $L_{loc}^2\left(\R^3,\R^3 \right)$, 
and $\vec{u}$, $\vec{v}$ be in 
$L^\infty_{loc}([0, \infty);L^2_{loc}(\R^3,\R^3))$, the latter being given by 
\begin{multline}
		\bigl\lbrace \vec{w} \bigm\vert 
				\forall t \in [0,\infty) \, 
					\forall \, \text{compact} \, 
					 K \subset \R^3 \colon 
					 \vec{w}(t, \, . \, ) \in 
					 L^2(K,\R^3) 
					 \\
					 \quad \text{and} \quad 
					 \left( t \mapsto \norm{\vec{w}(t, \, . \, )}_{L^2(K,\R^3)} 
					 \right) \in L^\infty_{loc}([0, \infty),\R)
			\bigr\rbrace 
			\, .
\end{multline}
It can be checked by a straightforward 
estimate of the integrals,%
	\footnote{For $\varphi \in C^\infty_0 
					\left([0,\infty) \cross 
					\R^3,\R \right)$ denote by $I$ the 
					(compact) support 
					of the function 
					$t \mapsto \norm{\varphi(t, \, . \,)
									}_{L^1
								\left(\R^3,\R \right)}$. 
					Then for all 
					$\vec{w} \in L^\infty_{loc}([0, \infty);L^2_{loc}(\R^3,\R^3))$ 
					we have 
					\begin{align*}
					\abs{\int_0^\infty \d t \, \int_{\R^3} \d^3 r \,  
					\vec{w}^2 \, \varphi }
					& \leq \int_0^\infty \d t  \, 
							\norm{\vec{w}^2(t, \, . \,) \, 
									\varphi(t, \, . \,)}_{L^1
								\left(\R^3,\R \right)}
					\\
					& \leq \int_{I} 
								\d t \, 
								\norm{\varphi(t, \, . \,)}_{L^\infty
								\left(\R^3,\R \right)} \, 
								\norm{\vec{w}(t, \, . \, )}
									_{L^2(\supp 
										\varphi(t, \, . \,),\R^3)} ^2
					\\
					&  \leq \left( \int_{I} 
								\d t \right) 
							\left(\sup_{t \in I} \norm{\varphi(t, \, . \,)}_{L^\infty
								\left(\R^3,\R \right)} \right) \\
					&	\quad \quad \quad \quad \quad \quad 
						\left(\esssup_{t \in I} \left( \norm{\vec{w}(t, \, . \, )}
									_{L^2(\supp 
										\varphi(t, \, . \,),\R^3)} \right) 
										\right)^2
							\, .
					\end{align*}
					The other integrals can either be estimated in a similar 
					manner or via standard estimates.  
				} 
that Eqs. \eqref{eq:Nelsonvar} are mathematically sensible in that instance.

\begin{Remark}[Quasi-irrotational 
				solutions are not in $L^2_\text{loc}$]
	\label{Rem:noirrothydrogen}
\begin{subequations}
	The alert reader may object, that by using the identity 
	\eqref{eq:Webercurlv0} we implicitly disposed of 
	solutions such as the ones in  
	Eqs. \eqref{eq:varrhostandardhydrogen} and 
	\eqref{eq:vstandardhydrogen}. 
	
	Indeed, those 
	solutions (including their `non-quantized' siblings) 
	are not admissible in the above variational 
	formulation of the Madelung equations, though for 
	other reasons:  
	
	It is easy to check that $\vec{v}$ from 
	Eq. \eqref{eq:vstandardhydrogen} is not in 
	$L^2_{loc}(\R^3,\R^3)$ (or 
	$L^\infty_{loc}$ $([0, \infty)$ $;L^2_{loc}$ $(\R^3,\R^3))$, respectively). 
	Therefore, there is no solution $(\vec{u},\vec{v})$ 
	of Eqs. \eqref{eq:Nelsonvar} for which $\vec{v}$ is given 
	by Eq. \eqref{eq:vstandardhydrogen}.  
	This problem was, however, not entirely caused by the use of the 
	identity \eqref{eq:Webercurlv0} in setting up the variational 
	formulation. It is also the $\vec{u}\cdot \vec{v}$ term in
	the variational formulation of Nelson's formulation of the 
	continuity equation, Eq. \eqref{eq:Nelsonvar2}, that 
	suggests use of the space $L^2_{loc}(\R^3,\R^3)$ 
	for fixed time. 

	Note that the variational formulation of 
	the Madelung equations by Gasser and 
	Markowich, Eqs. \eqref{eq:gassermarkovar} above, 
	does not suffer from this issue. 
	
	We also point out that there 
	may be a mathematical link between quasi-irrationality of 
	$\vec{v}$ from Eq. \eqref{eq:varrhostandardhydrogen} 
	and the fact that it is in $L^1_{loc}(\R^3,\R^3)$ 
	yet not in $L^2_{loc}(\R^3,\R^3)$. Investigating 
	this possible link may provide an answer to the 
	question, whether there are (quasi-)rotational 	
	solutions of Eqs. \eqref{eq:Nelsonvar} or not. 
	
	Regarding the physical (in)acceptability of `throwing out 
	solutions' such as the ones in 
	Eq. \eqref{eq:varrhostandardhydrogen} and 
	\eqref{eq:vstandardhydrogen}, we refer back to 
	Sec. \ref{ssec:vorticity}. 
\end{subequations}
\end{Remark}

In the variational formulation of Nelson's equations, 
Eqs. \eqref{eq:Nelsonvar} above, the 
initial data $\vec{u}_0$ for $\vec{u}$ is to be determined from 
the initial data $\varrho_0$ for $\varrho$. 
As $\varrho(t,\, .\,)$ is supposed to be a probability 
density for all $t$, $\varrho_0$, needs to be in $L^1(\R^3,\R)$ 
and (every representative in the equivalence class is) strictly 
positive almost everywhere. If we 
understand the derivative in Eq. \eqref{eq:defu} in the weak 
sense, we furthermore require that $\varrho_0$ is 
weakly differentiable and that the quantity 
$\nabla \varrho_0 / \varrho_0$ is in $L^2\left(\R^3,\R^3 \right)$ 
(as an equivalence class of vector fields) so that 
$\vec{u}_0$ is, too. 

Apart from providing a formulation involving the 
vector field $\vec{v}$ directly, Eqs. \eqref{eq:Nelsonvar}
can also easily be generalized to distributional potentials. 
If, for instance, we wanted to consider 
$V(t,\vec{r}) = \alpha \, \delta^3 (\vec{r})$ for some $\alpha \in 
\R \setminus \lbrace 0 \rbrace$, 
then we would make the following replacement in Eq. \eqref{eq:Nelsonvar1}: 
\begin{equation}
	\int_0^\infty \d t \int_{\R^3} \d^3 r \,  
	V  \, \nabla  \cdot 
				\vec{\varphi} 
	\quad \rightsquigarrow \quad 
		\alpha \int_0^\infty \d t  \, \left( \nabla  \cdot 
				\vec{\varphi} \right) (t , 0) \, . 
\end{equation}
It is not immediate how to generalize Eqs. 
\eqref{eq:gassermarko} above, so that it can also account for 
distributional potentials. 

Under appropriate 
conditions of regularity of the quantities involved, 
every solution of the Schrödinger equation will also be a 
solution of Eqs. \eqref{eq:Nelsonvar}---with the restriction 
expressed in Rem. \ref{Rem:noirrothydrogen} above. As, to our 
knowledge, Eqs. \eqref{eq:Nelsonvar} have not been 
considered in the literature before, it is unknown whether the 
initial value problem is well-posed. Accordingly, it is also 
not known how precisely the system of equations relates 
to the Schrödinger equation---though the condition 
that $\vec{v}(t, \, . \,) \in L^2(\R^3,\R^3)$ for almost all 
$t$ does introduce an additional restriction. 

To show that there are indeed non-trivial solutions to 
Eqs. \eqref{eq:Nelsonvar}, we shall consider a simple example. 
\begin{Example}
	\label{Ex:solution}
	For the (stationary) 
	ground state of the isotropic harmonic oscillator of 
	frequency $\omega$ in $3$ dimensions we have 
	\begin{equation}
		\vec{u} \left(t, \vec{r}\right) 
			= - \omega \, \vec{r} 
			\quad \text{and} \quad 
		\vec{v} \left(t, \vec{r}\right) = 0
	\end{equation}
	for all $t \in \R$ and $\vec{r} \in \R^3$. One easily checks that 
	$\vec{u} \in L^\infty_{loc}([0, \infty);L^2_{loc}(\R^3,\R^3))$. 
\end{Example}

Given a solution $\left(\vec{u},\vec{v} \right)$ of 
Eqs. \eqref{eq:Nelsonvar}, we may then call $\left(\varrho,\vec{v} \right)$
a solution of the Madelung equations, if for almost all $t \in \R$ 
the (equivalence class of) functions 
$\varrho(t, \, . \,)$ solves the boundary value 
problem%
	\footnote{Strictly speaking, $\varrho$ 
				is not a dimensionless quantity so that $\ln \varrho$ 
				is not well-defined. Yet this can be easily addressed 
				with the introduction of a constant `reference value' 
				(cf. p. 42 in Ref. \cite{delapenaEmergingQuantumPhysics2015}).
				} 
\begin{subequations}
	\label{eq:uBVP}
\begin{gather}
	\int_{\R^3} \d^3 r	\, 
		\left( \vec{u}(t, \, . \,) \cdot \vec{\varphi} + \frac{\hbar}{2 m} 
		\ln \varrho(t, \, . \,) \, \nabla \cdot \vec{ \varphi}
		\right) \\
	1 = \int_{\R^3} \d^3 r \, \varrho(t, \, . \,)
\end{gather}
\end{subequations}
for all $\varphi \in C_0^\infty \left(\R^3, \R \right)$. 
Such solutions $\varrho(t, \, . \,)$, if they exist, 
satisfy the same assumptions as stated for $\varrho_0$ above. If this boundary 
value problem does not have a solution $\varrho$, $\vec{u}$ 
itself needs to be discarded. 

That is, the full system of equations is given by
Eqs. \eqref{eq:Nelsonvar} with 
\begin{equation}
	\vec{u}_0 = \frac{\hbar}{2 m} \, \nabla \ln \varrho_0 
\end{equation}
in the weak sense, 
along with Eqs. \eqref{eq:uBVP}. 

The two distinct variational formulations of the 
Madelung equations given here should clarify what is 
meant by the somewhat vague term `distributional Madelung equations'. 

In the remainder of this section we will provide 
some general remarks that may be of use in formulating   
`the right' distributional Madelung equations. We shall 
not go into the question of whether (quasi-)rotational 
solutions are acceptable or not. For the latter question 
we refer the reader to Secs. \ref{ssec:vorticity} and 
\ref{sec:conclusion}. 

We shall begin by noting that for the above two variational 
formulations, we have assumed that there are no further 
boundary conditions given. That those can be of 
relevance is shown by the examples of the 
`infinite potential well' as well as the single/double slit 
experiment.%
	\footnote{In physics textbooks the double slit 
				experiment is often incorrectly depicted as a 
				stationary problem and, in turn, used to justify the 
				`superposition principle'. Viewing it as 
				a dynamical scattering problem is clearly 
				more physical. Such an approach was taken, 
				for instance, by Sanz and Miret-Artés in Ref. 
				\cite{sanzTrajectorybasedUnderstandingQuantum2008}---though 
				the authors used Gaussians for their analysis and 
				did not impose any additional boundary conditions. 
				}
Both are examples of homogeneous Dirichlet boundary conditions, 
which can be handled via the use of trace operators  
(see e.g. Sec. 5.5 in Ref. \cite{evansPartialDifferentialEquations1998}). 

This brings us to the more general problem of the support of 
$\varrho(t, \, . \, )$, which is defined by viewing $\varrho(t, \, . \, )$ 
as a regular distribution 
(cf. footnote \ref{fn:supportdist}). 

In the second variational formulation proposed here 
we implicitly require that $\supp \varrho(t, \, . \, )=\R^3$ 
for almost all $t \in [0,\infty)$, as otherwise 
$\ln \varrho(t, \, . \, )$ is not well-defined. The variational 
formulation by Gasser and Markowich does not suffer from this 
defect. 

The case that $\supp \varrho(t, \, . \, )$ is a proper subset 
of $\R^3$ should generally be allowed. 
The physical reason is that the bodies in the ensemble 
are typically localized in a small region of space. Commonly used  
probability densities such as Gaussians are in actuality 
an idealization, though in many 
quantum-mechanical problems of physical relevance 
stationary states are supported on the entirety of $\R^3$. 

Yet, if we want to allow for the case that 
$\supp \varrho(t, \, . \, )$ is a proper subset 
of $\R^3$, we have to ask whether we ought to place 
any further constraints on $\supp \varrho(t, \, . \, )$ 
or, better to say, on $\supp \varrho_0$. 

Indeed, physical considerations demand that the 
support of $\varrho_0$ (and of $\Psi_0$) is connected: 

Consider a simple experiment (in vacuum and without external forces) 
in which two mutually parallel 
particle guns are directed at a detector screen at an appropriate distance  
away. After each triggering of the apparatus, 
a single particle gets ejected from one 
of the guns, the chance of the first gun getting triggered being 
$a \in (0,1)$ and the chance of the other one getting triggered being $1-a$. 
For each run, $t=0$ is fixed to be the point in time 
after which the gun was triggered and a brief, given moment has passed 
to give the particle time to leave the apparatus. In this 
experiment, one can model $\varrho_0$ as the sum of 
two bump functions with mutually disconnected support and one can 
idealize $\vec{v}_0$ as a constant vector field on 
$\supp \varrho_0$ pointing in the direction of the detector and 
vanishing otherwise. 
Correspondingly, choose the values of the initial wave function 
$\Psi_0$ to be $\sqrt{\varrho_0 (\vec{r})} \, \exp(\iu \,\vec{k}_0 \cdot \vec{r})$ 
with $\vec{k}_0 = m \, \vec{v}_0 (\vec{r}_0) / \hbar$ for some 
arbitrary $\vec{r}_0 \in \supp \varrho_0$. 
The respective solution of the free Schrödinger equation  
yields that eventually the disconnected supports of $\varrho_0$ will 
merge and that there will be interference---thus contradicting the 
impact statistics observed on the detector. 

The failure of the Schrödinger theory to predict the correct behavior 
in this instance is due to a misapplication of the theory. The 
correct prediction is obtained by finding the Schrödinger evolution for 
each individual bump function $\varrho_{1,0}$ and $\varrho_{2,0}$ 
(both normalized to unity) with their associated drift fields, so that at 
each $t>0$ the probability density is given by 
\begin{equation}
	\varrho(t, \, . \, ) = a \, \varrho_1(t, \, . \, ) + 
							(1-a) \, \varrho_2(t, \, . \, ) \, .
\end{equation}
The determination of $\varrho(t, \, . \, )$ 
is analogous to the experiment being done with `classical' 
projectiles (whose statistics follows a different time evolution). The 
underlying reason is that, as in the case of projectiles, the time evolution of 
the individual sample in the ensemble is independent from the 
time evolution of the other samples. 

Mathematically, this physically incorrect prediction can be 
prevented by requiring that the support of any initial probability density 
$\varrho_0$, or initial wave function $\Psi_0$, is connected. 
In the variational formulation of  
Nelson's equations above, this is always the case. For this to 
be a sensible condition in the variational 
formulation by Gasser and Markowich, however, it needs to 
be shown that, at least under certain sufficiently 
general conditions on the potential $V$, this property is preserved under 
forward time evolution. Due to the `dispersive' nature of the 
Bohm force, there is some hope that this can be done. 

\begin{Remark}[The Wallstrom phenomenon]
	\label{Rem:Wallstromphenomenon}
\begin{subequations}
	The above requirement that the initial probability 
	density ought to have connected support is closely related to 
	Wallstrom's second objection 
	\cite{wallstromInitialvalueProblemMadelung1994} 
	that was outlined in Sec. 
	\ref{sec:intro}. We will explain the objection, its 
	relation to the so-called Pauli problem in quantum mechanics, and 
	provide some potential avenues for resolution. 
	
	In Ref. \cite{wallstromInitialvalueProblemMadelung1994} 
	Wallstrom considered a (sufficiently regular) initial 
	wave function $\Psi_0$ with associated density 
	$\varrho_0$ and drift field $\vec{v}_0$.
	He argued that, 
	if the set 
	\begin{equation}
		\Omega_0= \set{\vec{r} \in \R^n}{\varrho_0\left(\vec{r}\right) \neq 0}
	\end{equation}
	is disconnected, then changing the phase factors for $\Psi_0$ 
	on each connected component of $\Omega_0$ does not change 
	the initial conditions $\left(\varrho_0,\vec{v}_0\right)$
	for the respective Madelung equations while the corresponding 
	Schrödinger equation can in principle lead to different 
	solutions of the Madelung equations whenever 
	the components `merge' under time evolution.  
	Thus, Wallstrom argued, the Madelung equations 
	cannot provide unique solutions in general. 
	
	Indeed, in a recent article \cite{markowichNonuniquenessWeakSolutions2019}, 
	Markowich and Sierra independently rediscovered the described effect, 
	which we shall call the `Wallstrom phenomenon' hereafter. 
	While the assumptions of the 
	main mathematical 
	statement in Ref. \cite{wallstromInitialvalueProblemMadelung1994} 
	are too strong to allow for the phenomenon -- as the corresponding 
	time-dependent flow of $\vec{v}$ 
	is continuous -- Markowich's and Sierra's 
	results 
	leave no doubt that it exists 
	(cf. Thm. 3.2 and Cor. 1 in Ref. 
	\cite{markowichNonuniquenessWeakSolutions2019}). 	
	
	In the beginning of their article 
	Markowich's and Sierra \cite{markowichNonuniquenessWeakSolutions2019} 
	gave a $1$-dimensional example of an initial $H^2$-wave function 
	with connected support for 
	which this phenomenon occurs. Thus, the 
	assumption of connected support alone is insufficient 
	to prevent the occurrence of the Wallstrom phenomenon. Moreover, 
	while the above operation can change the regularity of 
	the initial wave function, it does also not seem to be sufficient to 
	impose more restrictive assumptions of regularity---with the important 
	exception that, if the flow of $\vec{v}$ exists and is continuous, 
	such a `merging' of the connected components of $\Omega_0$ 
	cannot occur (cf. Prop. 3.6 
	in Ref. \cite{reddigerMadelungPictureFoundation2017}). 
	
	While Wallstrom viewed the phenomenon as a problem of the Madelung 
	equations, we wish to point out that it is also a problem of the 
	Schrödinger equation: 
	With the exception of comparing the time evolution of the 
	probability density after the supports have merged, it is not 
	possible to physically discern the different initial states---all 
	predicted detection probabilities and expectation values of 
	observables%
	\footnote{The `operator' suggested in Eq. 23 of Ref. 
				\cite{weigertHowDetermineQuantum1996}
				is not an observable in the quantum-mechanical 
				sense, since there is no non-trivial domain 
				for which it is an operator 
				on $\mathcal{H}=L^2 \left(\R,\C\right)$. 
				}
	at $t=0$ are the same. Accepting all of those 
	initial states as physical would undermine the predictive 
	power of the Schrödinger theory. 

	As Antonelli and Marcati had already suggested in 
	Ref. \cite{antonelliFiniteEnergyWeak2008}, the 
	above problem is closely related to the `Pauli problem'. 
	As the name suggests, this refers to a historical 
	question initially posed by Pauli 
	\cite{pauliAllgemeinenPrinzipienWellenmechanik1990}: 
	Is it possible to 
	reconstruct a wave function $\Psi_0$ from its position probability 
	density $\abs{\Psi_0}^2$ and its `momentum probability 
	density' $\abs{\mathcal{F}\Psi_0}^2$? 
	This question was answered negatively, since a variety of 
	counterexamples had been found over the decades. The 
	first one was given by Bargmann in Reichenbach's book 
	\cite{reichenbachPhilosophicFoundationsQuantum1948}, for 
	later ones we refer to  
	Sec. 2 in Ref. \cite{weigertPauliProblemSpin1992} 
	as well as Ref. \cite{stulpeRemarksDeterminationQuantum1990}. 
	Therein the reader also finds a historical review of 
	the Pauli problem.  
	
	The Wallstrom phenomenon and the Pauli problem are connected by
	the overarching question of how to discern quantum-mechanical states 
	that are initially physically equivalent -- i.e. with regards to measurement 
	-- yet nonetheless lead to physically inequivalent states under time evolution. 
	In one dimension and for wave functions with first order strong 
	derivatives this question was studied by 
	Weigert \cite{weigertHowDetermineQuantum1996}, leading to 
	the conclusion that it is precisely Wallstrom's phenomenon 
	that leads to this problem.%
	\footnote{The author of Ref. \cite{weigertHowDetermineQuantum1996} 
				did not mention the connection to 
				Wallstrom's work \cite{wallstromInitialvalueProblemMadelung1994}.
				}
	
	The underlying problem exhibited by the Wallstrom phenomenon 
	and the negative answer to the Pauli problem 
	therefore seems to be the failure of quantum mechanics to 
	specify which (initial) states are physically 
	acceptable and which ones are not. Phrased differently, the 
	Schrödinger theory in $L^2$ is physically 
	incomplete without a further restriction on the set of (initial) 
	states. 

	We shall provide some considerations for how to resolve the problem. 
	
	While one may ask for the support of initial wave functions or 
	probability densities to be connected, one cannot simply ask that 
	the same holds for $\Omega_0$, for 
	that would exclude any stationary state with nodes and thus be 
	physically unacceptable. Moreover, for $L^1$- or $H^1$-wave functions 
	the nodal set is not even defined. For $H^2$-wave functions 
	the Sobolev embedding theorem does provide for a consistent 
	definition, however (cf. Rem. \ref{Rem:Takabayasi}). 

	Wallstrom's suggested resolution 
	\cite{wallstromInitialvalueProblemMadelung1994} is to ask 
	for JWKB-initial data 
	$\Psi_0= \sqrt{\varrho_0} \, \exp(\iu S_0 / \hbar)$ and 
	for $S_0$ to ``be continuous across the nodal boundaries.'' 
	This condition, however, seems arbitrary, for even for continuous 
	$\Psi_0$ the function $S_0$ cannot be assumed to be 
	continuous everywhere (cf. Sec. \ref{ssec:vorticity}). 
	He also suggests that one could 
	require $S_0$ to be computed from the respective 
	line integral over a given drift field 
	$\vec{v}_0$. That is, in $3$ dimensions we have 
	\begin{equation}
		S_0(\vec{r}) - S_0(\vec{r}_0) = m 
		\int_\gamma \vec{v}_0 \cdot \d \vec{r}
	\end{equation} 
	for any (smooth or $C^1$) curve $\gamma$ from $\vec{r}_0$ 
	to $\vec{r}$. Yet here $\vec{v}_0$ needs to be sufficiently 
	regular for the integral to be defined -- as it is the 
	case for Takabayasi's condition -- and the same issue occurs 
	whenever the connected components of the domain of $\vec{v}_0$ are 
	multiply-connected. 

	In the example Markowich and Sierra 
	\cite{markowichNonuniquenessWeakSolutions2019} provide, it is 
	intuitively clear that the wave function without the artificial 
	phase factor provides the physically correct time evolution. 
	As one can make this choice, neither 
	Wallstrom's phenomenon nor the Pauli problem undermine the  
	predictive power of the Schrödinger equation here. The mathematical 
	problem is therefore to formalize this intuitive choice for 
	the general case and to place appropriate assumptions 
	on $\varrho_0$ and $\vec{v}_0$ (or $\vec{j}_0$) to 
	assure that Wallstrom's phenomenon does not lead to non-uniqueness for  
	the given variational formulation of the Madelung equations. 
	
	Not only would this be an important contribution to the 
	mathematical theory of the  
	Madelung equations, it could also provide a satisfactory 
	conceptual resolution to Pauli's original question. 
\end{subequations}
\end{Remark}

We shall provide some 
general considerations on the regularity on $\varrho (t, \, . \,)$ and 
$\vec{v}(t, \, . \,)$ that one might want to impose in finding 
an appropriate variational formulation of the Madelung equations. 

As Teufel and Tumulka indicated 
(cf. p. 352 in Ref. \cite{teufelSimpleProofGlobal2005}), 
the hydrogen ground state in the $3$-dimensional 
Schrödinger theory places limits on the differentiability 
of $\varrho (t, \, . \,)$. Recalling Eq. 
\eqref{eq:varrhostandardhydrogen}, in spherical polar 
coordinates $(r,\theta,\phi)$ the respective probability density is 
proportional to $e^{-r/a_0}$. 
By integrating by parts in those 
coordinates, one checks that this density is twice 
weakly differentiable, but its third distributional 
derivative yields a Dirac delta at the origin. 

In $3$-dimensional models one should therefore 
not assume existence of weak derivatives of 
$\varrho(t\, . \,)$ of order 
higher than two. 

Regarding the constraints put on the time-dependent vector field 
$\vec{v}$, one may want to -- but need not -- impose 
the almost everywhere existence of its integral 
curves, 
in turn yielding respective assumptions on $\vec{v}$
itself. Berndl 
et al. \cite{berndlGlobalExistenceBohmian1995} 
as well as Teufel and Tumulka \cite{teufelSimpleProofGlobal2005} 
have studied this 
question in the context of the Bohmian theory, where it is of 
relevance to the conceptual structure of the theory. 
One of the major assumptions in Ref. \cite{berndlGlobalExistenceBohmian1995} 
is that the initial wave function should be a $C^\infty$-vector of the 
respective Hamiltonian $\hat H$, i.e. an element of 
$\dom \hat{H}^n$ for all $n \in \N$. This is a very strong assumption. 
In Ref. 
\cite{teufelSimpleProofGlobal2005} Teufel and Tumulka 
weakened this assumption, requiring $\vec{v}$ to be 
a (time-dependent) $C^1$-vector field. Furthermore, they 
argue 
that ``the general existence theory for first order 
ODEs that are not Lipschitz but only in some Sobolev 
space'' is not applicable, for the divergence of 
$\vec{v}$ ``typically diverges at the nodes of the 
wave function.'' We note that no examples were 
given to support this claim. 
Regarding this general mathematical theory, 
we refer the reader to the 2014 review article 
\cite{ambrosioContinuityEquationsODE2014} by 
Ambrosio and Crippa. Secs. 4 and 5 therein discuss 
the general link 
between the continuity equation and the integral curve 
equation for Sobolev vector fields 
and those of bounded variation, respectively. 	
The respective sections 
are mainly based on Refs. 
\cite{dipernaOrdinaryDifferentialEquations1989}
and \cite{ambrosioTransportEquationCauchy2004}.
Sec. 2.2 in Ref. \cite{klebanovApproximationPDEsUnderlying2016}
also provides a comprehensible summary of 
mathematical results related on the existence theory 
of integral curves for the vector field $\vec{v}$. 

\begin{Remark}[Hall's criticism and time-invariance of $\dom \hat{H}$]
	\label{Rem:Hall}
\begin{subequations}
	We shall make 
	some comments on Hall's criticism 
	\cite{hallIncompletenessTrajectorybasedInterpretations2004} 
	of the Bohmian theory. 
	With respect to the discussion in this 
	section, its importance lies in pointing out that 
	a mathematically precise relationship 
	between the Schrödinger equation and the respective  
	distributional Madelung equations need not respect 
	the Hilbert space formalism of quantum mechanics. 
	That is, one does not need to  
	constrain oneself to the respective $L^2$-space or a 
	closed (Hilbert) subspaces thereof---as the Dirac-von Neumann 
	axioms indirectly call for in this instance. The subsequent discussion 
	should clarify those words. 
		
	In Ref. \cite{hallIncompletenessTrajectorybasedInterpretations2004} 
	Hall gives an example that supposedly 
	illustrates `the breakdown of the velocity equation', Eq. 
	\eqref{eq:defv} 
	(cf. p. 9554 in Ref. 
	\cite{hallIncompletenessTrajectorybasedInterpretations2004}):  
	Based on a prior work by Berry 
	\cite{berryQuantumFractalsBoxes1996}, he considers 
	the $1$-dimensional single 
	`particle in a box' problem on the open interval $\Omega =
	(0,L)$ for some $L \in \R_+$
	and the (equivalence class of the) 
	constant wave function  
	$\Psi_0$ with $\Psi_0(x)=1 /\sqrt{L}$ for 
	all $x \in \Omega$. Berry \cite{berryQuantumFractalsBoxes1996} 
	has shown that the evolution of 
	$\Psi_0$ via the usual energy eigenfunction decomposition 
	yields wave functions at time $t$ whose graph 
	is a fractal for almost all $t \geq 0$. 
	
	As an attempt of resolution, both Sanz 
	\cite{sanzBohmianApproachQuantum2005} and later 
	Grübl and Penz \cite{grublNondifferentiableBohmianTrajectories2011}  
	have suggested that it may be possible to 
	construct `quantum trajectories' by considering a sequence of 
	so-called $C^\infty$-vectors, computing a sequence of such 
	trajectories for the time-evolved wave functions, and 
	then taking the limit in some appropriate sense. 
	While this might be a fruitful approach to the construction of 
	`quantum trajectories' 
	for general, `suitably irregular' drift fields $\vec{v}$, 
	we take a different perspective on the issue here. 
	
	We shall first recapitulate the mathematics of the  
	problem on the basis of the articles by 
	Grübl and Penz 
	\cite{grublNondifferentiableBohmianTrajectories2011} and 
	Bonneau et al. 
	\cite{bonneauSelfadjointExtensionsOperators2001}. 
	
	First one observes that one needs to satisfy 
	the homogeneous Dirichlet boundary condition at fixed times: 
	\begin{equation}
		\forall x \in \partial \Omega \colon \quad \Psi(x) = 0 
		\quad \iff \quad 
		\Psi(0)=\Psi(L) = 0
		\, . 
		\label{eq:Dirichletbc}
	\end{equation}
	In order to account for Eq. \eqref{eq:Dirichletbc}, 
	one considers the Sobolev spaces 
	$H^k_0(\Omega,\C)$. For $k \in \N$ those are defined as the 
	closure of $C^\infty_0(\Omega,\C)$ 
	with respect to the norm in $H^k(\Omega,\C)$, i.e. 
	$H_0^k(\Omega,\C)$ consists of 
	all elements $\Phi$ such that there 
	exists a sequence of $\left(\Phi_n\right)_{n \in \N}$ 
	in $C^\infty_0(\Omega,\C)$ 
	that converges to 
	$\Phi$ in the $H^k$-norm as $n \to \infty$. 
	Each $H^k_0(\Omega,\C)$ is a linear subspace of 
	$H^1_0(\Omega,\C)$, the latter consisting of all equivalence 
	classes of $\C$-valued, absolutely continuous functions satisfying 
	the boundary conditions (cf. Thms. 8.2 and 8.12 in Ref. 
	\cite{brezisFunctionalAnalysisSobolev2011}). 
	
	To account for the boundary condition, 
	Eq. \eqref{eq:Dirichletbc}, it is thus sufficient 
	that $\Psi \in H^1_0(\Omega,\C)$, viewing  
	$H^1_0(\Omega,\C)$ as a linear 
	subspace of $L^2 \left(\Omega, \C\right)$.%
	\footnote{We refer, for instance, 
					to the beginning of Sec. 8.4 
					in Brezis' book
					\cite{brezisFunctionalAnalysisSobolev2011} 
					for an example of how homogeneous 
					Dirichlet boundary conditions are treated 
					within the mathematical theory of PDEs.} 
	
	Furthermore, we may interpret the derivatives in the free-body Hamiltonian 
	$\hat{H}$ in the weak sense \cite{bonneauSelfadjointExtensionsOperators2001,
	grublNondifferentiableBohmianTrajectories2011}, leading us 
	to consider the linear subspace
	\begin{equation}
		\dom \hat{H} = H^1_0(\Omega,\C) \cap H^2(\Omega,\C)
			\label{eq:domHbox}
	\end{equation}
	of $L^2(\Omega,\C)$ as a natural domain for the Hamiltonian 
	$\hat{H}$ (cf. Sec. 3 in Ref. 
	\cite{grublNondifferentiableBohmianTrajectories2011}). 
	On this domain $\hat{H}$ is indeed self-adjoint, since it 
	is proportional to the 
	`Dirichlet Laplacian'  	
	(cf. Ex. 1 in Sec. X.3 of Ref. 
			\cite{reedMethodsModernMathematical1975}).%
		\footnote{Note that in Ref. \cite{bonneauSelfadjointExtensionsOperators2001} 
					the authors considered the case 
					$\dom \hat{H} = H^2_0(\Omega,\C)$ instead. However, 
					if $\Psi \in H_0^2(\Omega,\C)$, then its weak derivative $\Psi'$ 
					is an element of $H_0^1(\Omega,\C)$. Thus, we can choose an absolutely 
					continuous representative for which $\Psi'(0)=\Psi'(L)=0$. 
					By considering $H^2_0(\Omega,\C)$ we have therefore implicitly 
					imposed an additional boundary condition that is not derived 
					from any physical considerations. Accordingly, in Ref. 
					\cite{bonneauSelfadjointExtensionsOperators2001} the authors  
					find that $\hat{H}$ is not self-adjoint for this choice of 
					domain. If one considers Grübl's and Penz' choice of domain, 
					Eq. \eqref{eq:domHbox}, instead, one does therefore 
					not need to answer the question which 
					self-adjoint extensions of $\hat{H}$ on 
					$H_0^2(\Omega,\C)$ are physically preferred.  
					}
	In turn, it gives rise to a well-defined time-evolution operator 
	$U_t$ in 
	$L^2 \left(\Omega, \C\right)$ for all $t \in \R$
	(see e.g. Thm. VIII.7 in Ref. \cite{reedMethodsModernMathematical1972} or 
	Thm. 1.5 in Supplement 1 of Ref. \cite{berezinSchrodingerEquation1991}). 
	
	In particular, the time evolution $U_t \Psi_0$ of Berry's initially constant 
	wave function $\Psi_0$ is well-defined in this sense. 
	
	Having clarified the mathematics, let us now return to Hall's claim of the 
	`breakdown of the velocity equation' 	
	\cite{hallIncompletenessTrajectorybasedInterpretations2004}: 
	The author is correct to criticize that it is far from 
	clear how Eq. \eqref{eq:defv} is to be mathematically 
	interpreted for general $L^2$-wave functions. 
	Though Berry's example \cite{berryQuantumFractalsBoxes1996} 
	is particularly illustrative, it is a 
	well-known fact that not all $L^2$-functions are weakly 
	differentiable. Moreover, as neither $\dom \hat{H}$, 
	$H^1_0 \left(\Omega,\C \right)$, or even $H^1(\Omega,\C)$   
	are closed subspaces of $L^2(\Omega,\C)$, one cannot simply 
	resolve the issue by choosing a smaller Hilbert space. 
	Keeping the approaches suggested in Refs. 
	\cite{sanzBohmianApproachQuantum2005,
	grublNondifferentiableBohmianTrajectories2011} out of this
	discussion, 
	it is therefore not possible to address the problem 
	within the strict confines of quantum mechanics. 
	
	However, neither the Bohmian theory 
	nor theories based on the distributional Madelung equations
	aim to reproduce quantum mechanics in every detail. Rather, the 
	overarching question is whether they constitute internally consistent 
	physical theories in agreement with empirical data within their 
	domain of applicability. 
	
	Indeed, Berry's example exhibits one major 
	defect: Since $\Psi_0$ does not lie in 
	$H^1_0 \left(\Omega,\C \right)$, it does not 
	satisfy the boundary condition 
	for the problem, Eq. \eqref{eq:Dirichletbc}. 
	There is thus a physical argument for its 
	rejection as an acceptable initial condition. 
	
	On the other hand, Grübl and Penz 
	\cite{grublNondifferentiableBohmianTrajectories2011} 
	have observed that $U_t$ 
	preserves the domain of $\hat{H}$ in the sense that 
	for any initial wave function $\Phi_0 \in \dom \hat{H}$ 
	and any 
	$t \in \R$  we have $U_t \Phi_0 \in \dom \hat{H}$ here 
	(cf. Prop. \ref{Prop:ReedSimondomaininvariance} in Appx. A).
	
	Instead of considering the time evolution on 
	$L^2(\Omega,\C)$ -- 
	which ignores the boundary conditions -- we may therefore 
	restrict ourselves to the time evolution on the 
	natural domain of the Hamiltonian $\hat{H}$. In that case 
	the velocity equation, Eq. \eqref{eq:defv}, is well-defined 
	on the (distributional) support of all wave functions $\Phi_0 
	\in \dom \hat{H}$, as $\dom \hat{H}$ is a linear 
	subspace of $H^1(\Omega,\C)$. At least with 
	respect to the drift field itself, there is thus 
	no such `breakdown of the velocity equation'. 
	Note that the Sobolev embedding theorem does, however, 
	not imply that the drift field is 
	Lipshitz continuous in $x$ 
	(cf. Thm. 5.4, Part II, Case C' in Ref. 
	\cite{adamsSobolevSpaces1975}), so that the Picard-Lindelöf 
	theorem is not applicable to its integral curves. 
	
	It also worthy of note that $\dom \hat{H}$ is dense 
	in $L^2 \left(\Omega,\C \right)$, since this is the 
	case for $C_0^\infty \left(\Omega, \C\right)$. Thus, 
	for an arbitrary $\varepsilon>0$ there exists a  
	$\Psi'_0 \in \dom \hat{H}$ such that 
	$\norm{\Psi_0 - \Psi_0'} < \varepsilon$. That is,  
	$\Psi'_0$ approximates Berry's initial wave function to 
	arbitrary precision $\varepsilon$. As 
	$\norm{\Psi_0 - \Psi'_0} = \norm{U_t \Psi_0 - 
	U_t \Psi'_0}$, the same holds for all times 
	$t$. It is thus impossible to make an empirical 
	argument against the restriction to the subspace 
	$\dom \hat{H}$. 
\end{subequations} 
\end{Remark}

To summarize this section, the current state of research does not provide 
a satisfactory answer to the question of what precisely is 
meant with the term `distributional Madelung equations'. 
Nonetheless, we hope that the   
general considerations expressed here contribute to 
further research, so that more results 
will be established on the general 
relationship between such Madelung equations and the Schrödinger 
equation. 

\section{Some explicit `non-quantized' strong solutions of the 2-dimensional 
Madelung equations}
\label{sec:nonquantumsol}

In Ref. 
\cite{wallstromInequivalenceSchrodingerEquation1994} Wallstrom claimed 
that the isotropic harmonic oscillator in $2$ dimensions provides 
an explicit example in support of the claim that there are 
solutions of the Madelung equations that are not solutions 
of the Schrödinger equation. In the following section we will 
explicitly construct such strong solutions for the Madelung equations. 
As we shall see, their existence is not trivial. 
Thereafter, we analyze them from a distributional point 
of view. 

\subsection{Construction and analysis of strong solutions}
\label{ssec:strongnonquantumsol}

To construct such `non-quantized' strong solutions of the Madelung equations, 
consider again the time-independent Schrödinger equation in Eq. 
\eqref{eq:oscillatortindepSE}. As above, we apply the separation ansatz, 
Eq. \eqref{eq:sepansatz}, in order to obtain an angular 
equation, Eq. \eqref{eq:angularSE}, and a radial equation, 
Eq. \eqref{eq:radialSE}. However, we shall not constrain 
$\mu$ to the integers a priori. 

We first consider the question of how the `quantization' of 
$\mu$, that is the condition $\mu \in \Z$, may be obtained when one looks 
for strong solutions of the Schrödinger equation. 

To recapitulate the relevant statements in Ref. \cite{wallstromInequivalenceSchrodingerEquation1994}, Wallstrom argues    
that in this instance $\Psi$ has to 
be `single-valued': Supposedly, the condition 
\begin{equation}
	\Psi(\rho, \phi) = \Psi(\rho, \phi + 2 \pi k)
	\label{eq:svalued}
\end{equation}
has to hold for all $k \in \Z$. Then, since 
$\Phi (\phi) = e^{\pm \iu \mu \phi}$ yields two linearly independent 
fundamental solutions to the angular equation \eqref{eq:angularSE}, 
he then states that condition \eqref{eq:svalued} implies that 
$\mu \in \Z$. The author then goes on to claim  
that, without the above condition -- or another 
`quantization condition' such as the one in Eq. 
\eqref{eq:wallstrom} -- one obtains (stationary) 
solutions of the Madelung equations that are not solutions of the 
Schrödinger equation. 

Upon mathematical inspection, condition \eqref{eq:svalued} turns 
out to be invalid:  

In looking for strong solutions of Eq. \eqref{eq:oscillatortindepSE}, 
it is important to keep in mind that $\Psi$ is actually a function defined on 
a subset of $\R^2$. Yet in this instance we consider its coordinate representation 
in polar coordinates, the latter being only defined on the 
coordinate chart codomain $(0, \infty) \times (0, \phi)$. While the choice 
of this codomain is a matter of convention, there is no choice of 
convention in which Eq. \eqref{eq:svalued} is sensible. That is, 
Eq. \eqref{eq:svalued} cannot be justified mathematically. 

One may, however, ask for the global function to be continuous on 
the entirety of $\R^2$, so 
that for the coordinate representation $\Psi$ we require 
\begin{equation}
	\lim_{\phi \to 0} \Psi(\rho, \phi) = \lim_{\phi \to 2 \pi} \Psi(\rho, \phi) 
	\label{eq:contphi}
\end{equation}
for all $\rho > 0$. Indeed, Eq. \eqref{eq:contphi} then implies that 
$\mu \in \Z$. 

The problematic 
Eq. \eqref{eq:svalued} may therefore 
be replaced by the condition of continuity, Eq.  
\eqref{eq:contphi}, to yield the desired result. Note that this 
is a weaker condition than the one considered in Prop. 
\ref{Prop:C1quantization}. 

But is there an a priori justification for imposing continuity 
on the positive 
$x$-axis, Eq. \eqref{eq:contphi}? Quantum mechanics does indeed provide 
such a reason: 

For the sake of argument, let the domain $\dom \hat{H}$ 
of the Hamiltonian $\hat{H}$, as defined by Eq. 
\eqref{eq:oscillatortindepSE} in polar coordinates, be a linear subspace of 
the space of square integrable functions in 
$C^2 \left(\R_+ \times (0,2\pi),\C \right)$. This way, we can understand 
the derivatives in the strong sense as well as satisfy the condition that 
$\varrho$ is a probability density. 
Without worrying about self-adjointness of $\hat{H}$ here and, again, 
focusing on strong solutions only, 
we nonetheless require $\hat{H}$ to 
be symmetric: For all $\Psi$, $\Phi$ $\in \dom \hat{H}$ we need  
\begin{equation}
	\int_0^{2\pi} \d \phi \int_0^\infty \rho \,  \d \rho \,  
	\Psi^*(\rho,\phi) \, \bigl( \hat{H} \Phi \bigr) (\rho,\phi)
	= \int_0^{2\pi} \d \phi \int_0^\infty \rho \,  \d \rho \, 
	\bigl( \hat{H}\Psi \bigr)^* \negthinspace (\rho,\phi) \,  \Phi (\rho,\phi) \, . 
	\label{eq:Hsymmetric}
\end{equation}
Looking at the angular part of $\hat{H}$ in Eq. \eqref{eq:Hsymmetric} 
only, integration by parts shows that any $\Psi \in \dom \hat{H}$ does 
indeed need to satisfy 
Eq. \eqref{eq:contphi} as well as 
\begin{equation}
	\lim_{\phi \to 0} \partd{\Psi}{\phi}(\rho, \phi) = \lim_{\phi \to 2 \pi} 
	\partd{\Psi}{\phi}(\rho, \phi) 
	\label{eq:contdphi}
\end{equation}
for all $\rho \in \R_+$. 

Therefore, symmetry of $\hat{H}$ 
implies Eq. \eqref{eq:contphi} and thus `quantization' of 
$\mu$. 

Let us move on to Wallstrom's claim that `explicit solutions to the 
Madelung equations not satisfying the Schrödinger equation may be constructed' 
\cite{wallstromInequivalenceSchrodingerEquation1994}.  

We first observe that $\mu$ has to be real for 
wave functions 
satisfying the separation ansatz, Eq. \eqref{eq:sepansatz}, if we also ask 
for the respective probability density 
to be continuous on $\R^2 \setminus \lbrace 0 \rbrace$: 
$\Phi(\phi)= e^{\pm \iu \mu \phi}$ for $\mu \in \C$ are the only 
linearly independent solutions of the angular equation, Eq. 
\eqref{eq:angularSE}. Due to the separation ansatz, 
Eq. \eqref{eq:sepansatz}, $\phi \mapsto \abs{\Phi}^2= 
e^{\mp 2 \Im(\mu) \phi}$ has to extend to a continuous 
function in Cartesian coordinates. Thus 
$\mu \in \R$, indeed. 

Similarly, $E$ has to be real, as otherwise the factor 
$e^{- \iu E t / \hbar}$ in the ($t$-dependent) wave function 
would contradict probability conservation. 

Using reality of $\mu$, one easily checks that the corresponding 
drift field $\vec{v}$ can be smoothly extended to $\R^2 \setminus \lbrace 0 
\rbrace$ and for $\Phi(\phi)= e^{+\iu \mu \phi}$ the field 
$\vec{v}$ is given by Eq. \eqref{eq:vstandardoscillator} above, with 
an additional minus sign for $\Phi(\phi)= e^{- \iu \mu \phi}$. 

In the absence of other constraints, 
this is the mathematical argument that underlies Wallstrom's claim 
above. 

We remark that, even if one looks for strong solutions of the Madelung 
equations, the drift field $\vec{v}$ is `irregular' in the sense of 
having a singularity at the origin. Moreover, as all 
other singularities and nodes of the density and the drift field $\vec{v}$ 
are radial, none of the connected components of $\dom \vec{v}$ are simply 
connected (unless we make the cut at e.g. $\phi=0$). 
The topological condition required in the statement of Thm. 3.2  
in Ref. \cite{reddigerMadelungPictureFoundation2017}, which shows 
the (local) equivalence of the Schrödinger equation and the Madelung equations 
in the strong sense, is therefore violated here. 

Still, the situation is not quite as simple as Wallstrom suggested (cf. 
Sec. III in Ref. \cite{wallstromInequivalenceSchrodingerEquation1994}): 
In order to solve the radial equation \eqref{eq:radialSE}, we also need to 
satisfy the radial integrability condition 
\begin{equation}
	\int_0^\infty \abs{R(\rho)}^2 \, \rho \,  \d \rho < \infty \, .
	\label{eq:radialintegrability}
\end{equation}
As we will show in the proof of Thm. 
\ref{Thm:nonintegersol} below, it is not true that all 
solutions of the radial equation with 
$\mu \in \R$ automatically satisfy Eq. \ref{eq:radialintegrability}. 
In particular, that non-trivial solutions exist for non-integer 
$\mu$ is not given a priori, but needs to be checked explicitly. This 
point does not seem to have been taken into account in Ref. 
\cite{wallstromInequivalenceSchrodingerEquation1994}. 

Nonetheless, the following theorem shows that strong solutions of the radial equation 
with $\mu \in \R \setminus \Z$ that also satisfy the integrability 
condition do indeed exist. 

\begin{Theorem}
	\label{Thm:nonintegersol}
\begin{subequations}
	For arbitrary $\mu \in \R \setminus \Z$, $E \in \R$ and 
	up to a constant factor, all non-trivial (strong) solutions $R$ of Eq. 
	\eqref{eq:radialSE}, 
	that are integrable in the sense of 
	Eq. \eqref{eq:radialintegrability}, fall under one of the 
	following two cases: 
	
	\begin{enumerate}[i)]
		\item 	\label{itm:nonintegersol1}
				For $a \in \R \setminus - \N_0$ and 
				$\mu \in (-1,0) \cup (0,1)$ we have 
					\begin{equation}
						R (\rho) = \rho^{\mu} \, 
									U\left(a, \mu+1;  \frac{m \omega}{ \hbar} \, \rho^2 
									\right)
									\, e^{- \frac{m \omega}{ \hbar} 
									\, \frac{\rho^2}{2}} 
					\end{equation} 
				with 
					\begin{equation}
						E = \hbar \omega \left( - 2 a + \mu + 1 \right)\, . 
					\end{equation}
		\item 	\label{itm:nonintegersol2}
				For $n \in \N_0$ and 
				$\mu \in (-1,\infty) \setminus \Z$ we have 
					\begin{equation}
						R (\rho) = \rho^{\mu}  \, 
									L^\mu_n \left(  \frac{m \omega}{ \hbar} \, \rho^2 
									\right) 
									\, e^{- \frac{m \omega}{ \hbar} 
									\, \frac{\rho^2}{2}}
					\end{equation} 
				with 
					\begin{equation}
						E = \hbar \omega \left( 2 n + \mu + 1 \right)\, . 
					\end{equation}
	\end{enumerate}
\end{subequations}
\end{Theorem}

	It is important to keep in mind that, 
	due to linearity of the Schrödinger equation, superpositions of wave 
	functions of same energy $E$ constructed from the solutions in  
	Thm. \ref{Thm:nonintegersol} also need to be considered. We shall not 
	consider the question here in which cases solutions constructed in this manner  
	also give rise to strong solutions of the Madelung equations, but only 
	look at those arising from wave functions 
	satisfying the separation ansatz, Eq. \eqref{eq:sepansatz}. 
	
\begin{Corollary}
	\label{Cor:Madelungsol}
\begin{subequations}
	The functions $\varrho$ and $\vec{v}$, defined as follows, 
	provide smooth, stationary solutions to the $2$-dimensional Madelung 
	equations with potential 
	\begin{equation}
		V \colon \R^3 \to \R \quad \colon \quad (t,x,y) \mapsto 
		V(t,x,y)= \frac{m}{2} \, \omega^2 \, (x^2 + y^2) \, , 
	\end{equation}
		satisfying the normalization condition 
	\begin{equation}
		\forall t \in \R \colon \quad 
		1 = \int_{\R^2} \d x \, \d y \, \varrho(t,x,y) \, .
		\label{eq:2dnormalization}
	\end{equation}	
	\begin{enumerate}[i)]
		\item	For $a \in \R \setminus - \N_0$, $\mu \in (-1,0) \cup (0,1)$, 
				$(t,x,y)$ in 
					\begin{equation}
						\set{(t,x,y) \in \R^3}{(x,y) \neq 0 
								\quad \text{and} \quad 
								U\left(a, \mu+1;  \frac{m \omega}{ \hbar} \, 
									(x^2+y^2) 
									\right) \neq 0 
								} \, , 
					\end{equation}
				and $A_{a \mu} \in \R_+$ chosen such that Eq. 
				\eqref{eq:2dnormalization} is satisfied, set 
					\begin{equation}
						\varrho(t,x,y) = \frac{1}{A_{a \mu}} \, (x^2+y^2)^{\mu} \, \left( 
									U\left(a, \mu+1;  \frac{m \omega}{ \hbar} \, 
									(x^2+y^2) 
									\right) \right)^2 
									\, e^{- \frac{m \omega}{ \hbar} 
									\, (x^2+y^2)} 
								\label{eq:Wallrho1}
					\end{equation}
				and 
					\begin{equation}
						\vec{v} (t,x,y) = \pm \frac{\mu \hbar}{m} \, 
								\frac{1}{ x^2 + y^2}
								\begin{pmatrix}
									- y \\ 
									x
								\end{pmatrix} \, . 
								\label{eq:Wallv}
					\end{equation}
		\item	For $n \in \N_0$, $\mu \in (-1,\infty) \setminus \Z$, 
				$(t,x,y)$ in 
					\begin{equation}
						\set{(t,x,y) \in \R^3}{(x,y) \neq 0 
								\quad \text{and} \quad 
								L^\mu_n \left(  \frac{m \omega}{ \hbar} \, (x^2+y^2)
									\right) \neq 0 
								} \, , 
					\end{equation}
				and $A_{n \mu} \in \R_+$ chosen such that Eq. 
				\eqref{eq:2dnormalization} is satisfied, set 
				\begin{equation}
						\varrho(t,x,y)	= \frac{1}{A_{n \mu}}  \,	(x^2+y^2)^{\mu}  \, \left( 
									L^\mu_n \left(  \frac{m \omega}{ \hbar} \, (x^2+y^2) 
									\right) \right)^2
									\, e^{- \frac{m \omega}{ \hbar} 
									\, (x^2+y^2)}
								\label{eq:Wallrho2}
				\end{equation}
				and $\vec{v}(t,x,y)$ as above. 
	\end{enumerate}
\end{subequations}
\end{Corollary}

	As Wallstrom anticipated correctly 
	\cite{wallstromInequivalenceSchrodingerEquation1994}, from a purely 
	mathematical perspective  
	the solutions of the respective 
	Madelung equations in Cor. \ref{Cor:Madelungsol} are perfectly 
	ac\-cept\-able---at least if 
	we look at strong solutions only. Though the 
	densities $\varrho$, as given by Eq. \eqref{eq:Wallrho1} 
	or Eq. \eqref{eq:Wallrho2}, fail to be differentiable at 
	the origin, the origin is not in the domain of $\varrho(t, \, . \,)$ 
	due to the singularity of $\vec{v}$ there  
	and the requirement that the domains of 
	$\varrho$ and $\vec{v}$ coincide (cf. 
	p. 1351 in Ref. \cite{reddigerMadelungPictureFoundation2017}). 
	
	From a physical point of view, it would, of course, be more 
	appropriate to consider a physically more realistic model in 
	$3$ dimensions, such as the one of the `hydrogen-like atom' 
	considered in Sec. \ref{ssec:vorticity}. 
	For without the ultimate ability to argue on the basis 
	of empirical data, the discussion of whether the solutions in 
	Cor. \ref{Cor:Madelungsol} are physically acceptable is 
	misplaced. Do note, however, that such 
	data is commonly obtained from spectroscopy, while the 
	$3$-dimensional model in Sec. \ref{ssec:vorticity} 
	describes only hydrogenlike atoms in the absence of external 
	radiation. 
	
	Nonetheless, apart from the 
	fact that for the solutions in Cor. \ref{Cor:Madelungsol} $\mu$ is not 
	`quantized', it is noteworthy that there exists no minimum of 
	the energy $E$ in either one of the two sets of solutions. 
	For the solutions constructed from point \ref{itm:nonintegersol2} in Thm. 
	\ref{Thm:nonintegersol}, there is only an infimum of the energy, $E=0$, 
	but no minimum. For 
	solutions from point \ref{itm:nonintegersol1}, 
	$E$ is not even bounded from below. There is no doubt that 
	most physicists would find this fact alone objectionable. 
	That $E$ is indeed 
	the physical energy of the system, even in theories based on the 
	Madelung equations (see e.g. 
	Eq. 4.22 in Ref. \cite{reddigerMadelungPictureFoundation2017}), 
	can be easily derived from the 
	so called `Hamilton-Jacobi-Madelung equation'. 
	
	Whether analogous `non-quantized' 
	solutions exist in more physical models, as for instance 
	in the aforementioned model of hydrogen-like atoms, is 
	beyond the scope of this article. 
	
\subsection{Analysis of strong solutions from the point of view of distributions}
\label{ssec:distnonquantumsol}

In this section we analyze the solutions of Thm. 
\ref{Thm:nonintegersol} and Cor. \ref{Cor:Madelungsol} from a 
distributional perspective. Though, for a lack of an established 
variational formulation of the Madelung equations (cf. Sec. 
\ref{sec:distmadelung}), this analysis will be necessarily 
incomplete, the results here may nonetheless be of use for 
providing an answer to the precise relationship between the 
Schrödinger equation and the Madelung equations in this 
distributional framework. 

We begin by recalling that the drift fields 
in Cor. \ref{Cor:Madelungsol} -- which were computed 
using strong derivatives and smooth extensions -- 
do not satisfy the third 
Madelung equation, Eq. \eqref{eq:Madelung3}, in the 
distributional sense. We refer back to Sec. \ref{ssec:vorticity}, 
in particular Def. \ref{Def:dcurl} and Prop. \ref{Prop:wallstromv}. 
There we have discussed the option of requiring the third 
Madelung equation to hold in a distributional sense as well, 
with the caveat that this would require further physical 
justification. 

Clearly, if we do impose this condition on the strong solutions in  
Cor. \ref{Cor:Madelungsol}, they will have to be 
thrown out---just like the ones 
for integer $\mu$. 

For the case that $\mu$ is an integer, however, 
we noted that we can still get stationary solutions of the Madelung 
equations for every admissible energy by considering the respective 
real-valued wave functions instead: That is, we choose 
\begin{subequations}
		\label{eq:realPsisol} 
\begin{align}
	\Psi(\rho,\phi) 
			&= R(\rho) \, \sin( \mu \phi) \, \,   , \text{or} 
		\label{eq:realPsisol1}
		\\
	\Psi(\rho,\phi) 
			&= R(\rho) \, \cos( \mu \phi)  
		\label{eq:realPsisol2}
\end{align}
\end{subequations}
with the respective function $R$ and the trivial $t$-dependence omitted. 

Contrarily, if $\mu$ is real yet not an integer, we are faced with the 
following problem. 
 
\begin{Proposition}
	\label{Prop:weakderPsi}
	Let $R$ and $\mu$ be as defined by points 
	\ref{itm:nonintegersol1} or \ref{itm:nonintegersol2} in Thm. 
	\ref{Thm:nonintegersol}. Set 
	 	\begin{equation}
		\dom \Psi 
			= \set{(x,y) \in \R^2}{x<0 \, \, \text{whenever} \, \, y=0} \, . 
			\label{eq:domPsitcounterex}
	\end{equation}
	Then the following holds. 
	\begin{enumerate}[i)]
		\item 	\label{itm:Prop:weakderPsi1}
				The function $\Psi$ with values 
				\begin{equation}
					\Psi(x,y) = R\bigl(\sqrt{x^2+y^2}\bigr) \, 
								\exp\bigl(\iu \mu \arg(x+\iu y)\bigr) 
				\end{equation}
				for $(x,y) \in \dom \Psi$ 
				is not weakly differentiable. 
		\item 	\label{itm:Prop:weakderPsi2}
				If  
				$\Psi$ is given by 
				\begin{equation}
						\Psi(x,y) = R\bigl(\sqrt{x^2+y^2}\bigr) \, 
											\cos\bigl( \mu \arg(x+\iu y)\bigr)
				\end{equation}
				for any $(x,y) \in \dom \Psi$, 
				then it is also not weakly differentiable. 
		\item 	\label{itm:Prop:weakderPsi3}
				If  
				$\Psi$ is given by 
				\begin{equation}
						\Psi(x,y)  = R\bigl(\sqrt{x^2+y^2}\bigr) \, 
											\sin\bigl(\mu \arg(x+\iu y)\bigr)
				\end{equation}
				for any $(x,y) \in \dom \Psi$, 
				then it is weakly differentiable if and only if  
				$\mu$ is also half-integer.  
	\end{enumerate}
\end{Proposition}

We expect that $\Psi$ in Prop. 
\ref{Prop:weakderPsi}.\ref{itm:Prop:weakderPsi3} is not twice weakly 
differentiable (with respect to $y$), even if $\mu$ is 
half-integer. A rigorous  
proof would be rather laborious, however, as in some cases 
one simultaneously needs to handle 
the lack of integrability of the strong derivative on 
$\dom \Psi$ and the discontinuity due to the cosine. 

Regarding the respective densities $\varrho$ in points \ref{itm:Prop:weakderPsi2}
and \ref{itm:Prop:weakderPsi3}, the discontinuities in 
$\Psi$ are not removed by computing $\abs{\Psi}^2$, so that we 
also lose weak differentiability in this manner. For the 
special case of $\mu \in \Z /2$ in point \ref{itm:Prop:weakderPsi3}, 
we also expect that $\varrho$ is only once weakly differentiable. 

So requiring both irrotationality of $\vec{v}$ in the distributional 
sense and weak differentiability up to second order 
of the $L^1$-function $\varrho$ would exclude 
the aforementioned solutions, Eq. \eqref{eq:domPsitcounterex} and \eqref{eq:realPsisol}, for non-integer 
$\mu$, while still allowing for the real-valued solutions from Eq. \eqref{eq:realPsisol}  
for integer $\mu$. As discussed in 
Sec. \ref{sec:distmadelung}, requiring weak differentiability of $\varrho$ (and 
$\Psi$) up to second order is a reasonable request---as long as it is 
compatible with the respective dynamics (see also Rem. 
\ref{Rem:Hall}). 

We have to leave unaddressed the question of whether this result generalizes 
to general stationary superpositions of the wave functions determined by 
Thm. \ref{Thm:nonintegersol}. 

Surely, one may object to the exclusion of the quasi-irrotational, strong  
solutions of the Madelung equations for integer $\mu$, since these correspond to 
commonly accepted solutions of the respective Schrödinger equation. Yet the need 
for these solutions requires a physical argument. Moreover, 
the respective drift fields 
are `pathological' in the sense that they are not locally $L^2$-integrable 
(cf. Rem. \ref{Rem:noirrothydrogen}). 

In the remainder of this section we shall consider the question of 
what the mathematical theory of quantum mechanics 
has to say about the wave functions from Thm. \ref{Thm:nonintegersol}. 
Though we have already commented on this matter in Sec. 
\ref{ssec:strongnonquantumsol} to some degree, in this section we will not make 
any a priori assumption on the regularity of the wave functions 
apart from those imposed by the theory. 

We shall follow Berezin and Shubin in recalling the 
mathematical situation for the problem  (cf. Sec. 1.8 in Ref. 
\cite{berezinSchrodingerEquation1991}):  

For the $1$-dimensional problem the eigenfunctions 
$\Phi_n$ of the respective Hamiltonian 
belong to the space of Schwartz functions 
$\mathcal{S}(\R,\C)$, considered as a subspace of the 
Hilbert space $L^2(\R,\C)$. The corresponding energy eigenvalues 
are  
\begin{equation}
	E_n = \hbar \omega \left(n + \frac{1}{2}\right) \, . 
	\label{eq:oscillatorE1d}
\end{equation}
Due to Eq. \eqref{eq:oscillatorE1d}, the Hamiltonian is 
well-defined on the domain 
\begin{equation}
	\set{\Psi \in L^2 \left(\R,\C\right)}{\sum_{n=0}^\infty 
		\abs{\inp{\Phi_n}{\Psi} n }^2  
		\, \, \text{converges}
				} \, . 
\end{equation}
For this choice of domain the operator is indeed self-adjoint. 

For the $2$-dimensional case the appropriate Hilbert space is 
$\mathcal{H} = L^2 \left( \R^2, \C \right)$. As, roughly speaking, 
the Hamiltonian 
		\begin{equation}		
			 	\hat{H} = - \frac{\hbar^2}{2 m}	\Delta + 
				 	\frac{m}{2} \, \omega^2 (x^2+y)^2 
				 \label{eq:2isooschatH} 
		\end{equation} 
is the sum of two Hamiltonians for the 
$1$-dimensional problem, we consider the energy 
eigenfunctions 
\begin{equation}
		\Phi_{n_1 n_2} \colon 
		(x,y) \mapsto \Phi_{n_1 n_2} (x,y)= \Phi_{n_1}(x) \Phi_{n_2}(y) 
		\label{eq:2isooschatHeigenfunctions}
\end{equation}
with corresponding energy eigenvalue $E_{n_1 n_2} = E_{n_1} + E_{n_2}$. 
Accordingly, on the domain 
	\begin{equation}
		\dom \hat{H} = 
			\set{\Psi \in L^2 \left(\R^2,\C\right)}{\sum_{n_1,n_2 =0}^\infty 
			\abs{\inp{\Phi_{n_1 n_2}}{\Psi} (n_1 + n_2) }^2  
					\, \, \text{converges}
					} 
			\label{eq:2isooschatHdom} 
	\end{equation}
the operator $\hat{H}$ is self-adjoint. 

Since the Schrödinger time evolution is well-defined for any 
$\Psi \in \mathcal{H}$, Smolin was correct to note that 
solutions to the 
Schrödinger equation need not be continuous but only 
square-integrable (cf. Sec. IV in 
Ref. \cite{smolinCouldQuantumMechanics2006}).%
	\footnote{ 	In footnote 2 of Ref. 
				\cite{valentiniInflationaryCosmologyProbe2010a} 
				Smolin's observation was criticized, implying  
				that one should not allow for discontinuous 
				wave functions. The necessity to look at the 
				problem from 
				the perspective of the mathematical theory of 
				quantum mechanics resolves 
				those concerns (cf. Sec. \ref{ssec:distributions}). 
				} 
Still one has to ask if the wave functions from 
Sec. \ref{ssec:strongnonquantumsol} provide (necessarily 
stationary) solutions of the time-dependent Schrödinger equation. 

The next proposition answers this question in the negative. 
\begin{Proposition}
	\label{Prop:solutionSE}
\begin{subequations}
	Let $R$, $\mu$, and $E$ be given by points 
	\ref{itm:nonintegersol1} or \ref{itm:nonintegersol2} in Thm. 
	\ref{Thm:nonintegersol}. 
	Let $\Psi$ have values  
	\begin{equation}
		\Psi_t(x,y) = R \bigl(\sqrt{x^2 + y^2} \bigr) 
			\, e^{\iu \left( \mu \arg(x+ \iu y) - E t / \hbar \right)}
			\label{eq:Prop:solutionSEPsit}
	\end{equation}
	for all $t \in \R$ and $(x,y)$ in the set defined by 
	Eq. \eqref{eq:domPsitcounterex}. 
	
	Then 
		\begin{enumerate}[i)]
			\item \label{itm:Prop:solutionSE1}
					for all $t \in \R$ 
					(the equivalence class of) $\Psi_t$ is an element of  
					$\mathcal{H} =L^2(\R^2,\C)$, and 
			\item \label{itm:Prop:solutionSE2}
					(the family of equivalence classes of) $\Psi$ is not 
					an $L^2$-solution of the Schrödinger equation 
					for the operator $\hat{H}$  on $\dom \hat{H}$, 
					as given by Eqs. \eqref{eq:2isooschatH} and 
					\eqref{eq:2isooschatHdom}, respectively. 
		\end{enumerate}
\end{subequations}
\end{Proposition}

In a more colloquial language Prop. \ref{Prop:solutionSE} states the following: 
Even though the wave functions from Sec. \ref{ssec:strongnonquantumsol}
are admissible as `initial' wave functions, the corresponding 
`stationary solutions' are not quantum-mechanical solutions of the 
respective Schrödinger equation. Wallstrom's 
argument, stated in the beginning of Sec. \ref{ssec:strongnonquantumsol}, 
therefore requires the following modification to hold true: 
The solutions from Sec. \ref{ssec:strongnonquantumsol} do not satisfy the 
Schrödinger equation in the quantum-mechanical sense, while still yielding  
strong solutions of the respective Madelung equations. 

\begin{Remark}[Further clarification of Wallstrom's first objection]
	\label{Rem:fritsche}
\begin{subequations}	
	We shall comment on two works by Fritsche and Haugk 
	\cite{fritscheNewLookDerivation2003,
	fritscheStochasticFoundationQuantum2009} that are of
	relevance in this context. 
	
	According to 
	Ref. \cite{delapenaEmergingQuantumPhysics2015}, those 
	provide `[d]etailed rebuttals of Wallstrom's argument'. We 
	find that this is not the case. 
	
	In Ref. \cite{fritscheNewLookDerivation2003} the authors 
	state the following: 
	\begin{quote}
		As $\varrho(\vec{r},t)$ and $\vec{v}(\vec{r},t)$ will in general 
			be unique (and smooth) functions of $\vec{r}$ 
			we require $\Psi(\vec{r},t)$ to have the same property. 
			
			[Notation adapted]
	\end{quote}
	This clearly does not address Wallstrom's criticism, for his 
	argument is precisely that there exist strong solutions 
	of the Madelung equations that are not solutions of the Schrödinger 
	equation (in the quantum-mechanical sense). 
	
	The relevant argument in Sec. VIII of 	
	Ref. \cite{fritscheStochasticFoundationQuantum2009} is 
	more substantial: Therein the authors consider the 
	$3$-dimensional $1$-body Schrödinger equation with 
	spherically symmetric potential and try to argue from 
	the linearity of the Schrödinger equation 
	that the respective magnetic quantum numbers have 
	to be integer. While their argument is incomplete -- 
	it fails for the case that the functions $F$ they consider 
	are orthogonal -- it may be viewed as a precursor 
	to Prop. \ref{Prop:solutionSE}.\ref{itm:Prop:solutionSE2} 
	above. However, even if the respective statement in 
	Ref. \cite{fritscheStochasticFoundationQuantum2009} were 
	corrected and appropriately generalized, 
	it would fail to address Wallstrom's criticism. 
\end{subequations}
\end{Remark}

However, as argued in Sec. \ref{ssec:distributions} and elaborated 
upon in Rem. \ref{Rem:Hall}, the question is 
not how the Schrödinger equation in the quantum-mechanical sense 
relates to the Madelung equations in the strong sense. Rather, the 
question is how the Schrödinger equation is related to an 
appropriate variational 
formulation of the Madelung equations within the modern theory of PDEs. 

As we observed in Rem. \ref{Rem:Hall} above, the quantum-mechanical 
space of initial 
wave functions,  
$L^2 \left(\R^2,\C \right)$ in this instance, may be too large to obtain 
an equivalence between the two systems of PDEs and we might need to restrict 
ourselves to the domain of the 
Hamiltonian instead (cf. Prop. \ref{Prop:ReedSimondomaininvariance} 
in Appx. \ref{appx:A}). 

We may therefore ask, whether the solutions $\Psi$ found in 
Sec. \ref{ssec:strongnonquantumsol} are contained in $\dom \hat{H}$. 
To answer the question, one would need to compute all 
coefficients $\inp{\Phi_{n_1 n_2}}{\Psi}$ and check whether the 
respective series ``$\hat{H} \Psi$'' converges or not. 
Instead we shall provide the reader with the following 
sufficient condition (which might even be necessary). 
\begin{Lemma}
	\label{Lem:domHoscillator}
	Let $\hat{H}$ with domain  $\dom \hat{H}$ be 
	given by Eqs. \eqref{eq:2isooschatH} and 
					\eqref{eq:2isooschatHdom}, respectively.
	Then the set of equivalence classes 
	\begin{equation}
		\set{[\Psi] \in H^2 \left(\R^2,\C\right)}{\forall \Psi' \in [\Psi] \colon  
				\left[(x,y) \mapsto \left(x^2 + y^2 \right) \Psi'(x,y) \right] 
				\in L^2\left(\R^2,\C\right)}
				\label{eq:domHoscillatorprime}
	\end{equation}
	is a linear subspace of $\dom \hat{H}$. 
\end{Lemma}

By Prop. \ref{Prop:weakderPsi}.\ref{itm:Prop:weakderPsi1}, the 
`non-quantized' $\Psi$ found in Sec. \ref{ssec:strongnonquantumsol} 
do not meet the 
sufficient condition of Lem. \ref{Lem:domHoscillator} 
for being contained in $\dom \hat{H}$. Since 
the condition is arguably weak, it is at least doubtful 
whether the necessary condition is met. 

From Prop. \ref{Prop:weakderPsi}.\ref{itm:Prop:weakderPsi1} 
it does follow, however, 
that the (initial values of the) `non-quantized' solutions are not in 
$H^1(\R^2,\C)$. They are therefore not acceptable, if 
we consider the $2$-dimensional analog of the 
result by Gasser and Markowich 
\cite{gasserQuantumHydrodynamicsWigner1997} discussed in 
Sec. \ref{sec:distmadelung}. 

Furthermore, Prop. 
\ref{Prop:weakderPsi}.\ref{itm:Prop:weakderPsi1} shows that is 
not even clear how to define the drift field 
for the respective wave functions. The procedure 
of computing it from Eq. \eqref{eq:defv} almost everywhere 
in the strong sense and then smoothly extending the result to 
$\R^2 \setminus \lbrace 0 \rbrace$ is not acceptable 
in this context, for one has to consider 
distributional derivatives instead and, by Prop. 
\ref{Prop:weakderPsi}.\ref{itm:Prop:weakderPsi1},  
the gradient of $\Psi$ is not a regular distribution. 

As argued in Sec. \ref{ssec:distributions}, it is precisely this error 
in not approaching the relation between 
the Schrödinger equation and the Madelung equations from the 
perspective of the modern theory of PDEs that constitutes the 
biggest gap in Wallstrom's argument. 
Establishing a mathematically rigorous relationship between the two systems of 
PDEs requires a specification of the respective function 
spaces as well as a clarification of how the equations ought to be 
mathematically interpreted. Without such, the claim that the 
Madelung equations and the Schrödinger equation are inequivalent is 
devoid of mathematical content. 

\section{Conclusion}
\label{sec:conclusion} 

The central contributions of this work were already 
listed in Sec. \ref{sec:intro}, so we shall not repeat those here. 
Instead, we shall provide the reader with some general comments 
as well as a few clarifying remarks on the 
related literature. 

Most importantly, we wish to emphasize that this work does 
not provide a full resolution of Wallstrom's objections. What 
it does provide is a clarification to what extent those objections 
are justified as well as various possible avenues for finding such 
a resolution. 

In this respect, the central message is that 
the relation between the Schrödinger equation and the Madelung 
equations needs to be clarified in the context of the mathematical 
theory of distributions. Interpreting the respective 
PDEs in the strong sense only would already defy the theory of 
quantum mechanics, so we should not expect this 
approach to take us beyond established theory. 
In relating the two systems of PDEs the mathematical foundations of 
quantum mechanics need to be taken into account, even if it is not the 
aim of a given variational formulation of the Madelung equations 
to reproduce the quantum-mechanical Schrödinger theory in every aspect. 

Therefore, at this point in time the task of resolving the controversy is 
primarily one of functional analysis. Given the geometric nature of the 
equations, however, prior works that took a more geometric approach 
might be of use as well. We refer to Refs. 
\cite{kibbleGeometrizationQuantumMechanics1979,vonrenesseOptimalTransportView2012,
reddigerMadelungPictureFoundation2017,lesselShapeSpaceTerms2018,
khesinGeometricHydrodynamicsMadelung2018,khesinGeometryMadelungTransform2019,
khesinGeometricHydrodynamicsOpen2022}.  

Based on our research, the vast amount of articles citing 
Wallstrom's works on Takabayasi's condition 
\cite{wallstromDerivationSchrodingerEquation1989,
wallstromInequivalenceSchrodingerEquation1994} confirm 
that his objections have been viewed as a discreditation of 
theories based on the Madelung equations---thus confirming 
the statement by de la Peña et al. 
\cite{delapenaEmergingQuantumPhysics2015} 
quoted in the introduction. As Wallstrom's objections 
failed to properly account for the 
mathematical structure of quantum mechanics, 
we hope to have convinced the reader that casting such a 
judgment on the viability of the 
Madelung equations as fundamental laws of nature 
is premature at this point. 

Still, the view that Wallstrom effectively ended the 
scientific discussion deserves some sympathy, for 
there have been several works in the literature that 
have attempted to address Wallstrom's objections 
without giving a full convincing resolution: 

As stated in Sec. \ref{sec:intro}, addressing Wallstrom's 
objections on the level of stochastic processes only 
\cite{derakhshaniStochasticMechanicsAd2017,derakhshaniSuggestedAnswerWallstrom2019,
grossingClassicalExplanationQuantization2011,delapenaEmergingQuantumPhysics2015} 
is unlikely to convince the wider physics community.  
	
The proposals made in Refs. 	
\cite{carlenCorrespondenceStochasticMechanics1989,hennebergerWhenWaveFunction1994,
hennebergerWhenWaveFunction1994,
fritscheNewLookDerivation2003,fritscheStochasticFoundationQuantum2009,
catichaEntropicDynamicsTime2011} 
that attempt to address Wallstrom's objections on the phenomenological level 
we view as largely unsuccessful:  
The argument in Ref. \cite{carlenCorrespondenceStochasticMechanics1989} 
is misplaced, for the space $C^\infty_\alpha$ the authors 
define only corresponds to smooth functions on the circle, if 
$\alpha \in \Z$. Still, the authors deserve credit for 
discussing Wallstrom's objections in relation to the 
actual mathematical formalism of quantum mechanics. In Ref. 
\cite{hennebergerWhenWaveFunction1994} the respective authors 
state that 
``In bound states, the superposition principle by 
itself guarantees single-valuedness.'' and that this is supposedly 
`illustrated' in Ref. \cite{al-jaberTopologicalConsiderationsQuantum1992}. 
While no argument was given that could establish this general assertion, 
the argument seems to be similar to the one given in Ref. 
\cite{fritscheStochasticFoundationQuantum2009}. The 
latter we already addressed in Rem. \ref{Rem:fritsche}. Finally, 
in Ref. \cite{catichaEntropicDynamicsTime2011} the authors 
tried to use an (outdated) argument in a 1939 article by 
Pauli \cite{pauliUeberKriteriumFuer1939} to exclude wave functions such as the ones 
found in Sec. \ref{ssec:strongnonquantumsol}. The counterargument voiced in
Rem. \ref{Rem:fritsche} applies here as well. 

Apart from the suggestions of resolution provided in this article, 
potentially promising resolutions of the matter have also been given 
by Zak \cite{zakOriginRandomnessQuantum2014} and 
Loffredo and Morato \cite{loffredoLagrangianVariationalPrinciple1989}: 
Zak argued that the `solutions of the Schrödinger and Madelung equations may 
have different criteria of instability'. While the stability of solutions of the 
Schrödinger equation follows from linearity, the Madelung equations 
are inherently non-linear, independent of the chosen 
variational formulation. Stability is one of the criteria of 
well-posedness in the sense of Hadamard, so that Zak's suggestion may 
be of relevance in the mathematical study of a given variational 
formulation of the Madelung equations. The suggestion 
by Loffredo and Morato is more radical in that the authors suggest a 
modification of the first two Madelung equations, Eqs. 
\eqref{eq:Madelung1} and \eqref{eq:Madelung2}, for the case that the 
drift field has non-vanishing vorticity. We have discussed this 
suggestion in Sec. \ref{ssec:vorticity}, noting, in particular, that 
it is indeed difficult to justify the irrotationality 
condition \eqref{eq:Madelung3} on physical grounds. 

We shall finish with the remark that nowadays there is significant 
historical evidence that with regards to the 
Madelung equations the debate  
on the foundations of quantum theory has had a strongly ideological 
component 
\cite{cushingQuantumMechanicsHistorical1994,
bellerQuantumDialogueMaking1999,
freirejuniorQuantumDissidentsRebuilding2015}. We hope that our  
work contributes to shifting the debate more towards  
mathematical facts and a discussion of the 
actual empirical evidence.


\section[nonumber=true]{Appendix}
\newcounter{appendix}
\setcounter{appendix}{0}
\renewcommand{\theappendix}{\Alph{appendix}}
\refstepcounter{appendix}\label{appx:A}
\setcounter{equation}{0}
\renewcommand{\theequation}{\Alph{appendix}.\arabic{equation}}

\setcounter{Proposition}{0}
\renewcommand{\theProposition}{\Alph{appendix}.\arabic{Proposition}}

\markright{Appendix \theappendix}	

\subsection[nonumber=true]{Appendix \Alph{appendix}: 
		Invariance of domain of self-adjoint Hamiltonians 
		under time evolution}

	The following result is of potential importance 
	for determining the relevant function spaces of 
	a given variational formulation of the Madelung equations. 
	We make no claims of originality, since 
	it is commonly used as an argument in proving 
	Stone's theorem (cf. Thm. VIII.8 in Ref. 
	\cite{reedMethodsModernMathematical1972} and 
	Thm. 23.2 in Ref. \cite{blanchardMathematicalMethodsPhysics2015}) 
	and it was also used in Ref. 
	\cite{grublNondifferentiableBohmianTrajectories2011}. 
	Nonetheless, its potential relevance justifies an explicit statement. 
	\begin{Proposition}
		\label{Prop:ReedSimondomaininvariance}
		Let $\hat{H}$ be a self-adjoint operator with 
		domain $\dom \hat{H}$ in a Hilbert space 
		$\mathcal{H}$. Let 
		$t \mapsto U_t=\exp\bigl(- \iu t \hat{H}/\hbar\bigr)$ 
		be the corresponding strongly continuous one-parameter 
		unitary group. 
		
		Then for all $t \in \R$ we have 
		\begin{equation*}
				U_t \bigl(\dom \hat{H}\bigr) 
				= \dom \hat{H} \, . 
		\end{equation*}
	\end{Proposition}
	While one can use commutativity of $\hat{H}$ and $U_t$ to show the assertion 
	(cf. Cor. 3.16 in Ref. \cite{penzDensityPotentialMappingQuantum2016}),  
	we prove it here as a corollary of Thm. VIII.7 in Ref. 
	\cite{reedMethodsModernMathematical1972}. 
	\begin{Proof}
		Denote by $\norm{\, . \, }$ the norm on $\hat{H}$. 
		For any $\Psi_0 \in \dom \hat{H}$ and $t \in \R$ 
		set $\Psi_t = U_t \Psi_0$.
		Observe that for all $\varepsilon >0$ we have 
		\begin{equation}
			\norm{\frac{U_\varepsilon \Psi_0- \Psi_0}{\varepsilon}
			+ \iu \hat{H} \Psi_0 /\hbar}
			= \norm{\frac{U_\varepsilon \Psi_t- \Psi_t}{\varepsilon}
			+ \iu U_t \hat{H}\Psi_0 /\hbar} \, . 
		\end{equation}
		By point (c) in the aforementioned theorem, the limit on the left hand 
		side as $\varepsilon \to 0$ is $0$. By point (d) therein and equality 
		with the right hand side, 
		$\Psi_t$ is in $\dom \hat{H}$. 
	\end{Proof}
	

\refstepcounter{appendix}\label{appx:B}
\setcounter{equation}{0}
\markright{Appendix \theappendix}	

\subsection[nonumber=true]{Appendix \Alph{appendix}: Proofs}
	
\subsubsection*{Proof of Prop. \ref{Prop:C1quantization}}	

	Under the respective assumptions, $\vec{v}$ is well-defined and the 
	integral exists. Setting $\mathbb{S}^1 = \set {z \in \C}{1=\abs{z}}$, 
	consider the $C^1$-function 
	\begin{equation}
		Q \colon D \to \mathbb{S}^1 \colon x \mapsto Q(x) = \Psi(x)/ \abs{\Psi(x)} \, .
	\end{equation}
	
	Now restrict $Q \circ \gamma$ to $(a,b)$ to obtain the map $\xi$. $\xi$ 
	is a $C^1$-map between manifolds. Arguing as in the proof of Thm. 17.35 in Ref. 
	\cite{leeIntroductionSmoothManifolds2003}, we may apply Sard's theorem 
	to pick a regular value $z_0$ of $\xi$ in $\mathbb{S}^1$ 
	(cf. Thm. 1.5.18 in Ref. \cite{rudolphDifferentialGeometryMathematical2013}). 
	Then $\xi^{-1}(\lbrace z_0 \rbrace)$ is a (possibly empty) properly embedded 
	$0$-dimensional submanifold of $(a,b)$ 
	(cf. Cor. 1.8.3 in Ref. \cite{rudolphDifferentialGeometryMathematical2013}, 
	Prop. A.53 (c), as well as the 
	proof of Cor. 5.14 in Ref. \cite{leeIntroductionSmoothManifolds2003}). 
	Thus, $\xi^{-1}(\lbrace z_0 \rbrace)$ is a 
	compact $0$-dimensional manifold and, as such, it is finite (or empty). 
	So we may write $\xi^{-1}(\lbrace z_0 \rbrace)= \bigcup_{j =1}^N \lbrace t_j \rbrace$
	with $N \in \N_0$ and $t_j<t_{j+1}$ for all $j \in \lbrace 1, \dots, N-1 \rbrace$, 
	the case $N=0$ referring to $\xi^{-1}(\lbrace z_0 \rbrace)= \emptyset$. 

	We shall make a branch cut at $z_0 \in \mathbb{S}^1$ to redefine the complex logarithm: 
	For $\varphi_0 \in [0,2\pi)$ let $z_0 = e^{\iu \varphi_0}$ and then 
	\begin{equation}
	 	\ln \colon \mathbb{S}^1 \to  [\varphi_0,\varphi_0+2\pi) \iu \subset \C 	
	\end{equation}  
	is a (`single-valued') complex logarithm on the manifold $\mathbb{S}^1$, 
	discontinuous at $z_0$ and smooth otherwise. 
	
	Setting $t_0 = a$ and $t_{N+1}=b$, a 
	straightforward calculation shows that the left hand side of Eq. \eqref{eq:wallstrom} 
	is equal to 
	\begin{align}
		\frac{1}{2 \pi \iu} 
		\oint_\gamma \frac{\nabla Q}{Q} \cdot \d \vec{r} 
		&= \frac{1}{2 \pi \iu} \sum_{j=0}^{N+1} \int_{t_j}^{t_{j+1}} 
		\frac{\d}{\d t} \bigl( \ln (Q \circ \gamma(t)) \bigr) \, \d t \\
		&= \frac{1}{2 \pi \iu} \sum_{j=0}^{N+1} 
		\left( \lim_{t \to t_{j+1}^-} \ln (Q \circ \gamma(t)) 
					- \lim_{t \to t_{j}^+} \ln (Q \circ \gamma(t))\right) \, .
				\label{eq:ointsum}
	\end{align}	
	As $z_0$ is a regular value of $\xi$, for each 
	$j \in \lbrace 1, \dots, N \rbrace$ the derivative of 
	$Q \circ \gamma$ at $t_j$ is non-zero and its sign determines the 
	value of the respective one-sided limit in Eq. \eqref{eq:ointsum}. 
	If it is positive, the two summands with the limits 
	$t \to t_j^{\mp}$ together yield $2 \pi \iu$. Else they yield 
	$- 2 \pi \iu$. Regarding the endpoints, if 
	$(Q \circ \gamma)(a) = z_0$, we may apply the same argument since 
	$\gamma(a) = \gamma(b)$. 
	Otherwise, due to continuity 
	of the logarithm on $\mathbb{S}^1 \setminus \lbrace z_0 \rbrace$, 
	the respective limits will cancel each other. 
	The assertion follows by summing all terms. 
	
\subsubsection*{Proof of Prop. \ref{Prop:wallstromv}}	

	Using polar coordinates $(\rho,\phi)$, one easily shows that the 
	(Euclidean) components of $\vec{v}$ are locally integrable. 
	 
	For all $\varphi \in \mathcal{D}$ we find that 
	$\left( \operatorname{curl} \vec{v} \right) ( \varphi)$ equals 
		\begin{multline}
			- \frac{\mu \hbar}{m} 
			\int_0^\infty \rho \, \d \rho \int_0^{2 \pi} \d \phi 
				\left( 
						\frac{\cos \phi}{\rho} 
							\left(
								\cos \phi \, \partd{\varphi}{\rho}(\rho,\phi)
								+
								\frac{- \sin \phi}{\rho} \, 
								\partd{\varphi}{\phi}(\rho,\phi) 
							\right) 
				\right.
				\\
				+ 
				\left.
						\frac{\sin \phi}{\rho} 
							\left(
								\sin \phi \, \partd{\varphi}{\rho}(\rho,\phi)
								+
								\frac{\cos \phi}{\rho} \, 
								\partd{\varphi}{\phi}(\rho,\phi) 
							\right)
				\right) \, . 
				\label{eq:curlvcompute1}
		\end{multline}
	Thus 
	\begin{equation}
		\left( \operatorname{curl} \vec{v} \right) ( \varphi)
		= - \frac{\mu \hbar}{m} \int_0^{2 \pi} \d \phi \int_0^\infty \d \rho  \, 
		\partd{\varphi}{\rho}(\rho,\phi)
		= \frac{2 \pi \mu \hbar}{m} \, \varphi(0) 
		= \frac{2 \pi \mu \hbar}{m} \, 
		\delta_0 (\varphi) \, .
			\label{eq:curlvcompute2}
	\end{equation}
	The last statement follows from the fact that the Dirac delta 
	is not a regular distribution. 
		
\subsubsection*{Proof of Cor. \ref{Cor:hydrogenvdistder}}
	
	The first two components in Eq. \eqref{eq:dcurlvhydrogen} vanish, 
	due to the fact that $v^1$ and $v^2$ do not depend on $z$, $v^3=0$, and 
	due to the (generalized) fundamental theorem of calculus (cf. 
	Thm. 8.2 in Ref. \cite{brezisFunctionalAnalysisSobolev2011}). 
	The third component follows from direct comparison of the 
	(Cartesian) integral with the one from Prop. 
	\ref{Prop:wallstromv} above. 
	
	It remains to prove that there does not exist any 
	locally $L^1$-integrable function $\chi$ such that 
	for all $\vec{\varphi} \in \mathcal{D}$ we have 
		\begin{equation}
			\xi (\vec{\varphi}) = - \int_{\R^3} \chi \, \varphi^3 \, .
		\end{equation}
	To simplify notation, we write $\xi (\varphi)$ for the above expression, 
	so that $\xi$ is viewed as a distribution acting on an 
	$\R$-valued test function $\varphi$ instead. 
	
	Consider a sequence of bump functions 
	$(\zeta_k)_{k \in \N}$ in $C^\infty_0(\R,\R)$ 
	such that for all $k \in \N$ and $x \in \R$ we have  
	$0 \leq \zeta_{k+1}(x) \leq \zeta_{k}(x)$ and $\zeta_k(0)=1$. 
	Furthermore, as $k \to \infty$ we require that $\zeta_k(x)$ 
	tends to $1$ for $x=0$ and to $0$ else. Such a sequence exists. 
	Employing a standard line of reasoning, 
	we shall use this sequence to construct a contradiction. 
		
	Let $\eta$ be any real-valued, positive, nonzero function 
	such that for all $k \in \N$ the function 
	\begin{equation}
		\varphi_k \colon (x,y,z) \mapsto \varphi_k (x,y,z) = \zeta_k (x)  \, \eta(y,z) 
	\end{equation}
	is in $C^\infty_0(\R^3,\R)$. Then for all $(x,y,z) \in \R^3$ we find that  
	$\varphi_k(x,y,z)$ is bounded by 
	$\zeta_1(x,y,z) \, \eta(y,z)$. The respective bound is integrable and as $k \to \infty$ 
	the sequence $\varphi_k$ converges pointwise to the zero function almost everywhere. 
	Therefore, by Lebesgue's dominated convergence theorem, we have 
	\begin{equation}
		\lim_{k \to \infty} \int_{\R^3} \d^3 x \, \chi(x,y,z) \, 
			 \zeta_k (x)  \, \eta(y,z) = 0 \, .  
	\end{equation}
		Yet we also find that 
	\begin{equation}
		\lim_{k \to \infty} \xi(\varphi_k) = \int_{- \infty}^\infty 
		\d z \, \eta (0,z) > 0 \, , 
	\end{equation}
	thus yielding the contradiction. 	
	
\subsubsection*{Proof of Lem. \ref{Lem:curlvextenddomain}}

	We may drop the index and continuously extend $\varphi$ to $\R^3$. 

	For the first component of the distributional curl, we show that 
	$\int_{-\infty}^\infty \d z \, \partial{\varphi}/\partial{z}$ 
	vanishes. Denote by $\Id_{I}$ the indicator function on an interval 
	$I$ in $\subset \R$. For  $(x,y,z) \in \R^3$ and all $n \in \N$ 
	define  
	\begin{equation}
		f_n (x,y,z) =  \partd{\varphi}{z} (x,y,z) \, 
		\Id_{(-n,n)}(z) \, . 
	\end{equation} 
	Since $f_n (x,y, \, . \,)$ is also continuous at the origin, we find 
	\begin{equation}
		\abs{\int_\R \d z \, f_n (x,y, z) }
			= \abs{\varphi(x,y,n) - \varphi(x,y,-n)} 
			\leq 2 \sup_{n \in \Z} \abs{\varphi(x,y,n)} < \infty \, .  
	\end{equation}
	As for all $n \in \N$ we 
	have $f_n (x,y, \, . \,) \leq f_{n+1} (x,y, \, . \,)$, Beppo Levi's theorem 
	implies that $\partial \varphi/\partial z (x,y, \, . \,)$ is in $L^1$. Since 
	$\varphi$ vanishes at infinity, the respective integral of $\partial \varphi/\partial z$ 
	with respect to $z$ vanishes, too. 
	
	Then, by Tonelli's and Fubini's theorem, the first component of 
	the distributional curl vanishes. The same holds true for the second component. 
	
	Regarding the third component, 
	we observe that, if the integral exists, it is 
	equal to the corresponding integral in cylindrical 
	polar coordinates $(\rho, \phi,z)$. 
	Recalling 
	the expression in Eq. \eqref{eq:curlvcompute1}, we therefore consider 
	the formal integral 
	$\int_0^\infty \d \rho \, \partial \varphi /\partial \rho (\rho,\phi,z)$. 
	Upon defining 
	\begin{equation}
		g_n(\rho, \varphi, z) = \partd{ \varphi}{\rho} (\rho,\phi,z) 
			\Id_{(0,n)} (\rho) \, , 
	\end{equation}
	we may apply an analogous argument to the one above 
	to conclude that 
	\begin{equation}
		\int_0^\infty \partd{ \varphi}{\rho} (\rho,\phi,z)  
			= - \varphi(0,0,z) \, , 
	\end{equation}
	referring to the function in Cartesian coordinates 
	on the right hand side.
	
	Finally, we consider $\varrho$ from Eq. 
	\eqref{eq:varrhostandardhydrogen}. We have 
	$\varrho \in C^1(\R^3 \setminus \lbrace 0 \rbrace, \R)$. Moreover, 
	$\varrho$ is integrable and vanishes at infinity, independent of the values of 
	$n$, $l$, and $\mu$. 
	
	To show continuity at the origin, consider $\varrho$
	in spherical polar coordinates $(r,\theta,\phi)$. 
	For $n=1$ we find that  
	$l=0$, $\mu =0$, and thus $P_l^{\abs{\mu}}=1$. The limit 
	$r \to 0$ is independent of the angles and thus $\varrho$ 
	is continuous at the origin. For $n >1$ and $l=0$ the argument 
	is analogous. For $n>1$ and $l>0$ the function vanishes as 
	$r \to 0$ and is therefore continuous at the origin as well. 
	
\subsubsection*{Proof of Prop. \ref{Prop:Expectationvorthydrogen}}		
	
	As stated in Lem. \ref{Lem:curlvextenddomain}, 
	we may apply Eq. \eqref{eq:Corhydrogenvdistderequs}. For 
	$\mu =0$ the expression trivially vanishes. For $\mu \neq 0$
	the integrand has a factor of 
	\begin{equation}
		\left( P_l^{\abs{\mu}}(z/\abs{z})\right)^2= 
			\left( P_l^{\abs{\mu}}(\operatorname{sgn}{z})\right)^2 \, . 
	\end{equation} 
	Due to the Rodrigues' formula for the associated Legendre 
	polynomials (see e.g. Eq. 7.36 in Ref. 
	\cite{ballentineQuantumMechanicsModern1998}), the latter 
	expression vanishes for all $z \neq 0$, $l$, and $\mu \neq 0$. 
	
\subsubsection*{Proof of Thm. \ref{Thm:nonintegersol}}	

	We solve Eq. \eqref{eq:radialSE} by the known procedure of 
	reducing it to the confluent hypergeometric equation: 

	First, non-dimensionalize Eq. \eqref{eq:radialSE} by setting 
	$\mathcal{E} = E / (\hbar \omega /2)$ and 
	$r = \sqrt{ m \omega / \hbar} \, \rho$. This yields 
	\begin{equation}
	 r^2 \frac{\d^2 R}{\d r^2}(r) + r \frac{\d R}{\d r}(r) - r^4 R(r) 
	 + \mathcal{E} r^2 R(r) - \mu^2 R(r) = 0 \, . 
	 	\label{eq:nondimradialeq}
	\end{equation}
	As $r \in R_+$, we may set $x=r^2$ and formulate Eq. 
	\eqref{eq:nondimradialeq} in terms of the new variable $x$. 
	By an abuse of notation, denote the new function  
	by $R \colon x \mapsto R(x)$ as well. 
	Without loss of generality, we may now use the ansatz 
	\begin{equation}
		R(x) = x^{\mu/2} \, e^{-x/2} \, \xi (x) \, . 
	\end{equation}
	This indeed yields the confluent hypergeometric equation 
	\begin{equation}
		x \, \frac{\d^2 \xi}{\d x^2}(x) + (b-x) \, \frac{\d \xi}{\d x}(x) - a \, \xi(x)  = 0 
		\label{eq:Kummereq}
	\end{equation}
	with 
	\begin{equation}
		a = \frac{\mu + 1 - \mathcal{E}/2}{2} \quad \text{and} \quad b = \mu +1 \, . 
	\end{equation}
	By assumption, both $a$ and $b$ are real with $b \notin \Z$. 
	
	The remaining proof consists of looking up known solutions $\xi$ of 
	Eq. \eqref{eq:Kummereq} and assuring 
	that the integrability condition 
	\begin{equation}
		\int_0^\infty x^\mu \, e^{-x} \, \abs{\xi(x)}^2 \, \d x < \infty
		\label{eq:xintegrability}
	\end{equation}
	is satisfied. 
	The former requires a careful consideration of different sets of 
	linearly independent  
	solutions for the respective values of the parameters $a$ and $b$ 
	(cf. Table I in Ref. \cite{mathewsjr.PhysicistGuideSolution2021}), 
	the 
	latter a study of the asymptotic behavior of those solutions in the limits as 
	$x \to 0$ and $x \to \infty$ 
	(cf. Chap. 4 in Ref. \cite{slaterConfluentHypergeometricFunctions1960})---as 
	well as a proof of the con- or divergence of the respective integral. 
	\begin{enumerate}[i)]
		\item
				The first case we consider is $a \in \R \setminus - \N_0$ and 
				$b \in \R \setminus \Z$. 
	
				Denoting by ${}_1 F_1$ the confluent hypergeometric function of first kind, 
				we have 
				\begin{equation}
					\xi (x)  = c_1 \, {}_1 F_1(a,b;x) + c_2 \, U(a,b;x)
				\end{equation}
				for arbitrary $c_1$, $c_2$ $\in \C$. 
			
				As $x \to \infty$, the first solution satisfies 
				\begin{equation}
					{}_1 F_1(a,b;x) \sim \frac{\Gamma(b)}{\Gamma(a)} \, x^{a-b} \, e^x 
				\end{equation}
				(cf. Eq. 4.1.7 in Ref. \cite{slaterConfluentHypergeometricFunctions1960}), 
				and the second one obeys 
				\begin{equation}
					U(a,b;x) \sim x^{-a} \, ,
					\label{eq:Utoinfty}
				\end{equation}
				(cf. Eq. 4.1.11 in Ref. \cite{slaterConfluentHypergeometricFunctions1960}). 
				As we require 
				\begin{equation}
					\lim_{x \to \infty} x^\mu \, e^{-x} \, \abs{\xi(x)}^2 = 0
						\label{eq:xinftyintegrability}
					\end{equation}
				for the integral in Eq. \eqref{eq:xintegrability} to converge, 
				we find that $c_1 = 0$. 
				
				As $x \to 0$, the definition of $U$ in terms of 
				${}_1 F_1$, Eq. 1.3.1 in Ref. \cite{slaterConfluentHypergeometricFunctions1960}, 
				yields%
					\footnote{Note that $z \mapsto 1/\Gamma(z)$ is 
									holomorphic on the entire complex plane.}
				\begin{equation}	
					U(a,b;x) \sim 
							\begin{cases}	
								\frac{\Gamma(b-1)}{\Gamma(a)} x^{1-b}  
									& , 1-b < 0 \\
								\frac{\Gamma(1-b)}{\Gamma(1-b+a)} 
									& , 1-b >0
							\end{cases}
							\, . 
					\label{eq:Uto0}
				\end{equation}
				Recalling Eq. \eqref{eq:xintegrability}, for $\mu > 0$ 
				we thus obtain the restriction that $-\mu > - 1$, while 
				for $\mu < 0$ we similarly obtain $\mu > -1$. 
				
				In order to prove 
				convergence or divergence of the respective 
				integral, we need some estimates for $x\to 0$ and 
				$x \to \infty$: By Eq. \eqref{eq:Utoinfty} and 
				the definition of the limit, for every $\varepsilon > 0$ there exists 
				a point $x_\infty \in \R_+$ such that for all $x > x_\infty$ we have
				\begin{equation}
					\abs{U(a,b;x) - x^{-a}} < \varepsilon \, x^{-a} \, . 
					\label{eq:Utoinftyestimate}
				\end{equation}
				Analogously, for $\mu < 0$ Eq. \eqref{eq:Uto0} implies that 
				for our choice of $\varepsilon$ there exists $x_{-} \in \R_+$ 
				such that for all $x \in (0,x_-)$ we have 
				\begin{equation}	
					\abs{U(a,b;x) - d_-} < \varepsilon \, d_- \, , 
					\label{eq:Uto0estimate-}
				\end{equation}
				where we set $d_- = \Gamma(-\mu)/\Gamma(a-\mu)$. 
				For $\mu > 0$ Eq. \eqref{eq:Uto0} 
				implies that there exists $x_{+} \in \R_+$ such that for all $x \in (0,x_+)$ 
				we have 
				\begin{equation}	
					\abs{U(a,b;x) - d_+ \, x^{-\mu}} < \varepsilon \, d_+ \, x^{-\mu} \, , 
					\label{eq:Uto0estimate+}
				\end{equation}
				setting $d_+ = \Gamma(\mu)/\Gamma(a)$.
				
				We are now in a position to prove convergence of the integral in Eq. 
				\eqref{eq:xintegrability} for $\xi(x) = U(a,b;x)$ and 
				$\mu \in (-1,0)\cup (0,1)$: 	
				We split the integral into three, 
				going from $0$ to $d_{\pm}$, $d_\pm$ to $d_\infty$, and $d_\infty$ 
				to $\infty$, respectively. For the third integral we derive the estimate  
				\begin{align}
					\bigl( U(a,\mu +1;x )
					\bigr)^2 
					&= 
					\bigl( \abs{U(a,\mu +1;x ) 
					-x^{-a} + x^{-a}
					}\bigr)^2 
					\\
					&\leq 
					\bigl( \abs{U(a,\mu +1;x ) 
					-x^{-a}} + x^{-a}
					\bigr)^2 
					\\
					&< (1+\varepsilon)^2 x^{-2a} \, .
				\end{align}
				Thus, 
				\begin{equation}
					\int_{x_\infty}^\infty x^\mu \, e^{-x} \, \abs{U(a,\mu +1;x )}^2 \, \d x
					< (1+\varepsilon)^2 
					\int_{x_\infty}^\infty x^{\mu-2a} \, e^{-x} \, \d x \, , 
				\end{equation}
				so the former integral converges by the dominated 
				convergence theorem. 
				The first integral is handled analogously for each one of the two cases 
				$\mu \in (-1,0)$ and $\mu \in (0,1)$. In the second 
				integral we integrate a continuous function over a compact interval 
				$[x_\pm, x_\infty]$, so the integral converges. Finally, as the individual integrals 
				converge, their sum converges. 
				
				Formally, we also need to show divergence of the integral for 
				$\mu$ not in $(-1,0)\cup (0,1)$: For $\mu < -1$ and $0<x<x_-$
				Eq. 
				\eqref{eq:Uto0estimate-} yields 
				\begin{align}
					d_- 
						&= \abs{d_- - U(a,\mu+1;x) + U(a,\mu+1;x)} \\
						&\leq \abs{ d_- - U(a,\mu+1;x)} + \abs{U(a,\mu+1;x)} \\
						&< \varepsilon d_- + \abs{U(a,\mu+1;x)}
				\end{align}
				and thus 
				\begin{equation}
					(1-\varepsilon) d_- < \abs{U(a,\mu+1;x)} \, . 
				\end{equation}
				In turn, 
				\begin{align}
					\int_{0}^\infty x^\mu \, e^{-x} \, \abs{U(a,\mu +1;x )}^2 \, \d x
					&\geq \int_{0}^{d_-} x^\mu \, e^{-x} \, \abs{U(a,\mu +1;x )}^2 \, \d x
						\\
					&> (1-\varepsilon)^2 d_-^2\, \int_{0}^{d_-} x^\mu \, e^{-x}  \, \d x
					\, .
				\end{align}
				For $\mu <-1$ the latter integral diverges. The case 
				$\mu >1$ is handled analogously using the estimate 
				Eq. \eqref{eq:Uto0estimate+} for $0<x<x_+$. 
				
				Summing up, for $a \in \R \setminus - \N_0$ 
				Eq. \eqref{eq:Kummereq} only has nontrivial solutions
				whenever $\mu \in (-1,0) \cup (0,1)$. 
				Those solutions are proportional to $U(a,\mu+1; \, . \,)$. 
		\item
				There remains the case that $- a = n \in \N_0$ and 
				$b \in \R \setminus \Z$. 
				
				Following Table I in Ref. \cite{mathewsjr.PhysicistGuideSolution2021}  
				and using the definition of $L_n^\mu$ in 
				\S 5.5 in Ref. \cite{slaterConfluentHypergeometricFunctions1960}, 
				the general solution is given by 
				\begin{equation}
					\xi (x)  = c_1  L_n^\mu(x)
								+ c_2 \, x^{1-b} \, {}_1 F_1(1+a-b,2-b;x) \, . 
				\end{equation}
				for $c_1$, $c_2$ $\in \C$. 
				
				As $x \to \infty$, we conclude that $c_2=0$ by an argument analogous 
				to the one above. 
				
				As $x \to 0$, the polynomial $L_n^\mu(x)$ is asymptotic to 
				a constant. Hence Eq. \eqref{eq:xintegrability} requires that 
				$\mu > -1$. 
				
				For those values of $\mu$, the integral converges, because it 
				can be written as a finite sum of integrals over $x^{(k+\mu)} e^{-x}$
				for different $k \in \N_0$. For the other values of $\mu$ 
				we show divergence of the integral as we did for 
				$U(a,\mu+1;x)$ above. 
				
				In conclusion, for $- a = n \in \N_0$ we only obtain 
				nontrivial solutions for $\mu \in (-1, \infty) \setminus \Z$. 
				Those solutions are proportional to $L_n^{\mu}$. 
	\end{enumerate}

\subsubsection*{Proof of Prop. \ref{Prop:weakderPsi}}

	\begin{enumerate}[i)]
	\item 
	Set $\Phi(x,y) = R\bigl(\sqrt{x^2+y^2}\bigr)$. Extend 
	$\Phi$ to $\R^2 \setminus 
	\lbrace 0 \rbrace$. On this new domain, $\Psi$ is $C^1$. We shall use the notation 
	$\partial / \partial y$ for strong derivatives of functions on 
	the respective domain. 
	
	Similar to the proof of Cor. \ref{Cor:hydrogenvdistder} above, 
	we need to show that there does not exist a locally integrable 
	$\chi$ such that 
		\begin{equation}
			\int_{\R^2} \d x \, \d y \, \Psi(x,y) \, \partd{\xi}{y} (x,y)
				= -
			\int_{\R^2} \d x \, \d y \, \chi(x,y) \, \xi (x,y) 
				\,  
			\label{eq:weakderPsi}
		\end{equation}
	holds for all $\xi \in C_0^\infty\left(\R^2, \R\right)$. Again, the general 
	procedure consists of two main steps: First, integrate by parts to create non-vanishing 
	`boundary terms'. Second, assume existence of $\chi$ and use the boundary terms in order to 
	produce a contradiction. 
	
	The first main step is carried out by splitting the integral on the left hand side of 
	Eq. \eqref{eq:weakderPsi} into 
	integrals over $\Omega_1 = (-\infty,0) \cross \R$, $\Omega_2=(0,\infty)\cross (0, \infty)$ 
	and $\Omega_3=(0,\infty)\cross (-\infty,0)$. We then use Fubini's theorem 
	to rewrite each into an iterated integral 
	(cf. Thm. 4.5 in Ref. \cite{brezisFunctionalAnalysisSobolev2011}). 
	
	We shall only provide a full argument for the integral over $\Omega_3$ --
	which is the most illustrative one -- the 
	argument for the other ones is analogous: 
	
	We first show that for every $x \in (0, \infty)$ the 
	restriction of $\Psi$ to $\Omega_3$ evaluated at $x$ is in 
	$W^{1,1}( (-\infty,0), \C)$ 
	(cf. footnote \ref{fn:defSobolevW}).  
	Clearly, it is in 
	$L^1( (-\infty,0), \C) \cap C^1((-\infty,0),\C)$, so we only need 
	to show that the derivative is integrable over $(-\infty,0)$: 
	For $r >0$ denote by 
	$y \mapsto \Id_{(-r,-1/r)}(y)$ the indicator function for the interval 
	$(-r,-1/r)$. Now consider the function 
	\begin{equation}
				(-\infty,0) \mapsto \C \colon y \mapsto \partd{\Psi}{y} (x,y) \, 
				\Id_{(-r,-1/r)}(y) \, . 
	\end{equation}
	By the fundamental theorem of calculus, this function is integrable over $(-\infty,0)$. 
	Since $\Psi \evat{\Omega_3} (x, \, . \,)$ remains finite for both classes of solutions  
	as $y$ tends to $0$ from below and the function tends to 
	$0$ as $y \to \infty$, Beppo Levi's theorem yields the result (cf. 
	Thm. 4.1 in Ref. \cite{brezisFunctionalAnalysisSobolev2011}). 
	
	We may now integrate by parts using Cor. 8.10 in 
	Ref. \cite{brezisFunctionalAnalysisSobolev2011}: For all $x \in \left(0, \infty 
	\right)$ we find 
	\begin{equation}
		\int^0_{-\infty} \d y \, \Psi(x,y) \, \partd{\xi}{y} (x,y)
			=  - \int^0_{-\infty} \d y \, \partd{\Psi}{y}(x,y) \, \xi (x,y)
				+ \Phi(x,0) \, e^{2 \pi \iu \mu}  \, \xi (x,0) 
					\, .
		\label{eq:weakderPsiinteriorintegral3}
	\end{equation}
	
	One finds that the integral over $\Omega_2$ also yields two summands within the 
	integral with respect to $x$, while 
	the one over $\Omega_1$ only yields one. We wish to combine the expressions
	to conclude that  
	the left hand side of Eq. \eqref{eq:weakderPsi} equals 
	\begin{equation}
			- \int_{\R^2} \d x \, \d y \, \partd{\Psi}{y}(x,y) \, \xi (x,y) 
				\\
				+ \left( e^{2 \pi \iu \mu } - 1 \right)
				\int_0^\infty \d x  \,   \Phi(x,0) \, \xi(x,0)
				\, . 
				\label{eq:weakderPsiexplicit}
	\end{equation}
	This, however, requires the use of the linearity of $\int_0^\infty \d x$, 
	which is only admissible if the integral over either one of the two 
	summands converges---since we have almost everywhere convergence of the initial  
	integral by Fubini's theorem. 
	
	It is sufficient to show that $R \in L^1 \left((0,\infty),\R\right)$. 
	In full analogy to the respective arguments in the proof of Thm. 
	\ref{Thm:nonintegersol}, this is done by considering the asymptotic behavior 
	at $0$.  For the first class of solutions we use Eq. \eqref{eq:Uto0} 
	to find that, modulo constants, $R(x)$ is asymptotic to $x^{\mu}$ for 
	$\mu \in (-1,0)$ and to $x^{\mu - 2 \mu}=x^{- \mu}$ for $\mu \in (0,1)$. 
	For the second class of solutions the assertion is shown directly, thus proving 
	the above claim. 
	
	Since, all integrals in Eq. \eqref{eq:weakderPsiexplicit} 
	are convergent and equal to the left hand 
	side of Eq. \eqref{eq:weakderPsi}, the first main step is completed. 
	
	To tackle the second main step, we observe that $\chi$ on the right hand side of 
	Eq. \eqref{eq:weakderPsi} exists 
	if and only if there exists a locally 
	integrable function $\chi'$ such that 
	\begin{equation}
		\int_0^\infty \d x  \, \Phi(x,0) \,  \xi(x,0)
		= \int_{\R^2} \d x \, \d y \, \chi'(x,y) \, \xi(x,y)   
		\label{eq:weakderPsicontradictionequation}
	\end{equation}
	for all $\xi$. The aforementioned contradiction is now 
	obtained by a more or less standard argument: 
	
	First, consider a sequence of mollifiers 
	$x \mapsto \eta_n(x)$ in $C_0^\infty(\R,\R)$, as given on 
	p. 108 sq. in Ref. \cite{brezisFunctionalAnalysisSobolev2011}. 
	Second, take the sequence $(\zeta_k)_{k \in \N}$ of bump functions 
	from the proof of Cor. \ref{Cor:hydrogenvdistder}. 
	Then, for arbitrary $x_0 \neq 0$ set 
	$\xi_{n k}(x,y) = \eta_n(x_0-x) \, \zeta_k(y)$ so 
	that each $\xi_{n k}$ is in $C_0^\infty(\R^2,\C)$. 
	
	We will show that for all $x_0 \neq 0$ we have 
		\begin{equation}
		 	\lim_{n \to \infty} \lim_{k \to \infty} 
		 	\int_0^\infty \d x  \, \Phi(x,0) \,  \xi_{n k}(x,0)
		 	= \Phi(x_0,0) \, . 
		 \end{equation} 
	Due to $\zeta_k(0) \equiv 1$, the left hand side equals 
	\begin{equation}
		\lim_{n \to \infty} \int_{x_0}^\infty \d x \, R(x) \, \eta_n(x_0 -x) 
		= \lim_{n \to \infty} \int_{-\infty}^{x_0} \d z \, R(x_0-z) \, \eta_n(z) \, .
	\end{equation}	 
	We now follow Prop. 4.21 in Ref. \cite{brezisFunctionalAnalysisSobolev2011}:  
	Since $R$ is continuous at $x_0$, for every $\varepsilon >0 $ there 
	exists a $\delta>0$ such that for all $z$ with $\abs{z}$ less than $\delta$ 
	(and $x_0$) we have 
	\begin{equation}
		\abs{R(x_0-z)-R(x_0)} < \varepsilon \, . 
			\label{eq:continuityRx0}
	\end{equation}
	Now choose any natural number $n_0 > 1/\delta$. Then for any $n \geq n_0$ we find 
	\begin{align}
		\abs{\int_{-\infty}^{x_0} \d z \, R(x_0-z) \, \eta_n(z) - R(x_0)}
			&= \abs{\int_{-\infty}^{x_0} \d z \, \bigl( R(x_0-z)  - R(x_0) \bigr) \, \eta_n(z) } 
				\\
			&\leq  \int_{-1/n}^{1/n} \d z \, \abs{ R(x_0-z)  - R(x_0) } \, \eta_n(z)
			\, . 
	\end{align}
	Due to Eq. \eqref{eq:continuityRx0} and normalization of $\eta_n$, we conclude that 
	for every $\varepsilon > 0$ there exists an $n_0$ such that 
	for all $n \geq n_0$ the above expression is less than $\varepsilon$. 
	
	Returning to Eq. \eqref{eq:weakderPsicontradictionequation}, we will 
	also show that 
		\begin{equation}
		 	\lim_{n \to \infty} \lim_{k \to \infty} 
		 	\int_{\R^2} \d x \, \d y \, \chi'(x,y) \, \xi_{n k}(x,y) 
		 	= 0 \, . 
		 		\label{eq:weakderpsicontra2}
		 \end{equation} 
	It is enough to consider the limit $k \to \infty$ for fixed $n$. 
	Observe that the absolute value of the integrand is bounded by 
	$\abs{\chi'} \xi_{n 1}$ almost everywhere. 
	Furthermore, for almost every $(x,y) \in \R^2$ we have 
	\begin{equation}
		\lim_{k \to \infty} \chi'(x,y) \, \eta_n(x_0-x) \, \zeta_k(y)
		= 
		\begin{cases}
			\chi'(x,y) \, \eta_n(x_0-x) 
				&, y = 0 \\
			0 &, y \neq 0
		\end{cases} \, . 
	\end{equation}
	That is, in the pointwise limit 
	$k \to \infty$ the functions  $\chi' \, \xi_{n k}$ tend to 
	the zero function almost everywhere. Eq. \eqref{eq:weakderpsicontra2} 
	then follows from the dominated convergence theorem. 
	
	Finally, the contradiction results from choosing an $x_0>0$ for which 
	$\Phi( x_0 ,0) \neq 0$. 
	
	\item We use Eq. \eqref{eq:weakderPsiexplicit}.  
			Since it was the second term that prevented 
			existence of a weak derivatives in that instance, 
			the situations for the functions from 
			Eq. \eqref{eq:realPsisol1} and 
			\eqref{eq:realPsisol2} will be analogous, whenever 
			the respective factor in front does not vanish. 
			For $\Psi$ in Eq. \eqref{eq:realPsisol2} that 
			factor is $\cos (2 \pi \mu) -1$, thus yielding the 
			assertion. 
			
	\item As in point 
			\ref{itm:Prop:weakderPsi2}, 
			we need to look at the zeros of the factor 
			$\sin \left(2 \pi \mu \right)$. This vanishes for 
			$2 \mu \in \Z$. Following the chain of arguments in point 
			\ref{itm:Prop:weakderPsi1} up to Eq. 
			\eqref{eq:weakderPsiexplicit}, one shows that the weak derivative 
			of $\Psi$ with respect to $x$ exists as well. Thus $\Psi$ is 
			indeed weakly differentiable in those special cases and not 
			weakly differentiable otherwise. 				
	\end{enumerate}

\subsubsection*{Proof of Prop. \ref{Prop:solutionSE}}

	\begin{enumerate}[i)]
	\item 
	This is a direct corollary of Thm. \ref{Thm:nonintegersol}. 
	
	\item
	The general idea of proof is that one cannot write a discontinuous 
	function as a finite linear combination of continuous 
	eigenfunctions, which implies that the time evolution of 
	$\Psi_0$ cannot be given by Eq. \eqref{eq:Prop:solutionSEPsit}. 
	
	Define $\Phi_{n_1 n_2}$ as in Eq. \eqref{eq:2isooschatHeigenfunctions} 
	with corresponding energy eigenvalues $E_{n_1 n_2}$. Setting 
	\begin{equation}
		a_{n_1 n_2} = \inp{\Phi_{n_1 n_2}}{\Psi_0} \, , 
	\end{equation}
	the time evolution of $\Psi_0$ is given by 
	\begin{equation}
		\tilde{\Psi}_t = 
		\sum_{n_1, n_2 \in \N_0} a_{n_1 n_2} \, e^{- \iu E_{n_1 n_2} \,  t/ \hbar } \, 
		\Phi_{n_1 n_2} \, . 
	\end{equation}
	
	We first show that there are infinitely many $(n_1,n_2) \in \N_0^2$ 
	for which $a_{n_1 n_2} \neq 0$: 
	Aiming for a contradiction, assume there are only 
	finitely many. It follows that $\tilde{\Psi}_0$ 
	is continuous on $\R^2$. Yet, by construction, $\Psi_0$
	and $\tilde{\Psi}_0$ belong to the same equivalence 
	class in $L^2(\R^2,\C)$, and continuous representatives are 
	unique. Therefore, $\tilde{\Psi}_0$ is a continuous extension 
	of $\Psi_0$ to $\R^2$---an impossibility. 
	
	In the final main step of the proof we also aim for a contradiction: 
	Assume that for all $t > 0$ the functions $\Psi_t$ and 
	$\tilde{\Psi}_t$ belong to the same equivalence class. Consider 
	\begin{equation}
		\inp{\Phi_{n_1 n_2}}{\Psi_t} 
		= a_{n_1 n_2} \, e^{- \iu E t / \hbar} \, . 
	\end{equation}
	Due to linear independence of the $\Phi_{n_1 n_2}$, our assumption implies 
	that 
	\begin{equation}
		0 = \left( e^{- \iu E t / \hbar} - e^{- \iu E_{n_1 n_2} \, t /\hbar} 
		\right) a_{n_1 n_2}
		\label{eq:solutionSEaandErelation}
	\end{equation}
	for all $n_1$, $n_2$ $\in \N_0$. But unless $a_{n_1 n_2} = 0$, 
	we must have 
	\begin{equation}
		E \in E_{n_1 n_2} + 2 \pi \hbar \, \Z \, / t \, . 
	\end{equation}
	As this has to hold for all $t > 0$, we find that for any  
	$n_1$, $n_2$ $\in \N_0$ either $a_{n_1 n_2} = 0$ or 
	$E = E_{n_1 n_2}$. In particular, there must be infinitely many 
	$(n_1,n_2)$ for which $E = E_{n_1 n_2}$. 
	However, from the expression for $E_{n_1 n_2}$ we 
	find that $E$ is equal to $E_{n_1 n_2}$ at most finitely often---the 
	desired contradiction.
	\end{enumerate}

\subsubsection*{Proof of Lem. \ref{Lem:domHoscillator}}

	Denote the subspace in Eq. \eqref{eq:domHoscillatorprime} by $V$.  
	For $\Psi \in \dom \hat{H} + V =: \dom \hat{H}'$ we set 
	$\hat{H}' \Psi$ equal to $\hat{H} \Psi$ whenever $\Psi \in \dom \hat{H}$ 
	and we interpret $\hat{H}' \Psi$ in the weak sense whenever 
	$\Psi \in V$. 
	
	$\hat{H}'$ is well-defined: For 
	$\Psi \in \dom \hat{H} \cap V$ consider its energy eigenfunction 
	expansion 
	and apply $\hat{H}'$ as defined for vectors in $V$. 
	Upon recalling that the weak Laplacian is symmetric on 
	$H^2 \left(\R^2,\C\right)$, one finds that the two definitions of 
	$\hat{H}' \Psi$ coincide. 
	Thus $\hat{H}' $ is a well-defined linear extension of $\hat{H}$. 
	
	The assertion also follows from this calculation. 

\addsec*{Acknowledgements}

The authors would like to acknowledge support from The Robert A. Welch Foundation 
(D-1523). M.R. acknowledges additional financial support from the Department of 
Mathematics at Texas Tech University. M.R. would also 
like to thank Kazuo Yamazaki and Markus Penz for helpful discussion, 
Roderich Tumulka for pointing out Eq. 
\eqref{eq:omegaevolution}, as well as Antony Valentini and 
Wesley Mathews for correspondence. 


\printbibliography 

\end{document}